\documentclass[runningheads,natbib]{svjour2}
\usepackage{rawfonts}
\usepackage{longtable}
\usepackage{sidecap}

\smartqed  
\usepackage{graphicx}
\usepackage[figuresright]{rotating}

\journalname{Astronomy and Astrophysics Review}
\def\subsun{\mbox{$_{\normalsize\odot}$}}
\def\deg{\hbox{$^\circ$}}
\def\sun{\hbox{$\odot$}}

\def\lesssim{\mathrel{\hbox{\rlap{\hbox{%
 \lower4pt\hbox{$\sim$}}}\hbox{$<$}}}}
\def\gtrsim{\mathrel{\hbox{\rlap{\hbox{%
\lower4pt\hbox{$\sim$}}}\hbox{$>$}}}}

\makeatletter
\def\numberson{\def\@biblabel##1{##1.}%
\let\@bibsetup\NAT@bibsetnum}%

\newcommand\Bibpunct[7][, ]%
  {\gdef\NAT@open{#2}\gdef\NAT@close{#3}\gdef
   \NAT@sep{#4}\global\NAT@numbersfalse\ifx #5n\global\NAT@numberstrue
   \else
   \ifx #5s\global\NAT@numberstrue\global\NAT@supertrue
   \fi\fi
   \gdef\NAT@aysep{#6}\gdef\NAT@yrsep{#7}%
   \gdef\NAT@cmt{#1}%
   \global\let\bibstyle\@gobble
   \NAT@set@cites
   }

\renewcommand\tableofcontents{%
    \section*{\contentsname}%
    \@starttoc{toc}%
    \addtocontents{toc}{\begingroup\protect\small}%
    \AtEndDocument{\immediate\write\@auxout{\string\@writefile{toc}{\endgroup}}}%
    }
\makeatother

\begin{document}
\numberson

\title{Distant Radio Galaxies and their Environments
}


\author{George Miley      \and Carlos De Breuck
}


\institute{G. K. Miley \at
              Sterrewacht, Leiden University, Postbus 9513, 2300RA Leiden, The Netherlands \\
              Phone: +31 71 5275849\\
              \email{miley@strw.leidenuniv.nl}           
           \and
           C. De Breuck \at
              European Southern Observatory, Karl-Schwarzschild-Strasse 2 D-85748, Garching, Germany\\
              Phone: +49 89 32006613\\
              \email{cdebreuc@eso.org}
}

\date{Received: date}

\maketitle

\begin{abstract}
We review the properties and nature of luminous high-redshift radio galaxies (HzRGs, z $>$ 2) and the environments in which they are located. HzRGs have several distinct constituents which interact with each other - relativistic plasma, gas in various forms, dust, stars and an active galactic nucleus (AGN). These building blocks provide unique diagnostics about conditions in the early Universe. We discuss the properties of each constituent. Evidence is presented that HzRGs are massive forming galaxies and the progenitors of brightest cluster galaxies in the local Universe. HzRGs are located in overdense regions in the early Universe and are frequently surrounded by protoclusters. We review the properties and nature of these radio-selected protoclusters. Finally we consider the potential for future progress in the field during the next few decades. A compendium of known HzRGs is given in an appendix.

\end{abstract}

\newpage
\setcounter{tocdepth}{3}
\tableofcontents
\newpage

\section{INTRODUCTION}
\label{intro}

{\it Martin Harwit. ''Yes. I think that if there were less requirement for theoretical justification for building a instrument that is very powerful, you would then have a better chance of making a discovery."
\\
Bernard Burke. ''The record of radio astronomy is clear. Quasars were not found by the desire to find black holes or by the need to find long-distance cosmological probes."
\\

Proceedings of Greenbank Workshop on Serendipitous Discovery in Radio Astronomy, 1983}
\\
\\
Distant radio galaxies are among the largest, most luminous, most massive and most beautiful objects in the Universe. They are energetic sources of radiation throughout most of the electromagnetic spectrum. The radio sources are believed to be powered by accretion of matter onto supermassive black holes in the nuclei of their host galaxies.
Not only are distant radio galaxies fascinating objects in their own right, but they also have several properties that make them unique probes of the early Universe.

\subsection{Scope of this article - HzRGs}
\label{scope}
Our review will be restricted to galaxies that have redshifts, $z > 2$ and radio luminosities at 500 MHz (rest frame) L$_{500(rest)}$ $>$ 10$^{27.5}$W Hz$^{-1}$. Radio emission from such objects has a steep nonthermal spectrum, is collimated and is usually extended by tens of kiloparsec. There are very few known z $>$ 2 radio galaxies associated with compact flat-spectrum radio sources.

This definition is somewhat arbitrary. Luminous distant radio galaxies have properties that are different from less powerful radio galaxies at low redshifts and the properties of radio galaxies change gradually with luminosity and redshift. Likewise, z $>$ 2 quasars that are associated with steep-spectrum extended radio emission have many similar properties to our distant radio galaxies (see Section \ref{qso}) and are located in similar environments. However, without some restriction, our task would not have been tractable.

For conciseness, we shall frequently refer to such high-redshift radio galaxies as ''HzRGs". \citet{mcc93} gave an extensive review of this topic in 1993. Since then there have been several workshops on high-redshift radio galaxies, whose proceedings have been published, including Amsterdam \citep{rot99}, Leiden \citep{jar03b} 
and Granada \citep{vil06b}. Thousands of refereed papers dealing with distant radio galaxies and related topics have been published during the last fifteen years. Although our goal is to be comprehensive, some personal bias will inevitably have entered in selecting the topics and literature. Our apologies for this.

The structure of this review is as follows. We shall first set the scene by reviewing the history of the field. Then we shall describe techniques used for finding HzRGs and discuss how HzRGs are distributed in redshift. Following an overview of the diverse constituents of HzRGs, we shall discuss each HzRG component in detail. The nature of HzRGs and their role in the general evolution of galaxies will then be covered. We shall present evidence that HzRGs are the massive progenitors of dominant cluster galaxies in the local Universe and that they are located in forming galaxy clusters. The properties of these radio-selected protoclusters are then reviewed. Finally, we shall discuss some of the exciting prospects for future HzRG research. An appendix includes a list of known HzRGs at the time of writing (October 2007).

\subsection{History}
\label{history}

The study of distant radio galaxies has progressed in several phases:\\
\\
{\it (i) Infancy. Mid-1940s to mid-1960s:} After the discovery that Cygnus A is associated with a faint distant galaxy \citep{baa54}, radio astronomy became one of the most important tools of observational cosmology. Redshifts of up to z = 0.45 were measured for galaxies associated with radio sources \citep{min60}. Two fundamental breakthroughs were made towards the end of this period. First, quasi-stellar radio sources or ''quasars" were discovered, with much larger redshifts even than radio galaxies.  Secondly, it was demonstrated that the space density of radio sources varies with cosmic epoch, an observation that sounded the death knell for the Steady State cosmology \\
\\
{\it (ii) Childhood. Mid-1960s to mid-1980s:} This can be termed the "Spinrad era". Hy Spinrad devoted a considerable amount of his time, and that of the 120-inch telescope at Lick, to measuring the redshifts of faint radio galaxies by means of long photographic exposures. With considerable effort, he pushed the highest radio galaxy redshift out to z $\sim$ 1 \citep{spi76,spi77,smi80}. At that time, it was not realised that radio sources interact with their optical hosts and the different wavelength regimes were usually studied in isolation from one another. Radio sources were regarded as interesting high-energy exotica that were useful for pinpointing distant elliptical galaxies. They were not believed to play an important role in the general scheme of galaxy evolution. During this period the paradigm that radio sources are powered by accretion of matter onto rotating black holes was developed.\\
\\
{\it (iii) Teens. Mid-1980s to mid-1990s:} The replacement of photographic techniques by CCDs revolutionised optical astronomy. Large numbers of very distant radio galaxies were discovered with redshifts out to z $\sim$ 5 (HzRGs). The surprising discovery that the radio sources and the optical host galaxies are aligned showed that there must be considerable interaction between the radio sources and their host galaxies. The role that orientation could have in determining observed properties was realised.''Orientation unification" that explained differences in the observed types of active galaxies as merely due to the viewing angle of the observer became a popular interpretative ''do-it-all".\\
\\
{\it (iv) Maturity. Mid-1990s to the present:} New efficient techniques for finding distant galaxies were developed. These included photometric searches for
''dropout" objects with redshifted Lyman break features and narrow-band searches for objects with excess redshifted Lyman $\alpha$ fluxes. The advent of these new techniques placed the cosmological use of radio galaxies in a new perspective. Although radio astronomy lost its prime position as a technique for finding the most distant galaxies, it became clear that luminous radio galaxies play a highly important role in the evolution of galaxies and the emergence of large scale structure. The discovery of a relation between the masses of elliptical galaxies and the inferred masses of nuclear black holes led to the conclusion that all galaxies may have undergone nuclear activity at some time in their histories. This brought nuclear activity into the mainstream of galaxy evolution studies.
\\

\subsection{Hunting for HzRGs}
\label{findhzrg}

Appendix~A contains a compendium of 178 radio galaxies with z $>$ 2 known to the authors. This list is the result of more than two decades of hard work. The relatively small number of known HzRGs illustrates both their rarity and the difficulty of detecting them.

Finding distant radio galaxies involves a multistage process.

\begin {itemize}

\item {The first step \citep[e.g.][]{rot94,blu98,deb00,deb04a} is to filter out probable HzRGs from the huge number of radio sources contained in low-frequency radio surveys \citep[e.g.][]{con98}. The usual criteria used for filtering likely candidates are (i) radio colours - extremely steep spectra (Section \ref{alphaz}) and (ii) small angular sizes (Section \ref{sizez}).}

\item {In the second step, relatively bright nearby objects are discarded, by comparing the radio positions with existing wide-field shallow surveys at optical and infrared wavelengths.}

\item {The third step is to refine the positions of the remaining radio sources to arc-second accuracy, thereby facilitating their identification with faint galaxies. Until the 1990s, radio identifications were made using {\it optical} CCD data \citep[e.g.][]{rot95,mcc96}. However, after sensitive near-IR detectors became available, the near-IR $K-$band was found to be more efficient for identifying distant radio sources \citep[e.g.][]{jar01b,deb02,deb04a,jar04,cru06}. More than 94\% of bright radio sources are identified down to $K$=22. Furthermore, the $K-$band magnitude of an HzRG provides a first indication of its redshift, by means of the $K-z$ relationship (see Section \ref{oldpop}). }

\item {The fourth step is to carry out spectroscopic observations of the HzRG candidates with large optical and infrared telescopes. These are needed to determine redshifts from the wavelengths of HzRG emission lines \citep[e.g.][]{rot97,deb01,jar01a,deb06,bor07} (see Section \ref{warmgas}).}

\end{itemize}

The main limitation on finding HzRGs has been the scarcity of available observing time on large optical/IR telescopes for the last stage in the process. Because candidates are located too far apart in the sky to permit multi-object spectroscopy, tedious long-slit spectroscopy on many individual fields is required to determine the redshifts.

Although radio selection ensures that there is no a priori selection against dust properties (e.g. extinction), there are other observational selection effects that introduce bias into the redshift distributions of HzRGs and the list given in the appendix. The first obvious source of bias is that the determination of a spectroscopic redshift is dependent on being able to observe bright emission lines with ground-based telescopes. The primary line for such redshift measurement \citep[e.g.][]{mcc93} is Ly$\alpha$ $\lambda 1216 \AA$
(typical equivalent widths of several hundred $\AA$). Useful additional lines are CIV $\lambda 1549 \AA$, HeII $\lambda 1640 \AA$ and CIII] $\lambda 1909 \AA$ (equivalent widths in excess of $\sim 60 \AA$). There is a so-called ''redshift desert'' $1.2 < z < 1.8$, \citep[e.g.][]{cru06}), where [OII] $\lambda 3727 \AA$ is too red and Ly$\alpha$ $\lambda 1216 \AA$ is too blue to be easily observed from ground-based optical telescopes. Objects located within the redshift desert are thus under-represented in samples of HzRGs.

A second source of bias is redshift incompleteness. A small fraction of radio sources, ($\sim$ 4\%) are not identified to K $\sim$ 22 and about a third of those with K-band identifications do not show any emission or absorption lines, even after long exposures on 8 - 10m class optical/IR telescopes. These objects either (i) have emission lines that are hidden by substantial dust obscuration \citep[e.g.][]{deb01, reu03b}, (ii) are located at such large redshifts that Ly$\alpha$ and other bright emission lines fall outside the easily observable spectral windows, or (iii) are peculiar in that they radiate no strong emission lines.

\subsection{Redshift distribution of HzRGs}
\label{space}

HzRGs are extremely rare.
Luminous steep-spectrum radio sources associated with HzRGs have typical luminosities of L$_{2.7 GHz}$ $>$ 10$^{33}$ erg s$^{-1}$ Hz$^{-1}$ ster$^{-1}$. The number density of radio sources with this luminosity in the redshift range $2 < z < 5$ is a few times $10^{-8}$ Mpc$^{-3}$, with large uncertainties \citep{dun90,wil01,ven07}.

Although HzRGs are sparsely distributed in the early Universe, such objects are almost nonexistent at low redshifts.
The co-moving space density of luminous steep-spectrum radio sources increases dramatically by a factor of 100 - 1000 between $0 \leq z < 2.5$ and then appears to flatten out\citep{wil01,jar01}.  Although there are huge uncertainties in the evolution of the luminosity function at higher redshifts, no significant cut-off in space density has yet been observed.

The redshifts at which the space density of radio sources is maximum correspond to a crucial era in the evolution of the Universe. It coincides with the epoch when (i) luminous quasars also appear to have had their maximum space density  \citep[e.g][]{pei95,fan01b} (ii) star formation was rampant and more than an order of magnitude larger than the present \citep{mad98,lil96,sch05} and (iii) galaxy clusters were forming, but were not yet gravitationally bound structures.

The question of why the most luminous radio sources (and the most luminous quasars) became extinct in the local Universe is an intriguing one that is still not understood (Section \ref{dinosaur}).

\begin{table}[htbp]
\label{table: constituents}
\begin{center}
{\bf{Table 1 \\ [2ex]
Constituents of distant radio galaxies}}
\label{diagtable}
\\ [3ex]
\begin{tabular}{|p{80pt}|p{70pt}|p{95pt} p{20pt}|p{30pt}|}
\hline
\textbf{Constituent}&
\textbf{Observable}&
\textbf{Typical Diagnostics}&
\textbf{Refs.}\footnotemark[1]&
\textbf{Mass}
\par (M$_\odot$)
\\
[1ex]
\hline
\hline
Relativistic plasma&
Radio continuum&
Magnetic field, age, energetics, pressure, particle acceleration. Jet collimation and propagation&
1,2 &
\\
\cline{2-4}
 &
X-ray continuum&
Magnetic field, equipartition, pressures&
3,4,1 &
  \\
\hline
\hline
Hot ionized gas \par T$_{e}$ $\sim$ $10^{7}$--10$^{8}$K \par n$_{e}$ $\sim$ 10$^{-1.5}$cm$^{-3}$&
Radio (de)polarisation&
Density, magnetic field, &
1 &
$^{}$ \par 10$^{11 - 12}$
\\
\cline{2-4}
 &
X-rays&
Temperature, density mass&
&
  \\
\hline
Warm ionized gas \par T$_{e}$ $\sim$ 10$^{4}$--10$^{5}$K \par n$_{e}$ $\sim $ 10$^{0.5 - 1.5}$cm$^{-3}$&
UV-optical \par emission lines \par &
Temperature, density, kinematics, mass, ionisation, metallicity, filling factor&
5,6,7,8&
$^{}$ \par 10$^{9 - 10.5}$
 \\
\cline{2-4}
&
Nebular \par continuum&
SED contamination&
9,10&
\\
\hline
Cool atomic gas \par T$_{s}$ $\sim$10$^{3}$K \par n(HI) $\sim$ 10$^{1}$cm$^{-3}$&
HI absorption&
Kinematics, column densities, spin temperature, sizes, mass&
11,8&
$^{}$ \par 10$^{7 - 8}$
\\
\cline{2-4}
&
UV-optical \par absorption lines&
Kinematics, mass, column densities, metallicity&
8,12  13,14&
\\
\hline
Molecular gas \par T $\sim$ 50 - 500K \par
n(H$_{2}$) $>$ 10$^{2}$ cm$^{-3}$&
(Sub)millimeter lines&
Temperature, density, mass&
15
&
$^{}$ \par 10$^{10-11}$
 \\
\hline
\hline
Dust \par
T $\sim$ 50 - 500K &
UV-optical \par polarisation&
Dust composition, scattering, mass, hidden quasar&
16  17&
$^{}$ \par 10$^{8 - 9}$
\par
\\
\cline{2-4}
 &
(Sub)millimeter continuum&
Temperature, mass, heating source&
18&
  \\
\hline
\hline
Old stars \par t $>$ 1 Gyr&
Optical to near IR \par continuum&
Age, mass, formation epoch&
19 &
$^{}$ \par 10$^{11 -12}$
\\
\hline
Young stars \par t $<$0.5\,Gyr&
UV-optical &
Star formation rates, ages, history&
20,8&
$^{}$ \par 10$^{9 - 10}$
\\
\cline{2-4}
 &
Ly$\alpha $&
Star formation rate&
20 &
  \\
\hline
\hline
Quasar (hidden or dormant)&
UV-optical \par polarisation \par broad lines&
Luminosity&
21,22&
 \\
\hline
Supermassive black hole(SMBH)&
Extended radio, \par
Quasar &
Formation, evolution &
23,24&
$^{}$ \par $\sim$10$^{9}$
 \\
\hline
\end{tabular}
\end{center}
\footnote[1] RReferences:
{1 = \citet{mil80}, 2 = \citet{kla06}, 3 = \citet{fel66}, 4 = \citet{sch02}, 5 = \citet{ost06}, 6 = \citet{gro04a}, 7 = \citet{gro04b}, 8 = \citet{dop03}, 9 = \citet{all84}, 10 = \citet{dic95}, 11 = \citet{mor06}, 12 = \citet{van97}, 13 = \citet{bin00}, 14 = \citet{bin06}, 15 = \citet{dow93}, 16 = \citet{cim93}, 17 = \citet{ver01a}, 18 = \citet{reu04}, 19 = \citet{sey07}, 20 = \citet{mad98}, 21 = \citet{cim93}, 22 = \citet{ver01a}, 23 = \citet{bla01}, 24 = \citet{pea99}}
\label{table: component}
\end{table}

\subsection{Constituents of HzRGs}
\label{constituent}

Radio galaxies have several distinct emitting components which provide
diagnostics about various physical constituents of the early Universe.
A list of known HzRG building blocks is given in Table 1
, together with a summary of techniques used to study them. Also included are a list of the resultant diagnostics, some useful relevant references and our best estimate for the typical mass of the component in HzRGs. We caution the reader to thoroughly examine the assumptions inherent in deriving each of the diagnostics. Although, these assumptions are needed in order to reach some conclusions about the nature of HzRGs, it is important never to forget that they exist and to be sceptical about the ''houses of cards" that are frequently involved.

Because they are highly luminous and (unlike quasars) spatially resolvable from the ground, most components of HzRGs provide important diagnostic
information about the spatial distributions of processes within HzRGs and their environment. The fact that the different constituents are present in the same objects and that the {\bf {\it interrelationships and interactions between
them}}\ can be studied make distant radio galaxies unique laboratories
for probing the early Universe.

As can be seen in Table 1, several constituents of HzRGs are inferred to be extremely massive, including old stars (up to $\sim$ 10$^{12}$ M$_{\odot}$), hot gas (up to $\sim$ 10$^{12}$ M$_{\odot}$) and molecular gas (up to $\sim$ 10$^{11}$ M$_{\odot}$).

\begin{figure*}
\centering
\includegraphics[width=0.75\textwidth,angle=-90]{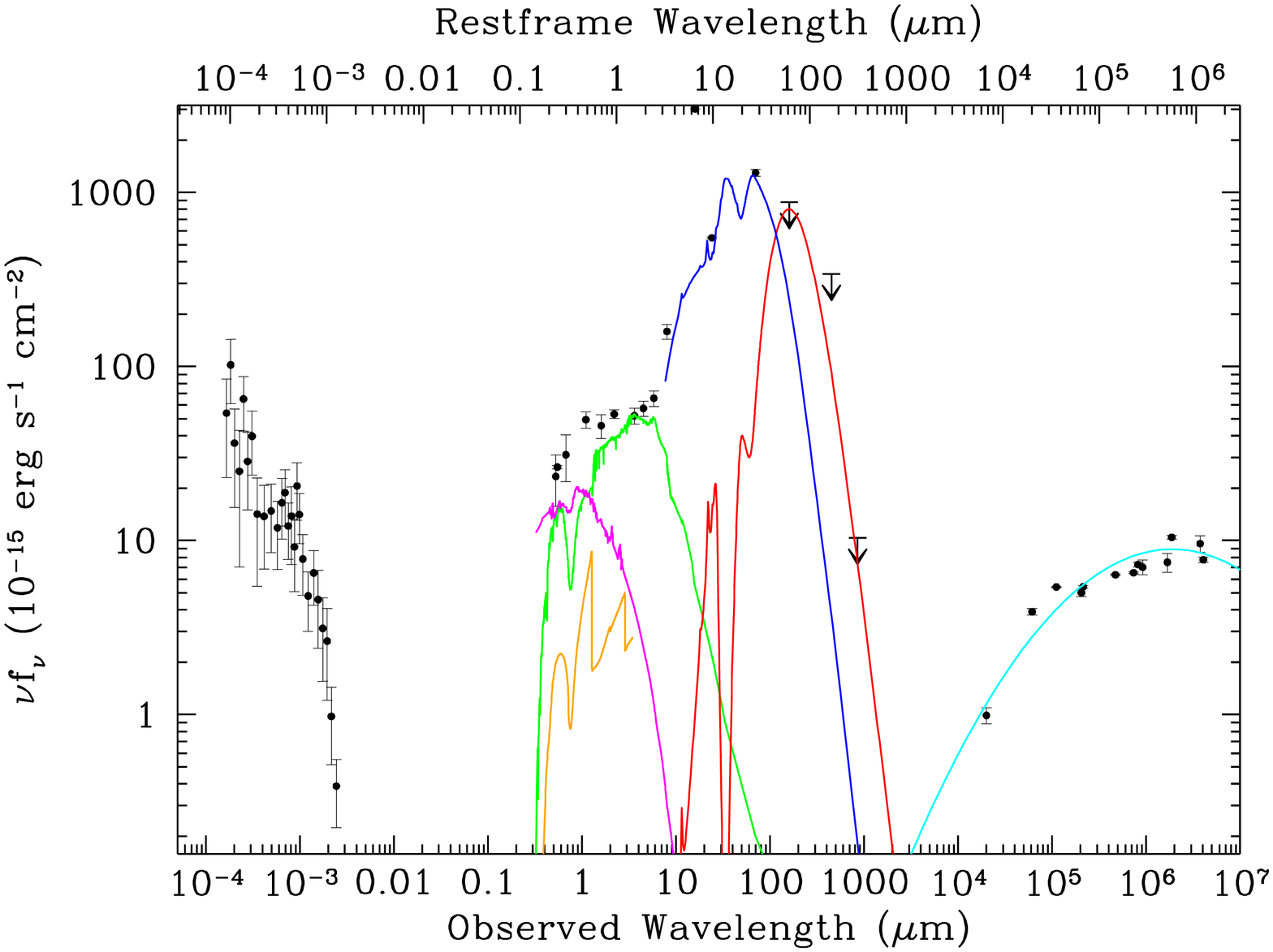}
\caption{Spectral energy distribution (SED) of the continuum emission from the HzRG 4C23.56 at z = 2.5, illustrating 
the contributions from the various constituents. [From De Breuck et al. in preparation]. Coloured lines show the decomposition of the SED into individual components, under many assumptions. 
Cyan = radio synchrotron (Section \ref{relativistic}). Black = Absorbed nonthermal X-ray AGN (Section \ref{ntx}). Yellow = nebular continuum (Section \ref{nebular}). Blue = AGN-heated thermal dust emission (Section \ref{dust}).  Red = Starburst-heated dust emission (Section \ref{dust}). Green = Stars (Section \ref{stars}). Magenta = scattered quasar (Section \ref{agn}). The addition of the overlapping modeled components fits the SED well. 
}

\label{fig: sed}       
\end{figure*}

Figure \ref{fig: sed} shows the spectral energy distribution (SED) of a typical HzRG from radio to X-ray wavelengths, together with a decomposition into various observable HzRG constituents - relativistic plasma, gas and dust, stars and the active galactic nuclei (AGN). We shall discuss each of these building blocks individually in Sections \ref{relativistic} to \ref{agn}. Note from Figure \ref{fig: sed} that disentangling the various components in the optical and the infrared is difficult and the results are often extremely model dependent and several alternative solutions may be equally consistent with the available multi-wavelength data.

\section{RELATIVISTIC PLASMA}
\label{relativistic}

{\it ''All things are double, one against another" Ecclesiasticus (xlvii. 24)
\\
''The problem is how much to stress the sameness of twins and how much to emphasise their differences.", Baby and Child Care, Benjamin Spock, 1946}
\\

Shortly after the discovery of extragalactic radio sources, their nonthermal spectra and polarisations led to the conclusion that their emission is produced by synchrotron radiation from a relativistic plasma. The large radio luminosities together with the lifetimes of the synchrotron-emitting electrons imply total energies of $\geq$ 10$^{60}$ erg \citep[e.g.][]{mil80}.
These huge energies, the collimation on sub-parsec scales and the similarity of orientation between the compact and overall radio structures led to the conclusion that the collimated relativistic beams are produced by rotating supermassive black holes (SMBHs) located at the centres of the host galaxies. The SMBHs are postulated to have rotation axes aligned with the radio source axes.  The radio sources are powered through gravitational energy from material accreting onto the SMBHs, that is converted into kinetic energy of the collimated relativistic jets (See Section \ref{mbh}).

\subsection{HzRGS and low-redshift radio galaxies}
\label{lowz}

Extragalactic radio sources can be classified according to their sizes.

\begin{itemize}

\item {Most radio sources detected in low-frequency radio surveys ($\nu$  $\lesssim$ 2 GHz) are {\it ''extended"}, with projected linear sizes ranging from several tens of kiloparsec to several megaparsecs, i.e. much larger than those of their optical galaxy hosts. Extended extragalactic radio sources have nonthermal spectra, with typical spectral indices at low-redshifts of $\alpha$ $\sim$ -0.7, with flux proportional to $\nu^{\alpha}$. An example is Cygnus A in Figure \ref{fig: specz}.
}
\item{Most radio sources detected in high-frequency radio surveys are {\it ''compact"}, with typical sizes of $<$ 1 pc and relatively flat spectral indices ($\alpha$ $>$ -0.4).
}
\item{A small fraction of sources have linear sizes between $\sim$ 100pc to a few kpc. These sources have peaked spectra that are self absorbed at low frequencies ($<$ a few GHz) and are variously known as ''peakers", gigahertz-peaked spectrum (GPS) sources or compact steep-spectrum (CSS) sources.}

\end{itemize}

Galaxies and quasars host extragalactic radio sources of all three classes, with extended radio sources predominantly identified with galaxies and compact sources mainly associated with quasars.

HzRGs generally host extended extragalactic radio sources and differ in several properties from low-redshift radio galaxies. At larger redshifts the typical radio luminosities increase, the typical sizes decrease (Section \ref{sizez}) and the typical radio spectra steepen (Section \ref{alphaz}). The host galaxies of HzRGs differ from the low-redshift radio galaxy hosts in (i) the presence of emission-line halos (Section \ref{warmgas}, (ii) increased clumpiness of the continuum emission (Section \ref{alignment}) and (iii) alignments of the radio sources with both the emission-line halos (Section \ref{jetgas}) and the UV/optical galaxy continuum emission (Section \ref{alignment}).

\begin{figure*}
\centering
\includegraphics[width=0.75\textwidth]{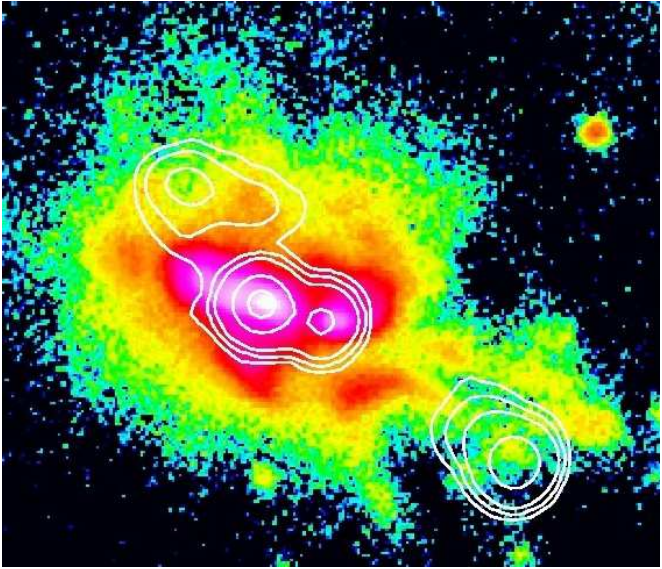}
\caption{Radio structure of 4C41.17 at z = 3.8. [From van Breugel and Reuland, private communication]. See also \citep{reu03a}. Contours obtained at 1.4 GHz with the VLA are superimposed on a Keck narrow-band image in redshifted Ly$\alpha$, showing the warm ionized gas (Section \ref{warmgas}. The radio angular size of $\sim$ 13 arcseconds corresponds to a projected linear size of $\sim$ 90 kpc. The radio spectrum of this object is shown in Figure \ref{fig: specz}c. Figure \ref{fig: 4117rot} illustrated the radio rotation measure of the brightest radio component and Figure \ref{fig: 4117jet} is a higher resolution radio contour map of the central region.
}
\label{fig: 4117radio}       
\end{figure*}

\subsection{Radio sizes and morphologies. Size as an evolutionary clock}
\label{morph}

The radio structures of HzRGs have been studied by \citet{car97} and \citet{pen00b}. Both articles describe observations of 37 radio galaxies with $z > 2$ using the VLA at 4.7 and 8.2 GHz, at resolutions down to 0.25 arcec. In accordance with their large radio luminosities, most HzRGs are ''Fanaroff-Riley Class II''  radio sources \citep{fan74}, with double structures, edge-brightened mophologies and one or more hot-spots located at the extremities of their lobes \citep{mil80,car94,car97,pen00b}. In general HzRGs do not have appreciable flat-spectrum core components at their nuclei, but the use of the ultra-steep spectrum criterion in searching for HzRGs discriminates against finding HzRGs with flat-spectrum cores.  Standard minimum energy assumptions \citep{mil80} give typical minimum pressures in these hotspots of a few x $10^{-9}$ dyn cm$^{-2}$ and corresponding magnetic field strengths of a few hundred $\mu$G.

There have been a few VLBI observations of fine--scale structure in the lobes of extended radio sources associated with HzRGs \citep{gur97,cai02,per05,per06}. A component of size $\sim$ 65 pc has been detected several kpc from the nucleus of 4C41.17 at z = 3.8 \citep{per06}. Consideration of the energetics suggests that the radio component is associated with a gas clump of mass $M_{B}\geq 1.5 \times 10{^{8}} M\odot$. Intriguingly, this is typical for the masses of ''predisrupted clumps" invoked as the progenitors of globular clusters \citep{fal77}. Investigating fine scale radio structure in HzRGs and using the radio jet interactions to probe the interstellar medium in the early Universe are likely to be an important field of study for the more sensitive long-baseline interferometers that are presently under construction, such as e-Merlin and the e-EVN.

There are several correlations between the sizes of radio sources and and other properties \citep{rot00}. First, for 3C sources at lower redshifts, there is a relation with optical morphology.  Smaller radio sources consist of several bright optical knots aligned along the radio axes, while larger sources are less lumpy and aligned \citep{bes98}. Secondly, there is a relation with emission line properties. Smaller radio sources have generally lower ionisations, higher emission line fluxes and broader line widths than larger sources \citep{bes99}. Thirdly, there is a connection with Ly$\alpha$ absorption. HzRGs with smaller radio sizes are more likely to have Ly$\alpha$ absorption than larger sources (Section \ref{lyaabs}) \citep{van97}. Taken together, these correlations are consistent with an evolutionary scenario in which radio size can be used as a ''clock" that measures the time elapsed since the start of the radio activity.

Note that the large sizes of radio sources (usually several tens of kpc) imply that the nucleus of host galaxies has been undergoing activity for at least $>$ 10$^{5}$ y (light travel time across the source) and up to $>$ 10$^{8}$ y (assuming that the jets advance at a few hundred km/s). Hence we know that AGN associated with extended radio sources {\it must be long-lived}. This is not the case for radio-quiet or compact radio quasars, that may only have been active quasars for a few $\times$ 10$^{2}$ y.

\subsection{Radio size vs redshift correlation}
\label{sizez}

It has been known since the nineteen sixties that there is a statistical decrease of the angular size of radio sources with redshift \citep{mil68}. More recent work includes measurement of the angular size-redshift relation for luminous extended radio sources \citep{nil93,nee95,dal02}, quasars \citep{buc98} and compact radio sources \citep{gur99}.

Over the years there have been many valiant attempts to use the angular-redshift relation to derive information about the geometry of the Universe (e.g. determination of ${q_0}$ and ${\Omega_0}$) and even to set constraints on dark energy \citep{pod03}. However, there are many observational selection effects involved. Furthermore, it is difficult to disentangle varying geometry of the Universe from effects due to physical evolution of the radio sources, their host galaxies and the surrounding ambient medium. For example, the sizes of radio sources can be expected to decrease at larger redshifts due to a systematic increase in density of the ambient medium. Also the energy density of the cosmic microwave background increases as $(1 + z)$${^4}$, substantially enhancing inverse Compton losses. This will tend to extinguish radio sources earlier in their lives, making them on average smaller.

In summary, the many non-cosmological effects that influence the angular size-redshift relation have made it impossible to draw robust conclusions about cosmology from such considerations.

\subsection{Radio spectral index vs redshift correlation}
\label{alphaz}

\begin{figure*}
\centering

\includegraphics[width=1.0\textwidth]{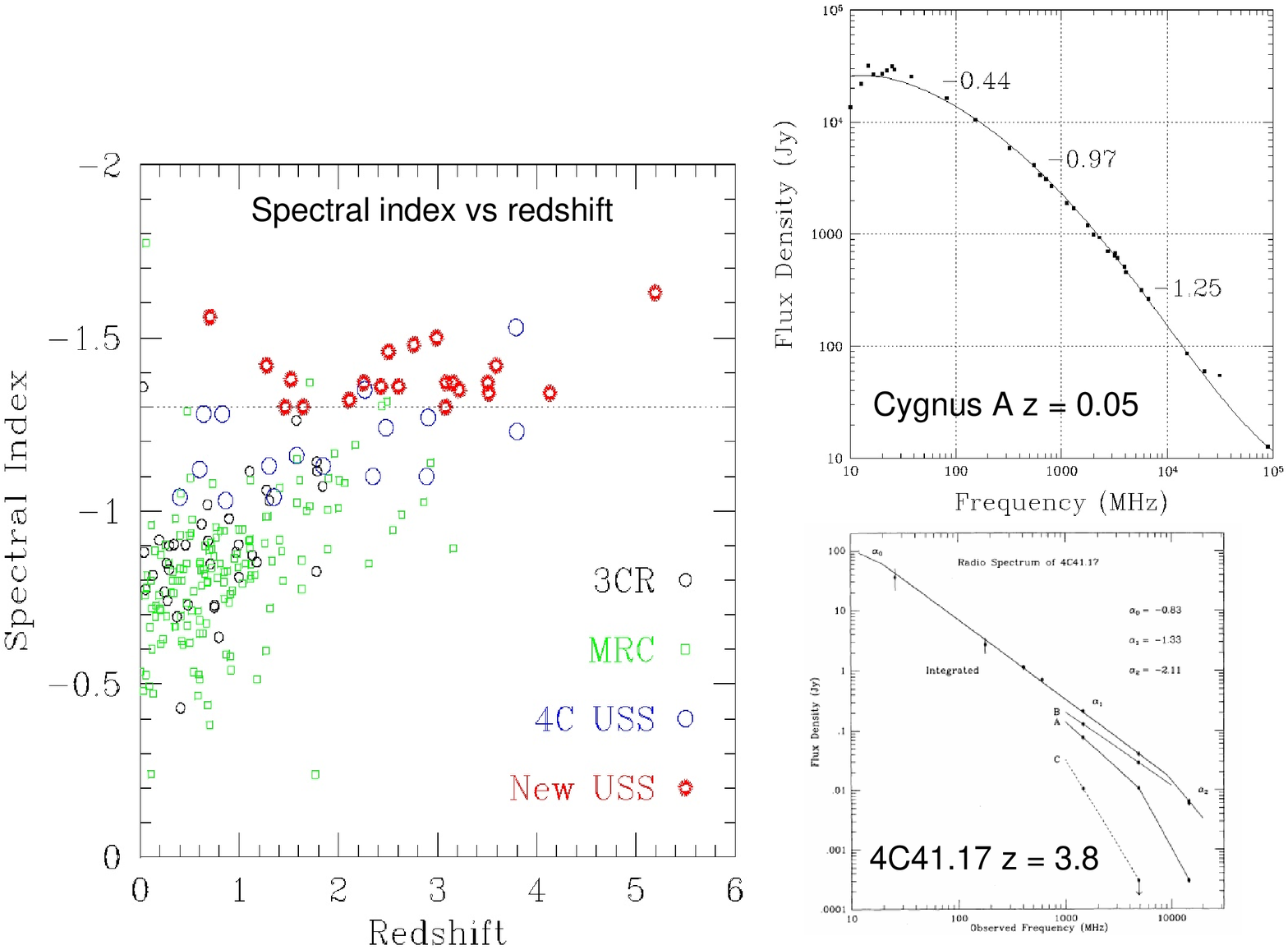}

\caption{Left. Plot of radio spectral index versus redshift, showing that more distant sources have steeper spectra. [From \citet{deb00}]. Above right. Radio spectrum of the luminous radio galaxy Cygnus A at z = 0.05, showing spectral curvature.
Bottom right. Radio spectrum of the HzRG 4C41.17 and its various components. [From \citet{cha90}]. This is still one of the most well determined spectra of a HzRG at low frequencies. Note the absence of significant spectral curvature.
}
\label{fig: specz}       
\end{figure*}

One of the most intriguing properties of the relativistic plasma in HzRGs is the strong correlation that exists between the steepness of radio source spectra and the redshift of the associated radio galaxies \citep{tie79,blu79}. Radio sources with very steep spectral indices at low frequencies $\lesssim$ 1 GHz tend to be associated with galaxies at high redshift (e.g. Figure \ref{fig: specz}). This empirical correlation between radio spectral steepness and redshift has proved to be an efficient method for finding distant radio galaxies. Most known HzRGs (Table 3) have been discovered through following up those radio sources with the steepest ten percentile of radio spectra (spectral index, $\alpha \sim $ -1).

The conventional explanation for the z $\sim \alpha $
correlation is that it is the result of a concave radio spectrum (see Cygnus A in Figure \ref{fig: specz}), coupled with a radio
K-correction. For higher redshifts, observations of sources at a {\it fixed observed}
frequency  will sample emission at {\it higher rest-frame} frequencies where the concave spectrum
becomes steeper.  The most important mechanisms for making the radio spectra concave are synchrotron and inverse Compton losses at high frequencies \citep[e.g.][]{kla06} and synchrotron self-absorption at low frequencies.

Although there is evidence from radio colour-colour plots that systematic concave curvature is present in the radio spectra \citep[e.g.][]{bor07}, such an explanation alone is insufficient to explain the observed correlation.  The radio spectra of many distant luminous radio
galaxies are {\it not} concave at the relevant frequencies. The radio source with the most accurately determined spectrum over a wide frequency range is  4C 41.17 at z = 3.8 (bottom right in Figure \ref{fig: specz}). This source has an extremely straight
spectrum between 40 MHz  and 5 GHz, the relevant frequency range
for the z $\sim \alpha $ correlation \citep{cha90}. Although the spectrum steepens above 5 GHz, this is too high a frequency to contribute to the z $\sim   \alpha $ correlation.
Furthermore, a
recent study by \citet{kla06} showed that 33 of 37 sources in their
SUMSS-NVSS sample have straight and not concave spectra between 0.8 and 18 GHz.

Two alternative effects have been proposed to explain the observed z $\sim $ $\alpha$ relation.
The first possibility is that the z $\sim $ $\alpha$ relation is an indirect manifestation of a luminosity, L $\sim $ $\alpha$ effect \citep{cha90, blu99}.

Classical synchrotron theory predicts that continuous particle injection will result in a
spectrum with a low-frequency cut-off, $\nu_{l}$, whose frequency depends on the source
luminosity, L according to $\nu_{l}$ $\sim$ L$^{-6/7}$.
For a flux-limited sample, Malmquist bias will cause sources at higher redshift to have preferentially larger radio luminosities. Over the relevant frequency range, the L$^{-6/7}$ vs $\nu_{l}$ effect would
therefore result in the observation of an z $\sim$ $\alpha$  relation.
However, it is unlikely that this luminosity - spectrum relation is the correct explanation, or at least  the whole story. \citet{ath98a}
showed that a z $\sim$  $\alpha $ correlation still persists even when samples are restricted to a limited range of L.

A second explanation is that some physical effect causes the spectral index to steepen
with higher ambient density and that the ambient density increases with redshift \citep{ath98a,kla06}.
For example, putting the radio source in a denser
environment would cause the upstream fluid velocity of the relativistic particles to decrease and a first-order Fermi acceleration
process would then produce a steeper synchrotron spectrum. Recently \citet{kla06} pointed out that such a mechanism would
(i) result in both z $\sim  \alpha $ and L $\sim   \alpha $
correlations and (ii) provide a natural physical link between high-redshift
radio galaxies and nearby cluster halos. 

However, it is difficult to produce the observed z $\sim$  $\alpha $ relation from such a simple density-dependent effect alone.
The clumpy UV/optical  morphologies (e.g Section \ref{spider}) indicate that the density of gas around HzRGs is highly non-uniform and that the density is larger close to the nucleus than in the outer regions. Furthermore, the internal spatial variations of spectral index {\it within} individual source is observed to be smaller
than the {\it source to source} variations of the integrated spectral indices \citep[e.g][]{car94}.
If the ambient medium is highly non-uniform,
how can one side of a radio source "know''
that the other side has an uncommon ultra-steep spectrum?
It is therefore likelier that the ultra-steep spectra are produced by some mechanism in which the spectral index is determined within the galaxy
nucleus rather than by the environment at the locations of the radio lobes.

In summary, the origin of the z $\sim$  $\alpha $ effect is still unclear and more detailed information is needed about the  dependence of the radio spectrum on redshift. In the near future accurate measurements of low-frequency radio spectra of HzRGs with new facilities such as LOFAR will be important for such studies.

\subsection{Nonthermal X-ray emission}
\label{ntx}

Although X-ray measurements of HzRGs are sensitivity-limited, significant progress in the field has been made in the last decade using the {\it{Chandra}} and {\it{XMM}} X-ray telescopes.
Extended X-ray emission has been detected from about a dozen high-redshift radio galaxies and radio-loud quasars \citep{car02,fab103,fab203,sch03,yua03,bel04,ove05,blu06,erl06,joh07}. The extended X-ray emission is typically elongated in the direction of the radio source, indicating that there is some physical link between the X-ray emission and the relativistic plasma.

Several mechanisms have been proposed for producing the extended X-ray emission. The one that is most widely suggested inverse Compton scattering (IC) of the cosmic microwave background \citep{fab203,sch03, bel04,erl06, joh07}. Because the density of CMB photons increase as $(1 + z){^{4}}$, IC scattering of the CMB becomes increasingly important at high redshift. Under the assumption that the X-ray emission is due to this process, comparison of the radio and X-ray luminosities \citep{fel69} yields magnetic field strengths consistent with equipartition \citep{bel04,ove05,joh07}. Because the radiative lifetimes of radio synchrotron-emitting electrons are shorter than the lifetimes of the X-ray emitting IC electrons, the IC emission traces older particles. In 4C 23.56 at z = 2.48, the X-ray emission is observed to extend by $\sim$ 500kpc \citep{joh07}, implying an energy in both relativistic and IC-emitting electrons of $\geq$ $10{^{59}}$ erg, an energy reservoir equivalent to $\sim$ $10{^{8}}$ supernovae.

Other processes that have been invoked to produce the extended X-ray emission include inverse-Compton up-scattering of synchrotron photons in the jet (synchrotron self-Compton emission - SSC) \citep{sch03} and thermal emission from shocks \citep{car02,bel04} (see Section \ref{hotgas}).

\section{GAS AND DUST}
\label{gas}
{\it ''Sometimes you can't stick your head in the engine, so you have to examine the exhaust"
Donald E. Osterbrock}
\\
\\
Gas and dust, in various states, are important constituents of HzRGs and must play a fundamental role in their evolution. Gas has been observed over a wide range of temperatures and having a variety of forms.

\subsection{Hot ionised gas - Radio depolarization}
\label{hotgas}

X-ray observations show that hot ionized gas is widespread at the centres of clusters and around radio galaxies at low redshifts.  Typical derived parameters are temperatures of $\sim$${10^{7.5}}$ K, densities of 0.05 cm$^{-2}$, and masses of a few x $10^{12}$ cm${{-3}}$, with pressures of $\sim$ $10^{-9}$ dyne cm${^{-2}}$. Such a gas would be sufficient to confine both the radio-emitting plasma and the Ly$\alpha$ halos (see Section \ref{warmgas}) of HzRGs. For low and intermediate redshift clusters, bubble-like structures in X-rays have been attributed to energy injection into such a hot intracluster medium by radio jets \citep[e.g.][]{mcn00}.

\begin{SCfigure*}
\centering
\includegraphics[width=0.6\textwidth]{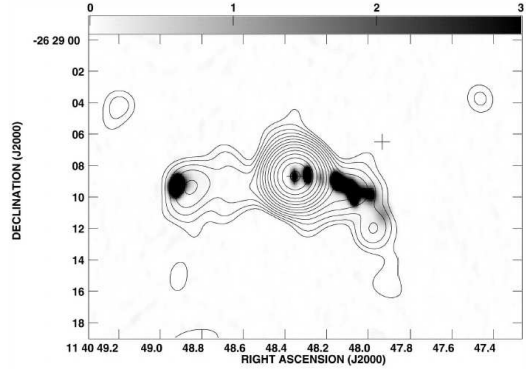}
\caption{X-ray emission from the Spiderweb Galaxy, PKS 1138-262 at z = 2.2, observed with the Chandra X-ray Telescope [From \citet{car02}]. X-ray contours are superimposed on a VLA gray scale representation of the 5 GHz radio continuum emission at 0.5'' resolution. The cross marks the position of the radio galaxy nucleus. Note that the X-ray and radio emission are aligned with each other. See also Figure \ref{fig: spider}.}
\label{fig: 1138x}       
\end{SCfigure*}

The sensitivity of the present generation of X-ray telescopes is only marginally sufficient for detecting similar hot gas at z $\sim$ 2. X-ray emission has been observed from the Spiderweb Galaxy at z = 2.2 (Figure \ref{fig: 1138x}) \citep{car02}. The X-ray emission is extended along the radio source axis and  attributed to thermal emission produced by shocks as the synchrotron-emitting relativistic jet that propagates outwards.

An alternative technique for detecting and studying this hot gas is through its effect on the polarisation of background radio emission \citep[e.g.][]{mil80}. As it propagates through a magnetoionic medium the polarisation of linearly polarised radiation is rotated through an angle $\chi$ that is proportional to the square of the wavelength, $\lambda$ (cm).

$\Delta\chi$ = 5.73 $\times$ 10$^{-3}$ RM $\lambda^{2}$, where R is the observed rotation measure, given by RM is RM = 812$\int{n_{t}}$(s)$B_{\parallel}$(s)ds $\sim$ 8.1 $\times 10^{8}$ n$_{t}$ $B_{\parallel}$s rad  m$^{-2}$.

Here s(kpc) is the path length through the medium, n$_{t}$ (cm$^{-3}$) is the density of thermal electrons and B$_{\parallel}$
is the component of magnetic field along the line of sight.
The intrinsic rotation measure is RM = RM$_{obs}$ $\times$ (1 + z)$^{2}$

\begin{SCfigure*}[][t]
\centering
\includegraphics[width=0.6\textwidth]{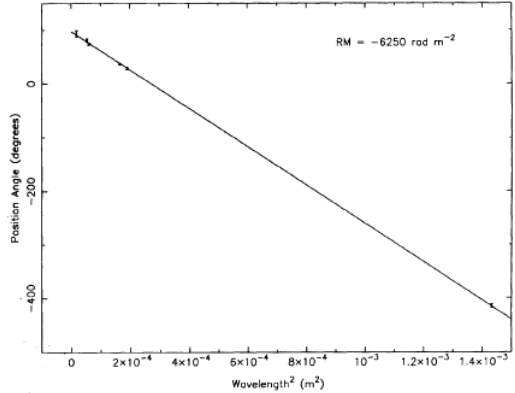}
\caption{The observed polarisation position angle as a function of the square of the wavelength$^{2}$ for a radio knot of 4C41.17 at z = 3.8. [From \citet{car94}]. The knot is the brightest of the 4 radio knots shown in Fig. \ref{fig: 4117radio}. Note the excellent fit to a $\lambda$$^{2}$ dependance and the large resultant rotation measure for the knot.
}
\label{fig: 4117rot}       
\end{SCfigure*}

Observations of radio polarisation data for $>$ 40 HzRGs \citep{car94,car97,ath98b,pen00b,bro07} revealed that several sources have extremely large rotation measures (RMs), up to more than 1000 rad m$^{-2}$ in the rest frame\citep{car97,ath98b} (e.g. Figure \ref{fig: 4117rot}), with considerable variation across the sources.  An obvious interpretation of the large observed rotation measures is that HzRGs are embedded in dense gas. At low-redshifts the largest rotation measures are observed for radio sources that are located in X-ray emitting clusters that are inferred to be ''cooling flows" \citep{ge94}. Hence the large rotation measures of HzRGs are strong circumstantial evidence that these objects are also located in cluster environments at high redshifts (Section \ref{Protoclusters}).

Assuming that (i) the rotation is produced over a line-of-sight path length that is comparable with the projected dimensions of the radio sources and (ii) the magnetic field strength is close to the equipartition values, the derived electron densities are $\sim$ 0.05 cm$^{-3}$, comparable to those obtained for low-redshift clusters from the X-rays.

\subsection{Warm ionized gas}
\label{warmgas}

The bright emission lines emitted by luminous radio galaxies allow their redshifts to be determined and have been an important reason why radio galaxies have been such important cosmological probes.  The emission lines are emitted by warm (10$^{4.5}$K) gas and provide a powerful tool for studying physical conditions within the HzRGs.

Emission lines observed from active galactic nuclei are both ''permitted" recombination lines and collisionally-excited ''forbidden" lines. When a free electron is captured by an atomic nucleus it cascades to the ground state emitting a series of recombination lines. The most prominent recombination lines are due to the most common elements, hydrogen and helium. However, some of the strongest lines in the spectra of emission nebulae correspond to so-called ''forbidden" transitions that have miniscule probability of occurring relative to the permitted transitions, but whose energies lie within a few kT of the ground levels and can therefore be easily populated by collisions. Each element has a critical density below which most de-excitations are radiative and a line is produced. Forbidden lines are only produced by gas with densities in the range $\sim$ 10 - 10$^{5}$ cm$^{-3}$ and the line ratios contain important information about the density, temperature, ionisation and abundances of the emitting gas \citep{ost06,pet97}.

A composite emission-line spectrum for HzRGs, and a list of emission line strengths were given by \citet{mcc93}. During the early nineties it was found that, for high-redshift radio galaxies, nuclear emission lines are generally accompanied by an additional component that is highly extended spatially.
The presence of giant luminous ionized gas nebulae are amongst the most remarkable features of HzRGs \citep[e.g.][]{reu03b}. These ''halos" have sizes of up to $\sim$ 200 kpc (e.g Fig \ref{fig: spider}) and their study provides a wealth of information about kinematics and physical conditions within and surrounding the HzRGs and on the origin of the gas.

During the last 15 years considerable progress has been made in studying the properties of Ly$\alpha$ halos, using large ground-based optical/IR telescopes. Observations of their emission line spectra have been extended into the infrared (optical rest-frame), where additional useful physical diagnostics can be obtained. Also, the kinematics of several HzRGs have been mapped in detail.
An excellent recent review of the emission-line properties of HzRGs has been given by \citet{vil07c}. For detailed information the reader is referred to that review and the references therein.

The ionised gas halos have Ly$\alpha$ luminosities of typically $\sim$ 10$^{43.5}$ erg s$^{-1}$ and line widths varying from a few hundred km/s in their outer parts to $>$ 1000 km s$^{-1}$, near the galaxy nuclei. The usual emission-line diagnostics \citep{ost06} show that the gas has
temperatures of T$_{e}$ $\sim$ 10$^{4}$K - 10$^{5}$K, densities of n$_{e}$ $\sim$ 10$^{0.5}$ - 10$^{1.5}$ cm$^{-3}$ and masses of $\sim$ 10$^{9}$ - 10$^{10}$M$_{\odot}$. The warm gas occupies a relatively small fraction of the total volume of the HzRGs, with filling factors estimated to be $\sim$ 10$^{-5}$ compared with unity for the hot gas.
Although the topology of the gas is not well determined, the properties of the emission lines and the covering factor deduced from statistics of absorption lines (Section \ref{lyaabs}) clouds led to a model \citep{van97} in which the halos are composed of $\sim 10^{12}$ clouds, each having a size of about 40 light days, i.e. comparable with that of the solar system. \citet{van97} speculate that the clouds might be associated with the early formation stages of individual stars.

The morphologies of the Ly$\alpha$ halos are clumpy and irregular. Their overall structures are often aligned with the radio axes and sometimes extend beyond the extremities of the radio source. There appear to be two distinct regimes in the halos that often blend together. The inner regions close to the radio jets are clumpy, with velocity spreads of $>$ 1000 km s$^{-1}$. They appear to have been perturbed by the jets. The outer regions are more quiescent, with velocity ranges of a few hundred km s$^{-1}$). They appear to be more relaxed than the inner regions (e.g. Fig. \ref{fig: 0943spec}).

The relative intensities of emission lines are, in principle, powerful diagnostic tools for studying physical conditions in the warm line-emitting gas. However, disentangling the effects of ionisation, abundances, density and temperature using the emission line ratios is complicated and requires detailed modeling, incorporating all facets of the HzRGs. We shall now discuss the physical conditions within the warm gas halos in more detail.

\subsubsection{Ionisation}
\label{ion}

Various mechanisms have been proposed for exciting the gas. These include (i) photoionisation from an AGN, (ii) photoionisation from stars, (iii) photoionisation by ionizing X-rays emitted by shocked hot gas
and (iv) collisional ionization from shocks.
Plots of optical-line ratios have been used extensively to study the ionisation of gas in nearby active galaxies, where evidence for both jet- and accretion-powered shocks and for photoionization by the central AGN has been found \citep[e.g.][]{vil97a,vil97b,bic00,gro04a,gro04b}.

Although these relationships have been most accurately calibrated in the rest-frame optical region of the spectrum, line diagnostic diagrams have also been developed for use in the ultraviolet \citep{all98,gro04b}, the spectral region of HzRGs sampled by optical observations. Interpretation of emission line ratios are complicated by effects of dust and viewing angles \citep{vil96a}. Also most HzRG spectra are spatially integrated over regions of 10 or more kiloparsec, where conditions and sources of excitation may change. In a recent comprehensive study using as many as 35 emission lines throughout the rest-frame UV {\it and} optical spectra \citet{hum06} concluded that photoionization is the dominant source of excitation in the quiescent gas. A harder source of photoionisation than stars is needed, consistent with photons from an AGN.

If a quasar is exciting the warm gas in the halo, why don't we see it? The usual explanation is  that the quasar emits radiation anisotropically. It is highly absorbed in the direction in which we view it, but not along the radio axis. Support for this idea is provided by the large optical polarisations measured for some HzRGs. (Section \ref{dustpol}). However, if highly anisotropic flux from a {\it hidden} quasar is indeed the dominant source of excitation, it is difficult to understand why many of the Ly$\alpha$ halos are approximately symmetric in shape.
An alternative explanation is that the quasar activity is isotropic and highly variable, with short sharp periods of intense activity and longer periods of relative passivity, when the quasar is dormant (see Section \ref{qso}).

Although photoionisation by a quasar is presently the "best bet'' for the dominant source of excitation, it is not likely to be the only culprit. Considerable variation is observed in the relative emission line strengths from object to object and within individual HzRGs. There is strong evidence that there is also collisional excitation from shocks, particularly close to the radio jets - see \citep[e.g.][]{bic00,bes00} and Section \ref{alignment}.

\subsubsection{Abundances and star formation}
\label{abundance}
Although Ly$\alpha$ is by far the brightest emission line emitted by the warm gas halos, other lines are also detected. In 4C41.17 at z = 3.8, kinematical structures in the Ly$\alpha$ line are closely followed in the carbon CIV and oxygen [OII] and [OIII] lines, with the [OIII] $\lambda$5007 emission extending by as much as 60 kpc from the nucleus \citep{reu07}. See also Fig. \ref{fig: 0943spec}

The chemical abundance of the halo gas is close to solar \citep{ver01a,hum06}, consistent with the HzRG having undergone prodigious star formation at earlier epochs. Further indications that the star formation rate in HzRGs was higher in the past come from measurements of the Ly $\alpha$ luminosities and Ly$\alpha$/HeII ratios that are both systematically larger for HzRGs with z $>$ 3, than for those with 2 $<$ z $<$ 3 \citep{vil07a,vil07b}.

The relative intensity of the NV~$\lambda$1240\AA\ line varies from being an order of magnitude fainter than the carbon and helium lines \citep{deb00b} to being as luminous as Ly$\alpha$ \citep[e.g.]{van94}. This has been interpreted as indicating that there are large variations in metallicity between HzRGs.

\subsubsection{Inner Kinematics: Outflows, jet interactions and superwinds}
\label{jetgas}

In considering the kinematics of the ionized gas halos, we shall separately consider the nuclear turbulent regions, where there is evidence for outflows and the outer passive regions where the most likely dominant systematic motions are inwards. The inner regions show shocked gas closely associated with the radio lobes. These display disturbed kinematics and have expansion velocities and/or velocity dispersions $>$ 1000 km s$^{-1}$. 

Besides synchrotron jets there is evidence that starburst ''superwinds" \citep{arm90,zir05} are also present in the inner regions of the Ly$\alpha$ halos. Both the jets and the superwinds will exert sufficient pressure on the warm gas to drive it outwards. Interactions between radio jets and the ambient gas is important in some low-redshift radio galaxies \citep[e.g.][]{hec82,hec84,van84,van85a,van85}. The radio sources are observed to excite, disturb and entrain the gas. Likewise, the gas can bend and decollimate radio jets and enhance the intensity of their radio emission through shock-driven particle acceleration. In general
small radio sources show more jet-gas interactions than large ones \citep{bes00}. At z $>$ 2, HzRG exhibit signatures of even more vigorous jet-gas interactions \citep[e.g.][]{vil98,vil99b,vil03,hum06}. The kinematics is more turbulent and the ionisation is higher in the region of the jet than in the quiescent outer halo \citep{hum06}.

\citet{nes06} recently carried out an important study of the Spiderweb Galaxy with an integral field spectrograph in the optical rest frame. They make a convincing case that there is an accelerated outflow of warm gas in this object. The only plausible source of energy for powering this outflow (few $\times$ 10$^{60}$ erg) is the radio jet and, even then, the coupling between the jet and the ISM must be very efficient to account for the observed kinematics. The pressure in the radio jets can drive gas outwards from the nuclei for tens of kiloparsecs and play an important role, together with starburst-driven superwinds, in ''polluting" the intergalactic medium with metals. There is spectroscopic evidence for the ejection of enriched material in 4C 41.17 at z = 3.8 up to a distance of 60 kpc along the radio axis \citep{reu07}. See also Fig. \ref{fig: 0943spec}.

In Section \ref{alignment} we shall discuss strong evidence that the jet-gas interactions can also trigger star formation in HzRGs.

\begin{figure*}
\centering
\includegraphics[width=0.75\textwidth]{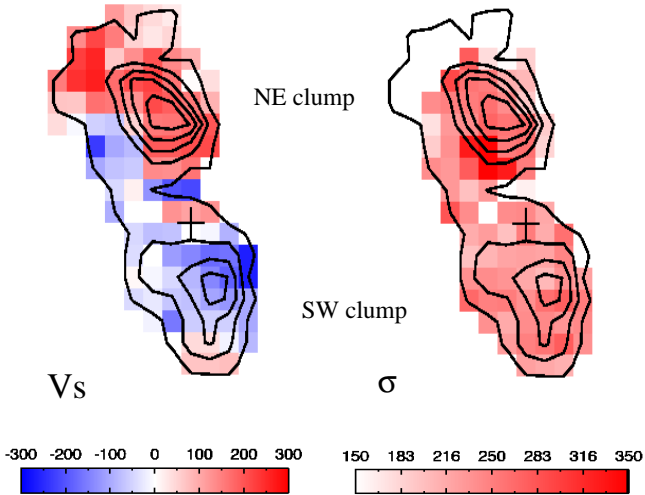}
\caption{
Velocities and velocity dispersions across MRC 2104-242 at z = 2.5 measured with
VIMOS on the VLT. [From \citet{vil07c}].
This HzRG is surrounded by a giant Lya nebula that extends by $\sim$ 120 kpc
The position of the radio core is indicated with a cross.
The velocity field appears symmetric and ordered implying
either rotation or radial motions. See also \citet{van96}.
}
\label{fig: 2104spec}       
\end{figure*}

\subsubsection{Outer kinematics: Infall of the quiescent halos}
\label{infall}

\begin{figure*}
\centering
\includegraphics[width=0.6\textwidth,angle=90]{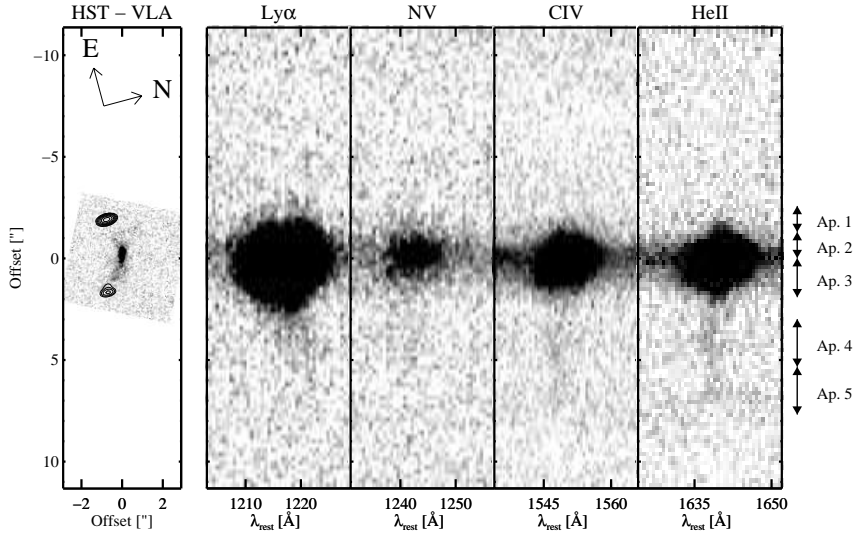}
\caption{
Long slit spectroscopy of the ionized gas halo in MRC 0943-242 at z = 2.9. [From \citet{vil07c,hum06}]. VLA radio map overlaid with the WFPC2 HST image (left), spatially aligned with the 2-dimensional Keck spectra of the main UV rest frame emission lines.
The impact of jet–gas interactions on the observed properties of the giant nebula (Size $\sim$ 70 kpc) can be seen in the much broader and brighter emission lines within the radio structures compared with the faint emission detected beyond.
}
\label{fig: 0943spec}       
\end{figure*}

An important diagnostic in tracing the origin of the warm gas is the kinematics in the outer region of the giant halos, where they are apparently unperturbed by the radio jet. The outer halos displays systematic velocity variations of a few hundred km/s (e.g. see Figure \ref{fig: 0943spec}).
Are these systematic velocity variations the result of rotation \citep{van96,vil03,vil06}, outflows \citep{zir05,nes06} or infalling motions?

It is difficult to discriminate between the various kinematic scenarios from velocity data alone, but comparison of spectroscopic and radio data provides additional relevant information. In a study of 11 HzRGs with redshifts 2.3 $<$ z $<$ 3.6, several correlated asymmetries were found between the halo kinematics and assymetries in the radio structures \citep{hum07,vil07b}. On the side of the brightest radio jet and hot spot, the quiescent nebula appears systematically redshifted (receding) compared with the other side.  On the assumption that the bright radio jet and hot spot are moving towards us and brightened by Doppler boosting \citep{ree67,kel03}, the quiescent gas must be moving {\it inwards}. The brightest radio hotspot is also the least depolarised one, as expected if it is on the closest side of the HzRG and consistent with the infall scenario.

Could the inflowing motion of the gas be a result of cooling flows, that have long been studied in clusters of galaxies at low redshifts, \citep[e.g.][]{fab94,kau06}? Recent XMM and Chandra observations have shown that the cooling rates are reduced by an order of magnitude below the simple
cooling flow models at temperatures $<$ $\times 10^7$K, probably through interaction of the gas with radio sources associated with galaxies in the cluster.
An alternative diagnostic of cooling flows that are more feasible for HzRGs than X-rays is the measurement of roto-vibrational lines from
H$_2$ molecules at $\sim 2000$K \citep{jaf97,jaf05}. At high redshifts these lines are shifted from the near to mid-IR bands, accessible with the Spitzer Telescope. The H$_2$  lines are a promising tool for studying both the cooling in the gas around HzRGs and
the excitation mechanisms.

To summarise, the kinematics of the warm gas is complex. There is evidence that gas in the outer regions of the halo is flowing inwards, providing a source of material for feeding the active nucleus (Section \ref{agn}) and that gas in the inner region is being driven outwards by pressure from jets and starbursts. The various motions in the ionized gas halos are likely to contribute to feedback processes between the AGNs and the galaxies invoked in current models of massive galaxy evolution (Section \ref{mass}). 

We end our discussion of the halo kinematics on a cautionary note. Most kinematic studies of HzRGs until now have been based on Ly$\alpha$, because of its relatively large equivalent width and its accessibility with optical telescopes. However, Ly$\alpha$ is a resonant line and subject to strong 
scattering and optical depth effects. Hence, the resultant kinematics may not be completely representative of the gas as a whole, particularly in the inner regions. 

\subsubsection{Relation to non-radio Ly$\alpha$ nebulae}
\label{blobs}
There may well be a connection between Ly$\alpha$ halos in HzRGs and the disembodied extended Ly$\alpha$ nebulae that have been discovered in recent years \citep[e.g][]{fyn99,ste00,fra01,mat04,dey05,col06,nil06}    These nebulae also have physical extents $\gtrsim$ 100 kpc and Ly$\alpha$ line fluxes of $\sim$ 10$^{-15}$ ergs s$^{-1}$ cm$^{-2}$.
Although in many respects, these Ly$\alpha$ blobs resemble the giant ionised gas halos around HzRGs, they have $<$ 1\% of the associated radio continuum flux and no obvious source of UV photons bright enough to excite the nebular emission. However, millimeter emission has been detected from several of these nebulae \citep{cha01,sma03,gea05} and \citet{mat04} suggest that the extended Ly$\alpha$ nebulae are also associated with dense environments in the early Universe. Just like the HzRG halos, these nebulae can be excited by quasars that are heavily obscured along the line of sight \citep{hai01,wei04,wei05}, or quasars that undergo recurrant flares (Section \ref{qso}). Alternatively, they may be associated with cooling-flow-like phenomena \citep{hai00,dij06a,dij06b}.

\subsubsection{Nebular continuum}
\label{nebular}

\citet{dic95} pointed out that the nebular continuum emission due to the line-emitting warm gas of HzRGs is significant and must be taken into account when computing the various contributions to the UV continuum. See also \citep{all84,ver01a,hum07}. However, in general the nebular component contributes $<$ 25\% of the continuum emission at 1500 \AA, and is much less important than starlight or emission from a hidden or dormant quasar. The contibution from nebular continuum can be quite accurately predicted by means of the strength of the HeII~$\lambda$4686\AA\ line, or when not available the HeII~$\lambda$1640\AA\ line \citep{all84}.

\subsection{Neutral Gas}
\label{neutgas}

There are two techniques for studying cool HI gas in HzRGs. One method is to measure redshifted absorption of the 21 cm hydrogen line against the bright radio continuum. The second method is to observe deep narrow absorption troughs that are often present in the Ly$\alpha$ profiles. In principle, both these techniques can be used to constrain properties of the neutral hydrogen such as spatial scales, mass, filling factor, spin temperature and kinematics \citep{rot99b}.

\subsubsection{HI Absorption}
\label{h1}

\begin{SCfigure*}
\centering
\includegraphics[width=0.6\textwidth]{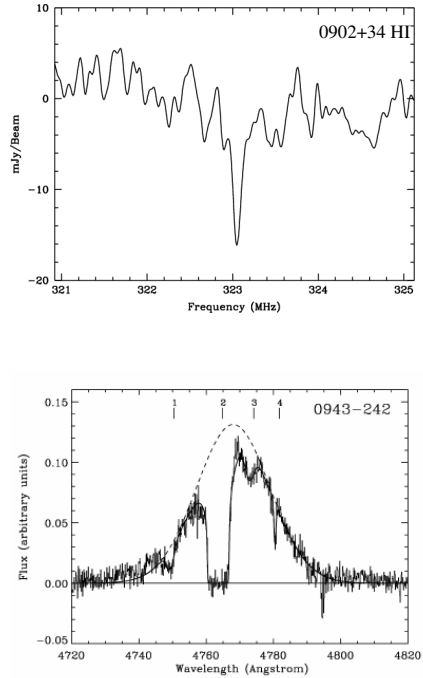}
\caption{Absorbing neutral hydrogen. 
Top: The redshifted HI absorption profile of 0902+34 at z = 2.4. [From \citet{cod03}].
Bottom: The redshifted Ly$\alpha$ absorption profile of 0943–242 at z = 2.9, with a model overlaid. [From \citet{jar03}].}
\label{fig: lyaabs}       
\end{SCfigure*}

Neutral hydrogen (HI) atoms are abundant and ubiquitous in low-density regions of the ISM. They are detectable by means of the hyperfine transition, emitting at 1420.405751 MHz ($\sim$ 21 cm). Observations of atomic neutral hydrogen by means of this line has been one of the most powerful tools of radio astronomy since its inception. Studying the 21cm HI line in absorption has been an important probe of HI around low-redshift radio galaxies see the recent review of \citet{mor06}. Limitations of such searches for HI in high-redshift radio galaxies include the availability of low-noise receivers that cover the observing frequencies dictated by the target redshifts and problems due to radio frequency interference. In 1991 HI absorption was detected in the HzRG, 0902+34 at z = 3.4, by \citet{uso91}. See Figure \ref{fig: lyaabs}. Since then progress in this field has been disappointing. For a review see \citet{rot99b}. Besides some follow-up work on 0902+34 \citep{bri93,deb96,cod03,cha04}, there have been only two tentative but unconfirmed additional detections of HI absorption in 0731+438 at z = 2.4 and 1019+053 at z = 2.8.

The HI absorption line provides a measure of the average column density of the absorbing material weighted by the flux density of the background source. The total column density is $\sim$ 4.4 $\times$ 10$^{18}$ T$_{S}$. Assuming a spin temperature of T$_{S}$ $\sim$ 10$^{3}$K, \citet{deb96} derives an HI column density of $\sim$ 2 $\times$ 10$^{21}$ cm$^{-2}$ for 0902+34 and a corresponding mass of neutral hydrogen of M$_{HI}$ $\sim$ 3 $\times$ 10$^{7}$ $M_\odot$.

Why has HI absorption not been detected in a larger number of HzRGs, despite extensive searches?
A possible explanation is that the HI absorption is caused by small $\sim$ 100 pc-sized disks or torus-like structures, aligned perpendicular to the radio source\citep{rot99b}. The radio emission of 0902+34 is more centrally concentrated than that for most HzRGs, consistent with the hypothesis that the absorption is produced by such a small disk. For the more extended radio sources associated with most HzRGs, the disk covering fraction would be very small and the disk would not produce significant absorption.

\subsubsection{Ly$\alpha$ absorption.}
\label{lyaabs}

More than a decade ago \citet{van97} discovered that
strong absorption features are common in the Ly$\alpha$ profiles of HzRGs. Such features were present in the majority of the 18 HzRGs that they studied with sufficient spectral resolution.
Derived column densities were in the range 10$^{18}- $–10$^{19.5}$ cm$^{-2}$.

The absorption, usually interpreted as being due to
HI surrounding the HzRG, provides an interesting diagnostic tool for studying
and spatially resolving neutral gas surrounding HzRGs. Because the spatial
extension of the absorbing region can be constrained, the Ly$\alpha$ absorption lines provide
information about properties of the absorbing gas
(e.g. dynamics and morphologies) that cannot be studied using quasar
absorption lines. Since in most cases the Ly$\alpha$ emission is absorbed
over the entire spatial extent (up to 50 kpc), the absorbers must have a
covering fraction close to unity. From the column densities and spatial
scales of the absorbing clouds, the derived masses of neutral hydrogen are typically $\sim 10^8$ M$_{\odot}$.

Additional information about the properties of the HI absorbers was obtained by \citet{wil04}. In a study of 7 HzRGs with 2.5 $<$ z $<$ 4.1, they identified two distinct groups of H I absorbers: strong absorbers with column densities of N$_{HI}$ $\sim$ 10$^{18}$ to 10$^{20}$ cm$^{-2}$ and weaker systems with N$_{HI}$ $\sim$ 10$^{13}$ – 10$^{15}$ cm$^{-2}$. They suggest that the strong absorbers may be due to material cooling behind the expanding bow shock of the radio jet and that the weak absorbers form part of the multiphase proto-intracluster medium responsible for the Ly$\alpha$ forest. Furhermore, \citet{kra05} carried out hydrodynamic simulations of a HzRG jet inside a galactic wind shell and showed that strong HI absorption could be produced. 

Absorption is occasionally observed in the profiles of other emission lines than Ly$\alpha$, such as
CIV - \citep{rot95}. \citet{jar03} studied the profiles of two of the most prominent absorbing HzRGs 0943-242 at z = 2.9 and 0200+015 at z = 2.2 with high spectral resolution. The data are consistent with a picture in which the absorbing gas has low density and low metallicity and is distributed in a smooth absorbing shell located beyond the emission-line gas. However, the metallicity, inferred from the C IV absorption, is considerably lower in 0943-242 than in the slightly larger source 0200+015. This difference in metallicity is explained as due to chemical enrichment via a starburst-driven superwind (Section \ref{jetgas}).
Further observations and modeling of the spectrum of 0943-242 by \citet{bin00} indicate that in this object the absorbing gas may actually be ionised - see also \citep{bin06}.
However, 0943-242 may be a special case. It has a relatively small radio size and one of the deepest Ly$\alpha$ absorption trough of all known absorbers.

\subsection{Molecular Gas}
\label{molgas}

Star formation is generally observed to occur in molecular clouds - cold dark condensations of molecular gas and dust, that are observable in the millimeter and near-IR. In these clouds atomic hydrogen associates into molecular hydrogen, H$_{2}$, a species that unfortunately does not emit easily observable spectral lines. The next most abundant molecule is carbon monoxide. Because the rotational transitions of the dipolar molecule $^{12}$CO are caused primarily by collisions with H$_{2}$, CO is an excellent tracer of molecular hydrogen. The most important redshifted CO transitions for the study of high-redshift objects are 
J=(1--0), (2--1), (3--2), (4--3) and (5--4) at 115.2712, 230.5380, 345.7960, 461.0407 and 576.2677\,GHz respectively. These lines are an important diagnostic for probing the reservoir of cold gas available for star formation.


Intensive searches for CO emission from HzRGs during the early 1990s were unsuccessful \citep{eva96,van97b}. Since then, the sensitivity of (sub)millimetre receivers has been improved and the high-redshift Universe has been opened 
to molecular line studies 
(see reviews by Solomon \& vanden Bout (2005)\nocite{sol05} and Omont (2007)\nocite{omo07}).


To convert the CO to H$_2$ mass, one often assumes a standard conversion factor. For high redshift CO studies,
this factor is calibrated based on observations of nearby ultra-luminous infrared galaxies (ULIRGS)\citep{dow93}. 
With the assumption that this value is also applicable to high redshift objects, one can use the 
strength of the (1--0) CO transition to derive the mass of the molecular gas. Because current centimetre 
telescopes do not allow observations of the (1--0) transition at $z<3.6$, one needs to observe higher order transitions shifted to the atmospheric windows at 3, 2 and 1.3\,mm.

\begin{table}
\begin{center}
{Table 2: HZRGs detected in CO}
\end{center}
\begin{tabular}{|lllllll|}
\hline
Name & $z$ & Transi- & $\Delta V_{\rm CO}$ & $S_{\rm CO}\Delta V$ & $M(\rm H_2)$&Ref.$^1$\\
&& tion  & km s$^{-1}$ & Jy km s$^{-1}$ & 10$^{10} M_{\odot}$ \\
\hline
53W002        & 2.390 & (3--2) &     420 & 1.2$\pm$0.2   &      1.2 & 1,2 \\
B3 J2330+3927 & 3.086 & (4--3) &     500 & 1.3$\pm$0.3   &        7 & 3 \\
TN J0121+1320 & 3.517 & (4--3) &     700 & 1.2$\pm$0.4   &        3 & 4 \\
6C 1909+72    & 3.537 & (4--3) &     530 & 1.6$\pm$0.3   &      4.5 & 5 \\
4C 60.06      & 3.791 & (1--0) &     ... & 0.15$\pm$0.03$^a$ &       13 & 6 \\
              &       & (4--3) & $>$1000 & 2.5$\pm$0.4   &        8 & 5 \\
4C 41.17      & 3.792 & (1--0) &     ... & $<$0.1        &      ... & 7 \\
              &       & (4--3) &    1000 & 1.8$\pm$0.2   &      5.4 & 8 \\
TN J0924-2201 & 5.197 & (1--0) &     300 & 0.09$\pm$0.02 & $\sim$10 & 9 \\
              &       & (5--4) &     250 & 1.2$\pm$0.3   &          & 9 \\
\hline
\end{tabular}
$^a$Only broad component; narrow component has $S_{\rm CO}\Delta V$=0.09$\pm$0.01\,Jy km s$^{-1}$.

$^1$References: 1 = \citet{sco97}, 2 = \citet{all00}, 3 = \citet{deb03}, 4 = \citet{deb03b}, 5 = \citet{pap00}, 6 = \citet{gre04}, 7 = \citet{pap05}, 8 = \citet{deb05a}, 9 = \citet{kla05}. 
\label{table: cotable}
\end{table}

The inferred masses of H$_2$ in the CO-detected galaxies are between $10^{10}$ and $10^{11} M_{\odot}$, indicating that there is a large mass of molecular gas in these objects and a substantial reservoir of material available for future star formation. However, the calculated masses should be treated with caution, because their derivation is based on a large number of assumptions \citep[e.g.][]{dow93}. Observations of higher CO transitions are biased to the detection of denser gas than ground-state observations and can result in an underestimate of the total molecular gas mass.

Observations of multiple rotational transitions allow the temperature and density of the molecular gas to be constrained
using large velocity gradient models.
The results indicate that the CO properties are heterogeneous. TN J0924-2201 is only detected in the ground-state (1-0) transition, but 4C~41.17 is only detected in CO(4-3), despite sensitive searches for the (1-0) transition \citep{ivi96,pap05}. This high excitation level in 4C41.17 implies large gas densities n(H$_{2}$) $>$ 1000 cm$^{-3}$, consistent with gas fueling a nuclear starburst. 

\begin{SCfigure*}
\centering
\includegraphics[width=0.5\textwidth,angle=-90]{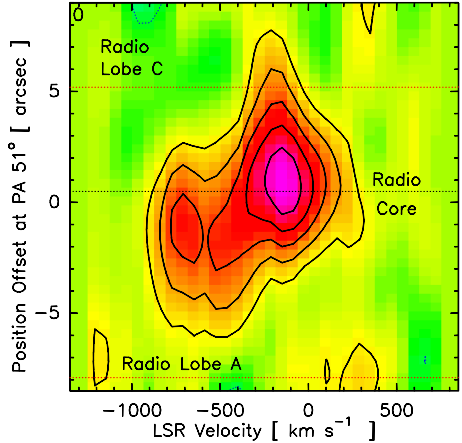}
\caption{CO in 4C41.17 at z = 3.8 with the Plateau de Bure interferometer. [From \citet{deb05a}]. Shown is the velocity plotted against position offset of CO(4-3), extracted along a PA of 51\deg (see radio morphology in Figure \ref{fig: 4117radio}).The central frequency at 96.093 GHz (z =3.79786) is based on the wavelength of the optical HeII$\lambda$1640\AA emission line. The position--velocity diagram shows the two components of the CO emission. One of these is coincident with the radio core, while a fainter component is spatially offset from it towards the southwestern lobe A.
}
\label{fig: 4c41CO}       
\end{SCfigure*}

In a few cases the CO appears spatially resolved \citep{pap00,gre04,deb05a}, extending over 10 - 20 kpc (e.g. see Figure \ref{fig: 4c41CO}), providing kinematic information about the molecular gas. \citet{kla04} have claimed that there are alignments between the molecular gas and radio morphologies in some of the detected CO HzRGs, as might be expected from jet-induced star formation (Section \ref{alignment}). However, higher resolution, larger signal to noise and more statistics are needed before any conclusions about possible alignments can be drawn.

Offsets between the velocities of the molecular gas (CO) and those of the warm gas (e.g. HeII$\lambda$1640) of up to 500 km/s have been measured \citep{deb03,deb03b,deb05a}. Because the H$_{2}$ masses exceed those of the warm ionized gas Ly$\alpha$ halos by an order of magnitude, the CO lines provide a better measure of the systemic redshift of the HzRGs than the UV and optical emission lines.



%
%
%
%


The CO detections listed in Table \ref{table: cotable} are not representative of the distribution of CO in HzRGs. There are several observational biases inherent in current CO studies:
 
\begin{itemize}

\item{Many searches for CO at high redshifts pre-select targets on the basis of prior dust detections, i.e continuum millimetre flux (see Section \ref{mm}). However, 850$\mu$m continuum emission was not detected for two of the CO-emitting HzRGs -  53W002 at z = 2.390 \citep{sco97,all00} and TN J0924-2201 at z = 5.2 \citep{kla04}. This indicates that the gas/dust ratios in high redshift CO samples could be be systematically underestimated due to bias in favour of dusty targets.
}

\item{Positive detections are more likely to be published than non-detections. Furthermore, the derived limits are often dubious in view of relatively large uncertainties in the targeted redshifts. These uncertainties are due to large offsets that are often present between the CO and optical redshifts, the unknown widths of the CO emission lines, and the limited bandwidth of correlators used for the detections.}

\item{For higher redshift objects progressively higher CO transitions need to be observed consistent with the atmospheric transmission bands acessible to millimetre arrays. Hence, measurements of CO in will tend to trace denser gas in HzRGs than at lower redshifts. However for $z>3.6$, the ground-state transition becomes observable with centimetre facilities, such as the VLA and ATCA, resulting in a discontinuity in the studies of multiple CO transitions needed to constrain the densities and temperatures of the gas.} 

\end{itemize}  

To investigate the statistics of CO in HzRGs and the evolution of molecular gas, it is important to carry out deep CO observations on well defined samples of HzRGs in multiple CO transitions, and without a pre-selection on their cold dust properties. The faintness of the CO lines and the uncertainties in their redshifts means that comprehensive searches for CO presently require large amounts of observing time. Limiting factors are the sensitivities of current facilities and the relatively limited instantaneous bandwidth covered by the receiver back-ends. 

Opportunities for observing molecular gas in HzRGs will be revolutionised during the next decade by ALMA and the EVLA. These facilities will provide an enormous improvement in sensitivity and discovery space throughout the sub-millimeter, millimeter and centimetre wavebands. Furthermore, their resolution will allow the spatial distribution of the molecular gas in HzRGs to be studied.  On a shorter timescale, the much wider tuning range and broad-band correlators that are presently coming on-line (e.g. at the EVLA and the ATCA) will provide new opportunities for investigating CO in high-redshift objects, particularly the ground state transition.

The increased sensitivity of ALMA will also open the study of fainter molecular and atomic lines in HzRGs. Lines such as HCN or HCO$^+$ probe at least an order of magnitude denser molecular gas than CO, and are therefore better tracers of the dense molecular cores in star-forming regions \citep{pap07}. 

\subsection{Dust}
\label{dust}

Dust is an important constituent of HzRGs and an additional diagnostic of star formation. It is both a major constituent of the molecular clouds from which stars generally form and an indicator that substantial star formation has already occurred. The presence of dust means that chemically enriched material is present. Thermal re-radiation from dust often dominates the spectral energy distribution of HzRGs at millimetre and sub-millimetre wavelengths (Fig \ref{fig: sed}) and dust is observable as a polarizing and absorbing medium in the optical and ultraviolet.

\subsubsection{Millimeter and sub-millimeter emission}
\label{mm}

Since the detection of 4C41.17 at z = 3.8 \citep{dun94,chi94,ivi95} a large number of HzRGs have been observed at millimeter wavelengths \citep{arc01, reu04}. \citet{reu04} analysed a sample of 69 radio galaxies with 1 $<$ z $<$ 5, detected at 850 and/or 450 $\mu$m. Isothermal fits to the submillimetre spectra give dust masses of a few $\times$ $10^{8}$ M$_{\odot}$ at temperatures of $\sim$ 50 deg \citep{arc01}.

Possible sources of heating for the dust are X-rays from an AGN (Section \ref{agn}) and UV photons from young stars. The typical submillimetre luminosity (and hence dust mass) of HzRGs strongly increases with redshift, with a (1+z)$^{3}$ dependence out to z $\sim$ 4. This is evidence that star formation rates were higher and/or the quasars brighter in HzRGs with z $>$ 3.

Disentangling whether the dust is heated by an AGN or young stars is difficult, because the two processes produce dust at partially overlapping temperatures.  Evidence for heating by young stars comes from the slope of the Rayleigh-Jeans part of the thermal dust emission - observed at millimetre and submillimetre wavelengths. This indicates that the dust temperatures are relatively cool, $\sim$ 50K, consistent with starburst-heated emission \citep[e.g.][]{arc01,ste03,reu04}.
However, there is also evidence for a warmer($\sim$ 300K dust) component in radio galaxies, consistent with AGN heating \citep{roc05}.  Furthermore, recent Spitzer observations at $5<\lambda_{\rm obs}<70\,\mu$m showed
that the Wien tail part of the dust SED is inconsistent with low dust temperatures. Dust with temperatures of $\sim$ 300K or higher is required, consistent with AGN heating. 
Although the submillimetre luminosity of the HzRGs is uncorrelated with radio luminosity, suggesting no strong dependence on the strength of the quasar/AGN emission. \citep{reu04}, the rest-frame 5\,$\mu$m luminosity does appear to be correlated with radio luminosity \citep{sey07}.

It is likely that young stars and AGN both play a role in heating the dust.
Clarification of their relative importance await accurate measurement of the SED near the peak of the thermal dust emission by the Herschel Telescope and high spatial resolution observations of the radial profiles of the thermal dust emission with ALMA.
Once properly isolated, starburst-heated dust emission will be a powerful tool for measuring the star-formation rates in radio galaxies. On the assumption that most of the submillimetre emission is heated by stars, derived star formation rates are up to a few thousand $M_{\odot}$\,yr$^{-1}$, consistent with those obtained from UV absorption line measurements (Section \ref{youngpop}).


\subsubsection{UV continuum polarisation}
\label{dustpol}

The rest-frame UV polarisation is an important probe of dust in the inner regions of HzRGs \citep{dis89, dis97,cim93,cim98,ver01a}. The fractional polarisation provides a unique tool for determining the contribution and nature of of scattered light, while the polarisation angle allows the location of the scattering medium to be pinpointed. However, polarisation measurements of HzRGs require both high sensitivity and high precision.

\begin{SCfigure*}
\centering
\includegraphics[width=0.6\textwidth]{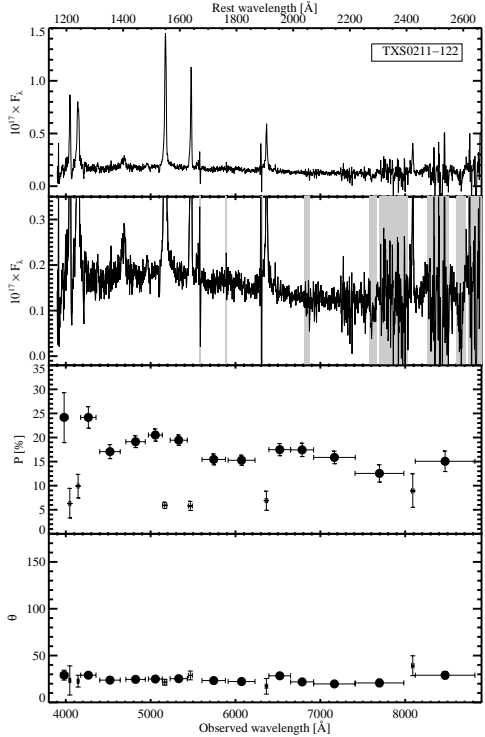}
\caption{Spectral and polarization properties of 0211-122 at z = 2.3 taken from observations with the Keck Telescope. [From
\citet{ver01a}]. Plotted from top to bottom are: the observed total flux spectrum in units of 10$^{-17}$
erg s$^{-1}$
cm$^{-2}$
\AA $^{-1}$
at two different scales, the first to show strong emission lines and the second to show the continuum, the
crosses respectively indicate continuum and narrow emission lines (including their underlying continuum). Shaded areas indicate regions of bright sky emission.
}
\label{fig: 0211pol}       
\end{SCfigure*}

Eight of 10 HzRGs  studied by \citet{ver01a},show high continuum polarization just redward of Ly$\alpha$, with fractions, f$_{p}$ ranging from 6\% to 20\% (e.g. Fig. \ref{fig: 0211pol}).
The shape of the polarised flux (= percentage polarisation $\times$ total intensity) is very similar to that of an unobscured quasar (Type~1 AGN) in both the slope of the continuum and the presence of broad emission lines \citep{ver01a}. This is strong evidence for the presence of an obscured quasar in the nuclei of HzRGs. The spectral energy distribution of a typical quasar is bluer than most other components that contribute to the SED of HzRGs between the UV and near-IR. The dilution by the other components is smaller in the rest-frame UV. At $z>2$, this waveband is redshifted into the optical, where the most sensitive (spectro-)polarimeters exist on large telescopes, allowing detailed studies despite the cosmological brightness dimming.


The polarisation angle is in most cases closely perpendicular to the radio structure, indicating that the scattering occurs within the cones traced out by the radio jets \citep{ver01a}. The spatially resolved narrow emission lines (see Section \ref{warmgas}) are generally not polarised (e.g. Fig. \ref{fig: 0211pol}), indicating that the scattering medium must be located 
between the broad and narrow line regions.

The wavelength dependence of the polarised emission is consistent with both grey dust or electron scattering. However, electron scattering can be excluded as it is much less efficient and would imply masses of the scattering medium which are close to the total galaxy mass \citep{man96}. A clumpy scattering dust medium can also produce the observed continuum slope. Using spectropolarimetric observations near the rest frame 2200~\AA\ dust feature, \citet{sol04} argue that the composition of the dust is similar to that of Galactic dust.

\citet{cim93} presented a dust-scattering model that explains the structure, the polarization properties and the spectral energy distribution of the ultraviolet aligned light with optically thin Mie scattering of quasar radiation emitted in a cone of $\sim$ 90\deg opening angle. The required amount of spherically distributed dust is (1-3) $\times$ 10$^{8}$ M$_{\odot}$, consistent with the estimate for the dust mass from the sub-millimeter data (Section \ref{mm}). However, more recent predictions \citep{ver01a} indicate that the dust responsible for scattering the AGN emission should have a sub-millimetre emission that is an order of magnitude fainter than observed.

Not all HzRGs have significant polarisation. Two of eight objects studied by \citet{ver01a} had fractional polarisation f$_{p}$ $<$ 2.4\%. There is no evidence that the nuclei of these HzRGs contain obscured quasars (Section \ref{qso}). We discuss information derived from the UV polarisation observations about the obscured or dormant quasar in Section \ref{qso}.

\section{STARS}
\label{stars}

{ \it Captain Boyle: "I ofen looked up at the sky an' assed meself the question -- what is the stars, what is the stars?"\\
Joxer Daly: "Ah, that's the question, that's the question -- what is the stars?"

Act 1 "Juno and the Paycock", Se\'{a}n O'Casey, Irish playright  (1924)}
\\

During the last fifteen years there has been mounting evidence that stars are a major constituent of HzRGs. Stellar signatures include (i) a plateau in the spectral energy distribution (SED) between 1 and 2\,$\mu$m characteristic of stars - the ''1.6 $\mu$m bump" \citep{sim99,sey07}(Fig. \ref{fig: sed}), (ii) the detection of rest-frame UV stellar absorption lines in a few HzRGs \citep{dey97} and (iii) measurement with the HST that the morphologies of several HzRGs follow de Vaucouleurs profiles. \citep[e.g.][]{van98,pen01,zir05}.

In principle, spectral energy distributions (SEDs) are a powerful diagnostic of stellar populations and galaxy evolution. Young ($<$0.5\,Gyr) stars dominate the SED in the UV, whereas the SED longward of $4000\AA$ gives information about populations older than $\sim$ 1 Gyr.  However, disentangling the stellar contribution to the SEDs from the nonstellar components is complicated (see Fig. \ref{fig: sed}). At rest-frame UV wavelengths, contributions from scattered quasar light and nebular continuum must be taken into account, whereas re-radiation by dust contributes to the SED at far infrared and millimeter wavebands (e.g Fig. \ref{fig: sed}).

\begin{SCfigure*}
\centering
\includegraphics[width=0.5\textwidth,angle=-90]{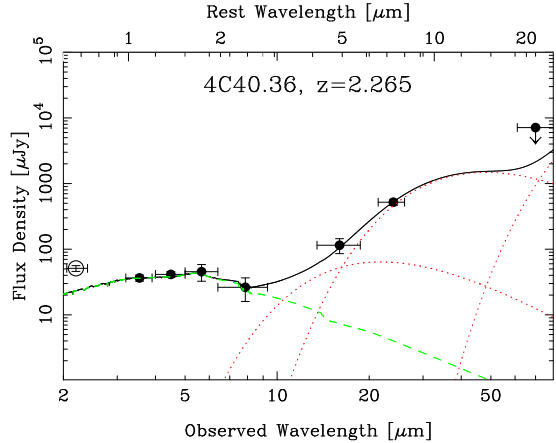}
\caption{The IR SED of 4C40.36 at z = 2.265, with modeling, using data taken with IRAC and MIPS on the Spitzer Telescope. [From \citet{sey07}]. The solid dark line indicates the total best-fit SED. The stellar and dust components are indicated by dashed and dotted lines, respectively. Note the plateau, corresponding to the stellar ''$\sim$ 1.6 $\mu$m bump" feature \citep{sim99}.}
\label{fig: 4036sed}       
\end{SCfigure*}

\subsection{Old stars: HzRGs as the most massive galaxies}
\label{oldpop}

The SEDs of old ($>$1 Gyr) stellar populations peak in the near-IR. The galaxies with the largest K-band luminosities in the early Universe are HzRGs. Because old stars are the best tracers of stellar mass, the brightness of the observed $K-$band emission has long been used to argue that distant radio galaxies are very massive, with masses of up to 10$^{12} M_{\odot}$\citep{roc04}. This places them on the upper end of the stellar mass function over the entire redshift range $0<z<4$ 
\citep[e.g.][]{roc04}.

\begin{figure*}
\centering
\includegraphics[width=0.8\textwidth]{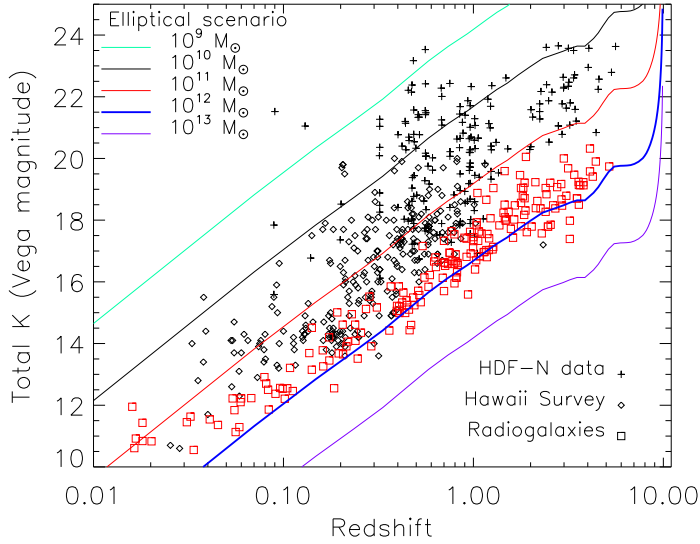}
\caption{Composite Hubble K-z diagram of radio and optically selected galaxies, slightly modified from \citet{roc04} and kindly provided to us by Brigitte Rocca. The radio galaxies are denoted by the red squares. The optically-selected galaxies are plotted in black (HDFN = crosses and Hawaii survey = dots). The radio galaxies trace the upper envelope of the K - z diagram, with HzRGs being amongst the brightest galaxies in the early Universe. Note that continuum fluxes derived from K-band or H-band magnitudes can be contaminated by bright emission lines in the band. Also plotted are K-magnitude tracks for elliptical galaxies formed from  various initial reservoir gas masses (baryonic). These were calculated using the P\'{e}gase galaxy evolution models as described in \citet{roc04}.}
\label{fig: kz2}       
\end{figure*}

A major tool in HzRG research is the Hubble $K-z$ diagram, introduced by \citet{lil84}. Despite significant spectral correction effects (K-corrections), radio galaxies form a remarkably narrow sequence out to to z $\sim$ 3, especially when compared with near-IR selected field galaxies \citep[e.g.][]{deb02}(Fig. \ref{fig: kz2}). The small scatter in the $K-z$ relation for HzRGs was found to be weakly correlated with radio luminosity \citep[e.g.][]{bes98}. Such a correlation can be understood if the radio power is dominated by the Eddington limiting luminosity of the nuclear supermassive black hole \citep{raw91} (Section \ref{mbh}) and thus a measure of the black hole mass. The well-established relation between black hole and bulge mass \citep[e.g.][]{mag98} then implies that radio power would be correlated with host galaxy mass, as appears to be the case for low-redshift less-luminous radio galaxies \citep{bes05}.

Caution should be exercised in determining the K-band continuum magnitudes for HzRGs and in interpreting the details of the K-z diagram in terms of galaxy masses. There are several reasons for this. First, it is often difficult to
disentangle the stellar component from the warm dust component of the K-band flux.  Secondly, depending on the redshift, there is also a contribution to the K-band emission from bright emission lines. Mass and age estimates derived from the uncorrected colors are therefore often overestimated \citep{eal96,roc04}. Thirdly, the stellar population models on which the masses are based are subject to considerable uncertainty. For example, incorporation of recent data on thermally pulsating asymptotic giant branch stars reduce the derived masses by large factors \citep{bru07}.
In addition the Initial Mass Function and the gas to dust content at early epochs is highly uncertain.

%
%
%
\begin{figure*}
\centering
\includegraphics[width=0.8\textwidth]{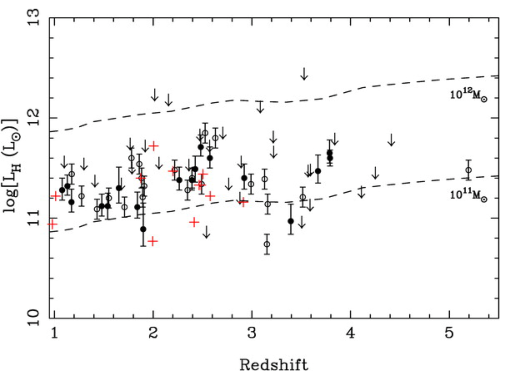}
\caption{Rest-frame H-band stellar luminosity vs. redshift for a sample of radio galaxies observed with the Spitzer Telescope. [From \citet{sey07}]. The masses of stars are derived from the best-fit models to the multiband photometry, including Spitzer fluxes. Solid circles indicate
HzRGs detected at 24 - 160 $\mu$m with MPIS. Open circles are luminosities
derived from HzRGs that were not detected by MIPS. Upper limits are indicated by arrows. Dashed lines correspond to elliptical galaxies formed at z$_{form}$ = 10.  Crosses mark the stellar luminosity of submillimeter galaxies.}
\label{fig: hz}       
\end{figure*}

Because of its ability to measure the rest-frame near-IR fluxes of HzRGs radio galaxies, the Spitzer Telescope, has facilitated considerable progress in this field during the last few years. \citet{sey07} surveyed 69 radio galaxies having 1 $<$ z $<$ 5.2 at 3.6, 4.5, 5.8, 8.0, 16 and 24 $\mu$m. They decomposed the rest-frame optical to infrared spectral energy distributions into stellar, AGN, and dust components and determined the contribution of host galaxy stellar emission at rest-frame H band. By consistently deriving the stellar luminosity at the same rest-frame wavelength near the peak of the stellar emission at $\lambda_{\rm rest}\sim 1.6\,\mu$m, effects of measuring the $K-z$ diagram with galaxies over a large range of redshift through fixed observing bands were eliminated and contamination of the fluxes by AGN and emission lines was minimised.

The Spitzer results confirm that radio galaxies have stellar masses mostly within the $10^{11}$ to $10^{12}$ M$_{\odot}$ range, with almost no dependence on redshift. The fraction of emitted light at rest-frame H band from stars was found to be $>$ 0.6 for $\sim$ 75$\%$ of the HzRGs. The weak correlation of stellar mass with radio power determined using ground-based data is only marginally significant in the Spitzer results.
Rest-frame near-IR studies of less luminous radio sources are needed to extend the range in radio luminosity and settle whether there is indeed a dependence of stellar mass on radio luminosity.

\subsection{Young stars - UV absorption lines}
\label{youngpop}

HzRGs undergo vigorous star formation. The most direct evidence for this is the detection of rest-frame UV photospheric stellar absorption lines and P-Cygni features driven by stellar winds. Because observation of these lines requires several hour exposures on 10m-class telescopes, their detection is limited to the S~V~$\lambda$1502 line in 4C~41.17 \citep{dey97}, tentative measurement of C~III~$\lambda$1428 in TN~J2007$-$1316 \citep{deb05b} and VLT spectroscopy of the Spiderweb Galaxy \citep{mil06,nes06,hat07}.
The presence of absorption lines are direct evidence for prodigious star formation in 4C41.17 (up to 1500 M$_{\odot}$yr$^{-1}$, after correction for extinction)\citep{dey97,reu04}.
For radio galaxies whose rest-frame UV continuum is polarised, scattered AGN continuum emission \citep{ver01}, make it difficult to measure the star formation rates directly.

\begin{SCfigure*}
\centering
\includegraphics[width=0.65\textwidth]{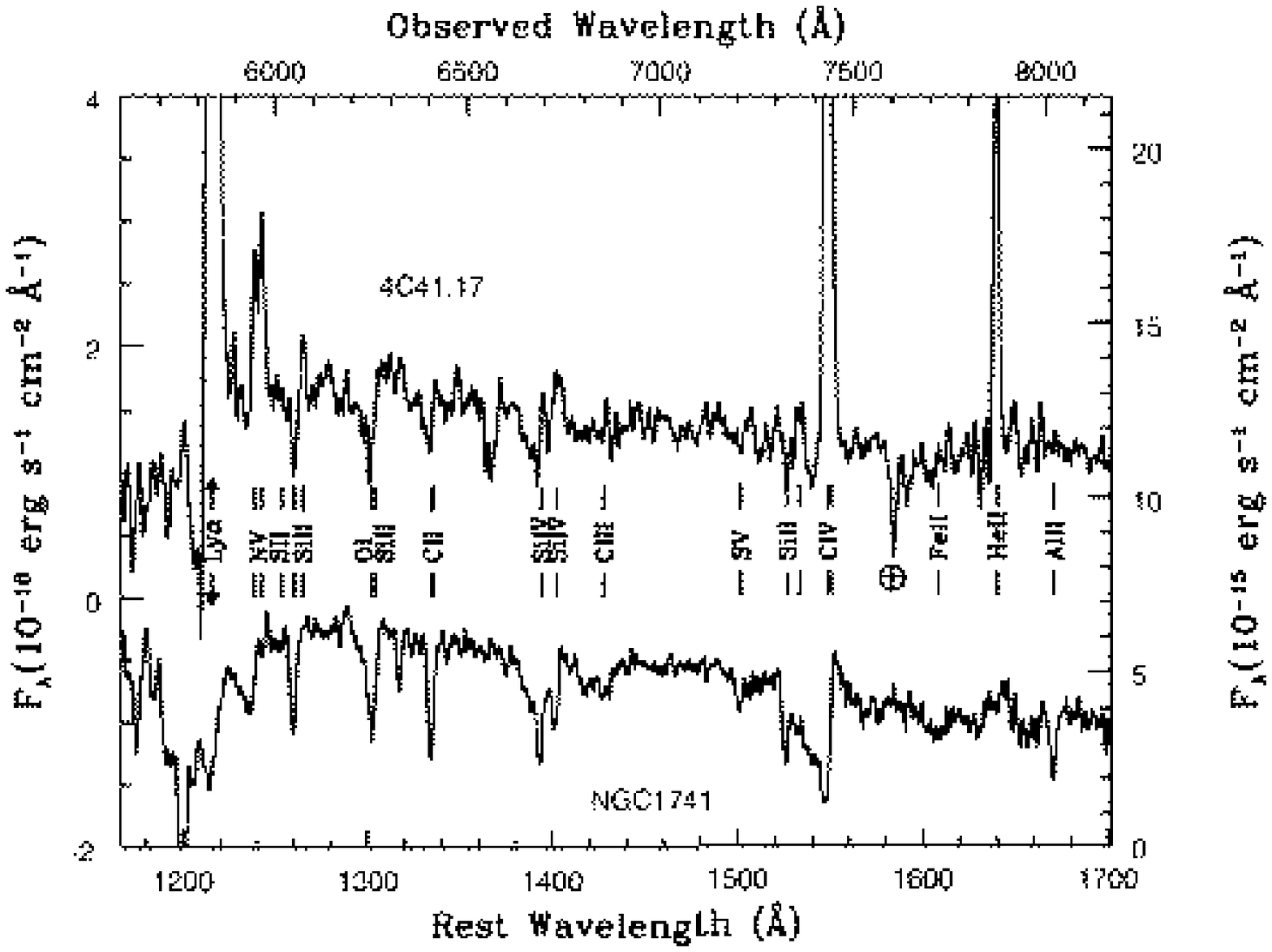}
\caption{Spectrum of the central 2 × 1 of the HzRG 4C 41.17 at z = 3.8. [From \citet{dey97}]. This is compared with 
the UV spectrum of the B1 star-forming knot in the nearby Wolf-Rayet 
starburst galaxy NGC 1741 \citep{con96}. 
The ordinate is labeled with the 
flux density scales for 4C 41.17 and NGC 1741B1 on the left and right axes respectively. 
The two spectra show many similarities in their absorption line properties. 
}
\label{fig: dey4117}       
\end{SCfigure*}

Additional evidence that substantial star formation is occurring in HzRGs is provided by the sub-millimeter observations of dust, for whose heating star formation rates of thousands M$_{\odot}$ per year are needed (but see Section \ref{dust}) and the systematically larger Ly $\alpha$ luminosities and Ly$\alpha$/HeII ratios for HzRGs with z $>$ 3, than for those with 2 $<$ z $<$ 3 \citep{vil07a,vil07b} (Section \ref{abundance}).

\subsection{The alignment effect - Jet-induced star formation}
\label{alignment}

One of the most remarkable properties of HzRG is the approximate alignment between their radio and optical continuum emissions. This phenomenon was completely unexpected when it was discovered two decades ago \citep{cha87,mcc87}. The ``alignment effect'' sets in for redshifts $z\geq 0.7$. Several models have been proposed or considered to
account for this alignments \citep[e.g.][]{mcc93}.
The two most promising models are (i) dust scattering of light from a hidden quasar at the nucleus of the HzRG \citep{tad98}
and (ii) triggering of star formation by the synchrotron jet
as it propagates outward from the nucleus \citep{dey89,ree89,beg89,bic00}.
We have already seen that radio jets frequently interact vigorously with the warm gas in HzRGs.

Since the extensive review by \citet{mcc93}, high resolution imaging with the {\it HST} has provided important new
information on the morphologies of HzRGs and on the alignment effect. \citep{pen99,bic00,pen01}. Images with the HST have revealed that HzRG hosts generally have clumpy optical morphologies \citep{pen98,pen99}.

Using {\it HST} observations of a sample of 20 HzRGs combined with similar resolution VLA radio maps, \citet{pen99,pen01} have for the first time examined the close relationship between the radio and rest-frame UV and optical morphologies. They find that the alignment effect extends into the rest-frame optical, with the $z>2.5$ radio galaxies displaying a much more clumpy structure than their low-redshift counterparts, which are well represented by de Vaucouleurs profiles.
At a slightly lower resolution, near-IR imaging with the Keck Telescope \citep{van98} also suggest that the morphology changes from aligned clumpy structures at $z>3$ to relaxed de Vaucouleurs profiles at $z<3$. However, the redshift-dependance of luminosity and the change in the observing wavelength with redshift complicates the interpretation of these results. We note that the IR structures of B2/6C galaxies at z $\sim$ 1 appear less aligned with their radio structures than are the brighter 3C galaxies at this redshift, implying that the strength of the alignment effect depends on radio luminosity \citep{eal97}.

It is now clear that no single mechanism can completely account for all
the observed properties of the aligned emission. The occurrence of optical
polarisation in HzRGs in some but not all HzRGs is evidence that dust scattering plays a role.
However, scattering cannot explain the observed detailed optical morphologies and
similar bending behaviour that has been observed
between the optical and radio structures. Neither can dust scattering account for the alignment of the continuum emission {\it redwards} of the rest frame 4000 A break.  The alignment effect is most likely due to a {\it combination} of star formation induced by the jet {\it and} scattering of hidden quasar light along the radio source.

The most comprehensive study of jet-induced star formation has been made for of 4C41.17 at z = 3.8 by \citet{bic00}.
The interaction of a high-powered ($\sim$ 10$^{46}$ erg s$^{-1}$) jet with a dense cloud in the halo of 4C 41.17 is shown to produce shock-excited emission lines and induce star formation. Such shock-initiated star formation could proceed on a timescale of a few $\times 10^{6}$ yr), i.e  well within the estimated lifetime of the radio source (few $\times 10^{7}$ yr) .

Studies of several low-redshift radio galaxies provide further evidence
that jet-induced star formation occurs. These include NGC 541/Minkowski's
Object (3C40) \citep{van85,cro06b}, 3C 285 \citep{van93} and the closest radio galaxy Centaurus A \citep{gra98}. Such effects
can be expected to be much more prevalent during the era of the
Universe at z $>$ 2, when every galaxy may well have undergone nuclear activity.

If the red aligned emission is indeed produced by stars, their ages would be $>$ 1 Gyr, implying that this mode of star formation has operated almost since the formation of the host galaxies. It would also require that the radio jet was active and aligned in approximately the same position angle for this length of time.
\citet{van98} suggested that the tendency of HzRGs to be clumpier at z $>$ 3 is due to the increased importance of jet-induced star formation at earlier epochs. We know little about how the first population of stars in galaxies were made. Although unconventional, it is tempting to speculate that jet-induced star formation played a significant role in producing stars during the first 3 Gyr after the Big Bang.
\citet{gop04} have suggested that a large fraction of all proto-galactic material within the cosmic web was impacted by the expanding lobes of radio galaxies during the quasar era, triggering star formation. 

\begin{figure*}
\centering
\includegraphics[width=0.75\textwidth]{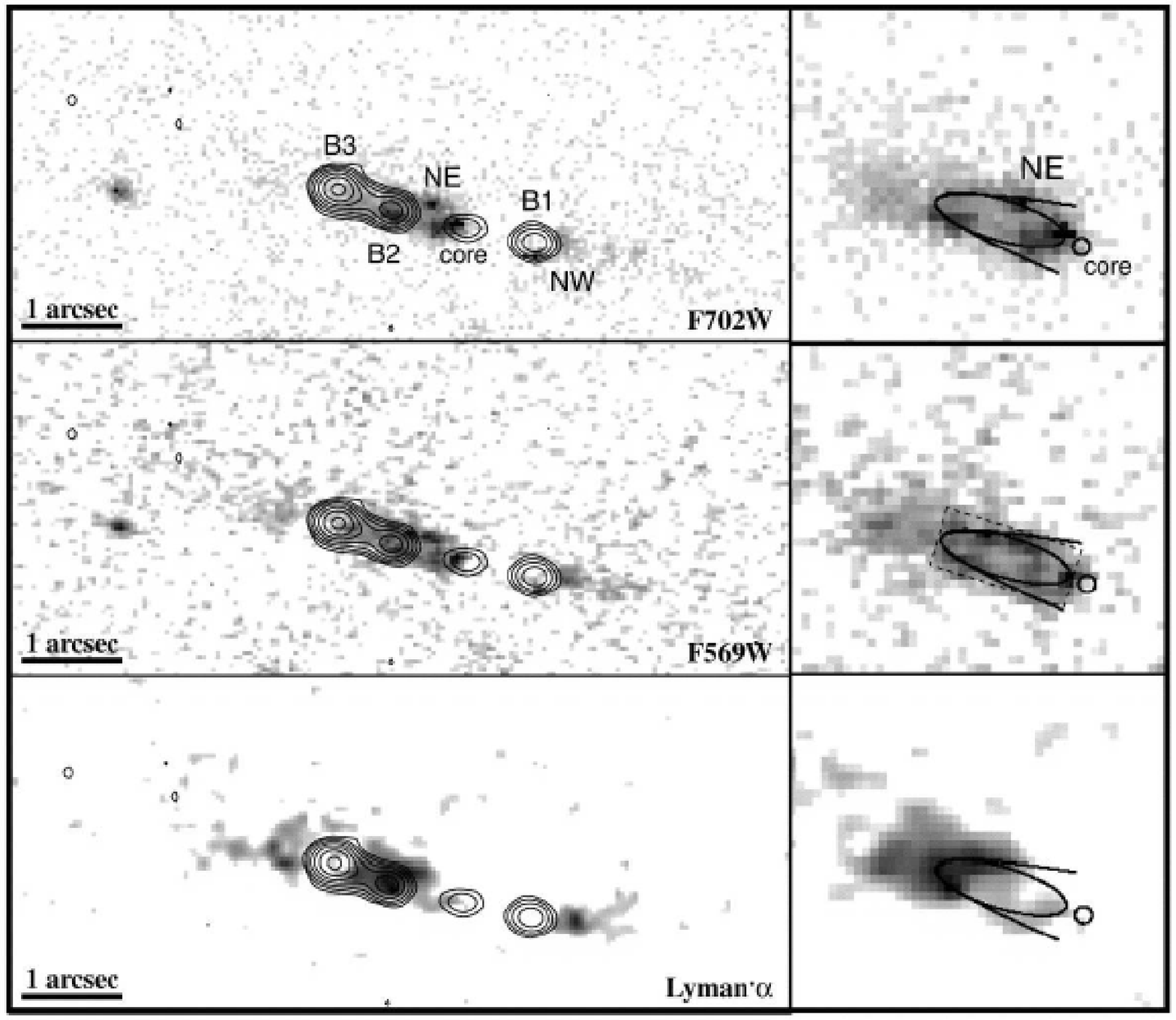}
\caption{Evidence for jet-induced star formation in 4C41.17 at z = 3.8 [ From \citep{bic00}]. Montage of three HST images taken through the F702W, F569W (continuum), and Ly$\alpha$ filters. The 8 GHz radio images of \citet{car94} are superimposed as contours. The absence of significant optical polarisation and the details of the optical-radio morphologies cannot be explained by the scattering models.
The Ly$\alpha$ image shows a bright arc-shaped feature near B2 at the apex of the edge-brightened UV structure, suggestive of a strong bow shock  at the location where the jet interacts with dense ambient gas (see also right hand panels). \citet{bic00} suggest that this is manifest through the associated shock-excited line emission and star formation in the bifurcated structure.}
\label{fig: 4117jet}       
\end{figure*}

\section{ACTIVE GALACTIC NUCLEUS (AGN)}
\label{agn}

{\it ''Power exercised in secret, especially under the cloak of national security, is doubly dangerous" US Senator, W. Proxmire}
\\
\\
Because the luminous collimated radio emission must have been generated in their cores, the nuclei of all HzRGs must have been active at some time in their histories. 

\subsection{Hidden or dormant quasar - The Unification Church}
\label{qso}

The most extreme form of nuclear activity is seen in quasars, for which the nuclei are so bright that the galaxies are observed as "quasi-stellar objects". There is strong evidence that radio galaxies and quasars are manifestations of the same parent population.  When we observe objects to be radio galaxies, the associated quasars are believed to be either (i) hidden due to our special viewing angle and/or (ii) highly variable and observed when they are in a dormant state.

The most widely accepted models for unifying quasars and radio galaxies postulate that their observed properties are dependent on the orientation of the radio source axes relative to us, the observers.
For a review of unification via orientation see \citet{ant03,urr95,urr04}. One of the key factors in establishing the validity of HzRG unification through orientation has been the signature of a hidden quasar detected in some radio galaxies in polarized light.
Polarimetry with the Keck Telescope \citep{ver01a} (See Section \ref{dustpol}) show that HzRGs are usually, but not always, highly polarised just redward of Ly$\alpha$ and that dust-reflected quasar light appears generally to dominate their rest-frame ultraviolet continua. Furthermore,  broad emission lines, with equivalent widths similar to those seen in quasars, are detected in the two most polarized HzRGs in the \citet{ver01a} sample of nine objects.

There is a convincing case that an obscured quasar exists in those HzRGs for which high fractional polarisations and broad polarised recombination lines have been measured. What about HzRGs with low polarisation? For these objects, it is possible that active quasars are present that are even more highly obscured than the polarised HzRGs. However an equally plausible explanation is that
the quasars are dormant during the epoch of observation \citep{van80}. We know that quasars are highly variable, varying by as much as 2 magnitudes during a day. Quasar variability has been studied on timescales of up to 50-years \citep{dev06}. The variability appears to be intrinsic to the AGNs, and occurs in flares of varying time-scales, possibly due to accretion-disk instabilities

During the lifetime of an extended radio sources (at least a few $\times$ 10$^{7}$ y), quasar activity can be expected to vary considerably. Knots in the radio morphologies can be interpreted as tracing bursts of quasar activity in the radio nuclei \citep{mil80}. According to such a dormant quasar  model, if the flare duration is substantially shorter than the quiescent periods the objects would be more likely to be observed as radio galaxies than radio-loud quasars. Also if the flares recurred on a timescale shorter than the recombination time ($\sim$ 10$^{4}$ - 10$^{5}$ y), the excitation of the gas would persist. A recurrently flaring quasar could also be a source of excitation for some of the non-radio Ly$\alpha$ halos (Section \ref{blobs}).

In summary, there is strong evidence that quasars exist in the nucleus of all HzRGs. Although
orientation and varying intrinsic activity both probably contribute to the observed properties, their relative importance is still unclear.

\subsection{Supermassive black hole - Powerhouse of the AGN}
\label{mbh}

Extragalactic radio sources, and all active galactic nuclei, are believed to be powered by gravitational energy produced by rotating supermassive black holes (SMBHs) in their nuclei \citep[e.g.][]{lyn69}. Material in the galaxies is accreted onto the SMBHs, converted into kinetic energy and ejected as collimated relativistic jets along the rotation axes of the SMBHs to produce the radio sources. For a theoretical review of this field, see \citep{bla01}. Although there is wide agreement about the general idea, many of the details are not well understood.

Observational evidence for the widespread occurance of SMBHs in galactic nuclei include (i) spectroscopy of the gigamaser in the nucleus of NGC 4258 \citep{miy95} (ii) spectroscopy and photometry of galactic nuclei with the HST leading to the correlation between the masses of galaxy bulges and SMBHs in early type galaxies \citep[e.g.][]{mag98,kor01,har04,nov06}, (iii) the small collimation scale of radio sources, measured with VLBI,  (10$^{17}$ cm \citep{jun99}) and (iv) the similarity of the large-scale and small-scale orientation of some giant radio sources, indicating that the 'memory" of collimation axis can persist for as long as $\sim$ $10^{8}$y \citep{fom75,sch79,mil80}.

Because of the interaction of the radio jet with the gas, line widths in the central regions of HzRGs cannot be used to derive the masses of the presumed SMBHs in their nuclei, as for radio-quiet galaxies. The empirical correlation of SMBH with the bulges of early type galaxies \citep{har04} would predict black hole masses of $\sim$ 10$^{9}$M$_{\odot}$ at the centre of a 10$^{12}$M$_{\odot}$ HzRG. \citet{mcl02} have derived similar masses for SMBHs in radio-loud quasars from the width of their emission lines. However, such estimates should be treated with caution. There is a possibility that the widths of the emission lines are widened due to pressure from the radio jets
Also, little is known about how the SMBHs assemble during the formation and evolution of a massive galaxy, such as a HzRG, so extrapolation of results from lower redshifts may not be valid.
\citet{sil98} have postulated that supermassive black holes form within the first sub-galactic structures that virialise at high redshift, and are in place before most galactic stars have formed.

\subsection {Extinction of the ''dinosaurs"}
\label{dinosaur} 

Luminous quasars and radio galaxies have undergone strong evolution in their space densities since z $\sim$ 2 (the peak of the ''quasar" era") and both species are now virtually extinct (Section \ref{space}). The reason why such extreme manifestations of nuclear activity are no longer with us is not fully understood. One obvious explanation is that the supply of ''food" for the SMBHs began to diminish as more and more of the the gas in massive galaxies was converted to stars \citep[e.g.][]{men04}.  

\section {NATURE OF HzRG HOSTS - MASSIVE FORMING GALAXIES}
\label{mass}

{\it Viktor Ambartsumian: ''It is easy to show applying statistical considerations, thta the galaxies NGC 5128 (Centaurus A) and Cygnus A cannot be the result of the collisions of two previously independent galaxies"\\
Rudolf Minkowski ''Results of a detailed investigation of NGC 1275 (Perseus A), reported at this symposium, admit to no other interpretation than that this system consists of two colliding galaxies", IAU Symposium No. 5, Dublin, 1955}
\\

Having considered the various building blocks of HzRGs in detail, we now address the nature of HzRGs and how these objects fit into general schemes of galaxy evolution.

There are several reasons for concluding that HzRGs are massive forming galaxies. First, we have seen that
their large near-IR luminosities imply that HzRGs are amongst the most {it massive} galaxies in the early Universe (Section \ref{oldpop}). Secondly, HzRG hosts have clumpy UV continuum morphologies (Section \ref{alignment}), as expected from galaxies {\it forming} through mergers and in accordance with hierarchical models of massive galaxy evolution\citep{lar92,dub98,gao04,spr05a}. Thirdly, the rest-frame UV spectra and millimeter SEDs indicate that HzRGs are undergoing
copious {\it star formation} \citep[e.g.][]{dey97} (Section \ref{youngpop}).

Not only are HzRGs massive forming galaxies, but there is also strong evidence that they are the progenitors of the most massive galaxies in the local Universe, i.e. the giant galaxies that dominate the central regions of rich local galaxy clusters. We have seen that HzRGs are generally embedded in giant (cD-sized) ionized gas halos (Section \ref{warmgas}. In (Section \ref{Protoclusters}) we shall show that they are often surrounded by galaxy overdensities, whose structures have sizes of a few Mpc (Section \ref{Protoclusters}).  For all these reasons, HzRGs are the probable ancestors of brightest cluster galaxies (BCGs) and cD galaxies.

What can studying the development of HzRGs tell us about the evolution of massive galaxies in general?
Recently, two processes have been invoked to explain discrepancies between the observed colour and luminosity statistics of faint galaxies and the model predictions. In contrast with what might be expected from simple hierarchical merging scenarios, the data imply that star formation
occurs earlier in the history of the most massive galaxies than for the less massive ones. This process is called
galaxy \textbf{\textit{``downsizing''}} \citep{cow96,hea04,tho05,pan07}. AGNs are preferentially found in the most massive galaxies. To quench star formation in these galaxies, it has been proposed that there is some sort of feedback mechanism, in which the AGN influences the star formation history of the galaxy. This {\it AGN feedback} \citep{dim05,bes05,spr05b,cro06,hop06,bes07} provides modelers with additional parameters that, not surprisingly, gives a better fit to the data. The potential importance of AGN feedback for galaxy formation processes was only realised after it was shown that every galaxy may host a supermassive black hole (Section \ref{mbh}), establishing a general link between galaxies and AGN.

As the most massive known galaxies in the early Universe and the seat of powerful AGN, HzRGs are important laboratories for studying the processes responsible for downsizing and AGN feedback. In practice,  feedback scenarios are complicated and involve several physical processes that occur simultaneously. We have seen in Sections \ref{relativistic} to \ref{agn} that there is considerable evidence for physical interaction between the various building blocks of HzRGs - relativistic plasma, gas, dust, stars and the AGN. As they merge, the satellite galaxies will interchange gas with the ambient medium in the system. The gas will move inwards through cooling flows and accretion and provide fuel for the supermassive black hole. The SMBHs generate quasars and relativistic jets. Hidden and/or dormant quasars heat the dust. The jets together with superwinds from starbursts \citep{arm90,zir05} drive gas outwards \citep{nes06} and can trigger star formation. Shocks will be rampant in the chaotic environments in which these processes are competing with each other. 

Two modes have been distinguished in theoretical models for coupling of the energy output of the AGN to the surrounding gas, with rather confusing names. ''Quasar-mode" feedback is defined as the situation when radiative coupling occurs that expels the gas and thereby quenches star formation in the forming galaxy \citep{hop06}. In ''radio-mode" coupling \citep{bes07}, star formation is slowed down due to the mechanical energy of the synchrotron radio jets that inhibits cooling of the ambient gas. However, such rigid distinction between feedback processes are highly simplified \citep[e.g.][]{fu07}.   

It is impossible to understand how these complicated AGN feedback processes influences the evolution of massive galaxies from a consideration of statistical data alone.

We shall now illustrate some of these processes and their relevance to galaxy formation by considering a specific example of a HzRG in more detail.

\subsection {The Spiderweb Galaxy - a case study}
\label{spider}

The Spiderweb Galaxy (MRC 1138--262) at a redshift of $z = 2.2$ is one of the most intensively studied HzRGs \citep{mil06}. This object provides a useful case study for illuminating several important physical processes that may occur generally in the evolution of the most massive galaxies. Because the Spiderweb Galaxy is (i) relatively close-by, (ii) one of the brightest known HzRGs and (iii) is the HzRG with the deepest HST optical image, it is an important laboratory for testing simulations of forming massive galaxies at the centers of galaxy clusters.

This large galaxy has several of the
properties expected for the progenitor of a dominant cluster galaxy\citep{pen97,pen98,pen00a,pen01}. 
The host galaxy is surrounded by a giant Ly$\alpha $
halo \citep{pen00a,kur04a,kur04b} and embedded in dense hot ionized gas with an ordered
magnetic field \citep{car98}. The radio galaxy is associated with a 3 Mpc-sized
structure of galaxies, of estimated mass $>$ 2 x 10$^{14}$ M$_{\odot}$, the
presumed antecedent of a local rich cluster (Section \ref{Protoclusters}).

\begin{figure*}
\centering
\includegraphics[width=0.75\textwidth]{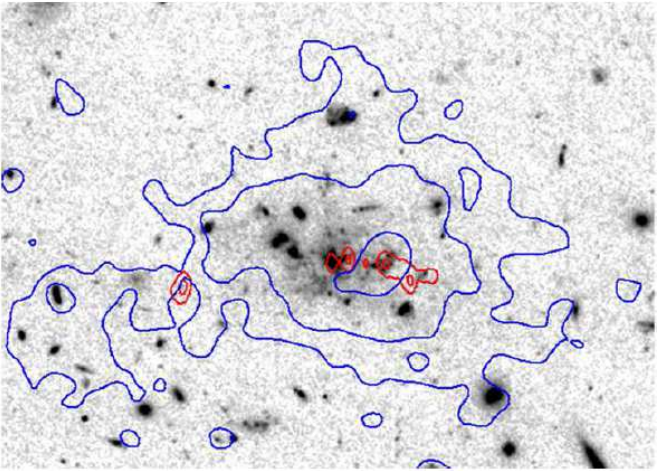}
\caption{The Spiderweb Galaxy. Deep Hubble image of the core of the MRC 1138-262
protocluster at z = 2.2 obtained with the Advanced Camera for Surveys. [From \citep{mil06}].
Superimposed on the HST image are contours of Ly$\alpha $ (blue, resolution
$\sim $1'') obtained with ESO's very Large Telescope (VLT), delineating the
gaseous nebula and radio 8GHz contours (red, resolution 0.3'') obtained with
NRAO's VLA, delineating the non-thermal radio emission. The gaseous nebula
extends for $>$200 kpc and is comparable in size with the envelopes of cD
galaxies in the local Universe.}
\label{fig: spider}       
\end{figure*}

The beautiful \textit{ACS} Hubble image of the Spiderweb Galaxy is shown in Figure \ref{fig: spider}, with
the radio source and Ly$\alpha$ halo superimposed. The figure illustrates the structures of the radio, warm gas and stellar components in a relatively nearby HzRG.

It also provides dramatic evidence that tens of
satellite galaxies were merging into a massive galaxy, $\sim$ 10 Gyr ago. The
morphological complexity and clumpiness agrees qualitatively
with predictions of hierarchical galaxy formation models \citep{lar92,kau93,bau98,dub98,gao04,spr05a}, and illustrates this process in unprecedented detail.  Ly$\alpha$ spectroscopy shows relative
velocities of several hundred km s$^{-1}$, implying that the satellite galaxies (''flies") will
traverse the 100 kpc extent of the Spiderweb many times in the interval
between z $\sim $ 2.2 and z $\sim $ 0, consistent with the merger scenario.

An intriguing aspect of the Spiderweb Galaxy is the presence of faint diffuse emission between the satellite galaxies
\citep{hat07}. Approximately $50\%$ of the ultraviolet light from the Spiderweb Galaxy is in diffuse ''intergalactic" light, extending over about 60 kpc diameter halo. The luminosity in diffuse light implies that the emission is dominated young stars with a star formation rate of $>$80 $M_\odot$ yr$^{-1}$. Under reasonable assumptions, the diffuse emission seen in the Spiderweb Galaxy could evolve into the CD envelopes seen in many dominant cluster galaxies seen at low redshifts.


The total mass of all the flies.
in the Spiderweb, derived from their UV
luminosities (assuming 1Gyr starbursts), is less than a tenth of the mass of the whole galaxy obtained
from its IR luminosity \citep{mil06}. Because the UV emission is produced by ongoing star formation and the IR emission by old stars, this implies that most of the galaxy mass may
already have assembled by z $\sim $ 2.2, consistent with downsizing
scenarios.

Merging, downsizing and feedback are all likely to be occurring simultaneously
in the Spiderweb Galaxy. Merging is a plausible fueling
source for the nuclear supermassive black holes that produce the radio
sources. Pressure from these radio sources is sufficient to expel a large
fraction of gas from the galaxies \citep{nes06}, thereby quenching star
formation \citep{cro06}.
Because radio lifetimes are relatively short (few $\times
$10$^7$yr), all massive ellipticals may have gone through a
similar short but crucial radio-loud phase during their evolution.

An unexpected feature of the HST image is that there is a significant excess of faint
satellite galaxies with linear structures \citep{mil06}.
These galaxies (linear ''flies") have similar morphologies (e.g. chains and tadpoles) to the
linear galaxies that dominate resolved faint galaxies (i$_{775 }>$ 24)
in the Hubble Ultra Deep Field (UDF)\citep{elm05,str06}. Although
linear galaxies must be an important constituent of the earliest galaxy
population, their nature is poorly understood.  Their presence in a merging system
is relevant for theories of their formation.
In the Spiderweb Galaxy the motions of the flies with velocities of
several hundred km s$^{-1}$ through the dense gaseous halo,
perturbed by superwinds from the nucleus \citep{arm90,zir05} and the radio jet,
would result in shocks. The shocks would then lead to Jeans-unstable clouds,
enhanced star formation along the direction of motion and to chain and tadpole morphologies \citep{tan01,mil06}.

\section{ENVIRONMENT OF HzRGS - PROTOCLUSTERS}
\label{Protoclusters}

{\it ''We have traced the broad outline of the development in regions of exceptional fruitfulness of the first settled village communities into more populous towns and cities" ''The city community arose around the altar of the seed -time blood sacrifice." The Outline of History, H. G. Wells, 1920}
\\

Within standard Cold Dark Matter (CDM) scenarios the first stars and stellar systems should form through
gravitational infall of primordial gas in large CDM halos \citep[e.g.][]{whi78}.
Simulations suggest that these halos merge and form web-like networks traced by young galaxies and re--ionized gas
\citep[e.g.][]{bau98,del07}. The most massive galaxies, and the richest clusters emerge from regions with the
largest overdensities.

\subsection{Finding protoclusters using HzRGs}
\label{hzrgprotocluster}

To investigate the emergence of large-scale structure in the Universe, it is important to find and study the most distant rich clusters of galaxies. Conventional methods for finding distant galaxy clusters usually rely on the detection of hot cluster gas using X-ray techniques. These are limited by the sensitivity of X-ray telescopes to $\lesssim 1.3$, i.e. much smaller than the redshifts of HzRGs.

Since HzRGs are amongst the most massive galaxies in the early Universe, they are likely to inhabit regions
that are conducive to the formation of rich galaxy clusters.
During the mid-nineties observational evidence emerged that HzRGs are located in dense cluster-type environments. First, the measurement of large radio rotation measures (RM) around some HzRGs, indicated that
the host galaxies are surrounded by a hot magnetized cluster gas \citep{car97,ath98b} (Section \ref{hotgas}). Secondly,  an excess of ''companion galaxies" was found near HzRGs \citep{rot96a,pas96,lef96,kno97}. These results together with the indications that HzRGs hosts are forming dominant cluster galaxies (Section \ref{mass}) prompted the initiation of direct searches for galaxy clusters in the vicinity of HzRGs.

Conventionally the term ''galaxy cluster" refers to a {\it bound} structure of several hundred galaxies. At z $\sim$ 3, the Universe is only about 2 Gyr old, i.e. insufficient time for a galaxy with a velocity of a few hundred km/s to have crossed cluster-scale structures. Hence any overdense structure of galaxies observed at high redshifts must still be forming and cannot be virialised and bound. Following \citet{ove06b}, we shall use the term ''protocluster" for an overdense structure in the early Universe ($z > 2$), whose properties are consistent with it being the ancestor of a local bound galaxy cluster.

During the last few years there have been several successful direct searches for protoclusters around HzRGs, with 8 to 10m-class optical/ infrared telescopes. One search technique is to use narrow-band imaging to detect emission line objects at redshifted wavelengths corresponding to those of the target HzRGs. A second technique is to carry out broad-band imaging with colours chosen to detect ''dropout objects" due to redshifted features in the continuum spectral energy distributions. The protocluster candidate galaxies detected in these imaging experiments were generally followed up by multi-object spectroscopy to confirm their redshifts and their membership of the protoclusters.
The most important emission line for establishing the redshifts of the protocluster galaxies is Ly$\alpha$ \citep{pen97,kur00,ven02,ven04,cro05,ven05,ven07}. Other relevant emission lines for such searches for z $>$ 3 are H$\alpha$ \citep{kur04a,kur04b} or [OIII]$\lambda 5007$. There are two relevant features in the continuum spectra of galaxies that are exploited for such protocluster searches - the Lyman break around the 912$\AA$ Lyman continuum discontinuity \citep{mil04,ove06,ove07,int06} and the 4000$\AA$ break or Balmer break close to 3648$\AA$ \citep{kur04a,kaj06,kod07}. The various galaxy detection techniques are complementary, because they tend to select stellar populations with different ages. For example, Ly$\alpha$ excess galaxies and Lyman break galaxies have young on-going star-forming populations, while the Balmer technique is sensitive to stellar populations that are older than a few $\times 10{^8}$y.

The measured overdensities of Ly$\alpha$ emitters in the radio-selected protoclusters are factors of 3 - 5 larger than the field density of Ly$\alpha$ emitters at similar redshifts \citep{ven05,ven07}. A recent study with the HST of 4000$\AA$ break objects in MRC1138-262 at z = 2.2, the protocluster that contains the Spiderweb Galaxy (Zirm et al. 2007, submitted to Ap.J) shows that the overdensity of red galaxies is 6.2, compared with non-protocluster fields. The photometric redshifts of galaxies in this field shows a significant "redshift spike" for 2 $<$ z $<$ 2.5.  Although such observations at $z > 5$ are sensitivity limited, a significant overdensity of both Ly$\alpha$ emitters \citep{ven04} and Lyman break galaxies \citep{ove06} has been established around TN J0924-2201 at z=5.2, the HzRG with the highest redshift known to date.

Possible protoclusters have also been detected in non-targeted optical surveys, with several overdense regions found at large redshifts in the form of ''redshift spikes" or ''filaments"  \citep[e.g.][]{ste98,kee99,ste00,mol01,shi03,hay04,mat05,ste05,ouc05}. Although caution must be exercised in deriving overdensities from the redshift distributions alone, because the effects of peculiar galaxy velocities can influence the apparent clumpiness \citep{mon05}.
However, the observed structures around the HzRGs have two additional ingredient expected of ancestors of rich clusters, namely (i) the presence of the HzRG hosts with properties expected of progenitors of dominant cluster galaxies and (ii) sharp peaks in the redshift distributions of Ly$\alpha$ emitters that lie close to the {\it predetermined} redshift of the radio galaxy. {\it Taken together}, the measured galaxy overdensities combined with the presence of HzRGs is strong evidence that the galaxy structures around HzRGs are indeed the ancestors of rich local clusters.

For some of the HzRG-selected protoclusters there is additional evidence for a dense environment from excess counts at millimeter \citep{ivi00,sma03,ste03,deb04b,gre07} and X-ray \citep{pen02,ove05} wavelengths. Also, the X-ray counts indicate an enhanced AGN fraction in protoclusters compared to the field.

\begin{SCfigure*}
 \centering
\includegraphics[width=0.6\textwidth]{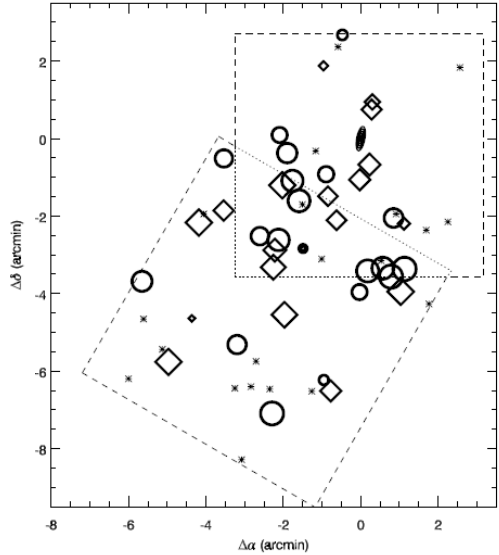}
\caption{HzRG-selected protocluster around TN J1338-1942 at z = 4.1. [From \citet{ven02,ven07}]. Shown are the spatial distribution of the spectroscopically confirmed z = 4.1 Ly$\alpha$ emitters (circles and diamonds), additional Ly$\alpha$ excess candidates from narrow-band imaging (stars) and the radio
galaxy (square).  The circles represent
emitters with redshift smaller than the median redshift and the diamonds with those larger than the median. The size of the circles is scaled according to the relative velocity of the object. Larger circles and diamonds represent a larger relative velocity compared with the median velocity.
The structure appears to be bound in the northwest of the image and unbound in the south.}
\label{fig: 1338fors}       
\end{SCfigure*}

\begin{SCfigure*}
 \centering
\includegraphics[width=0.6\textwidth]{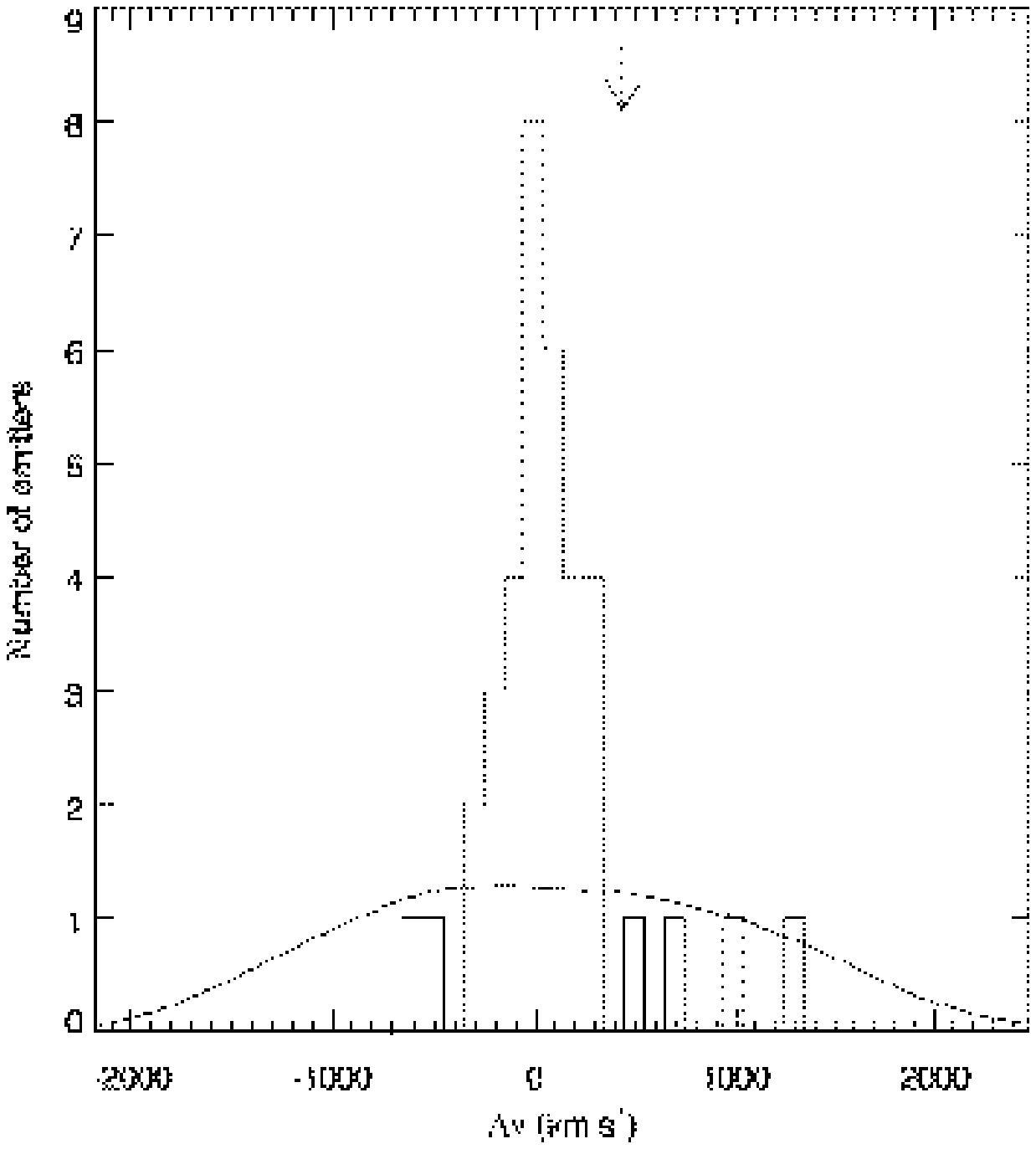}
\caption{HzRG-selected protocluster around TN J1338-1942 at z = 4.1 as in Figure \ref{fig: 1338fors}. [ From \citet{ven02,ven07}].
Velocity distribution of the confirmed Ly$\alpha$ emitters. The peak velocity of the radio galaxy is indicated by an arrow. The solid line represents the selection function of the narrow-band filter normalized to the total number of confirmed emitters.
Note that the velocity distribution of the detected emitters is substantially narrower than the filter width and centered within 200 km s$^{-1}$ of the redshift of the radio galaxy. Similar data are available for 6 protoclusters.}
\label{fig: 1338vel}       
\end{SCfigure*}

\begin{figure*}
\centering
\includegraphics[width=1.0\textwidth]{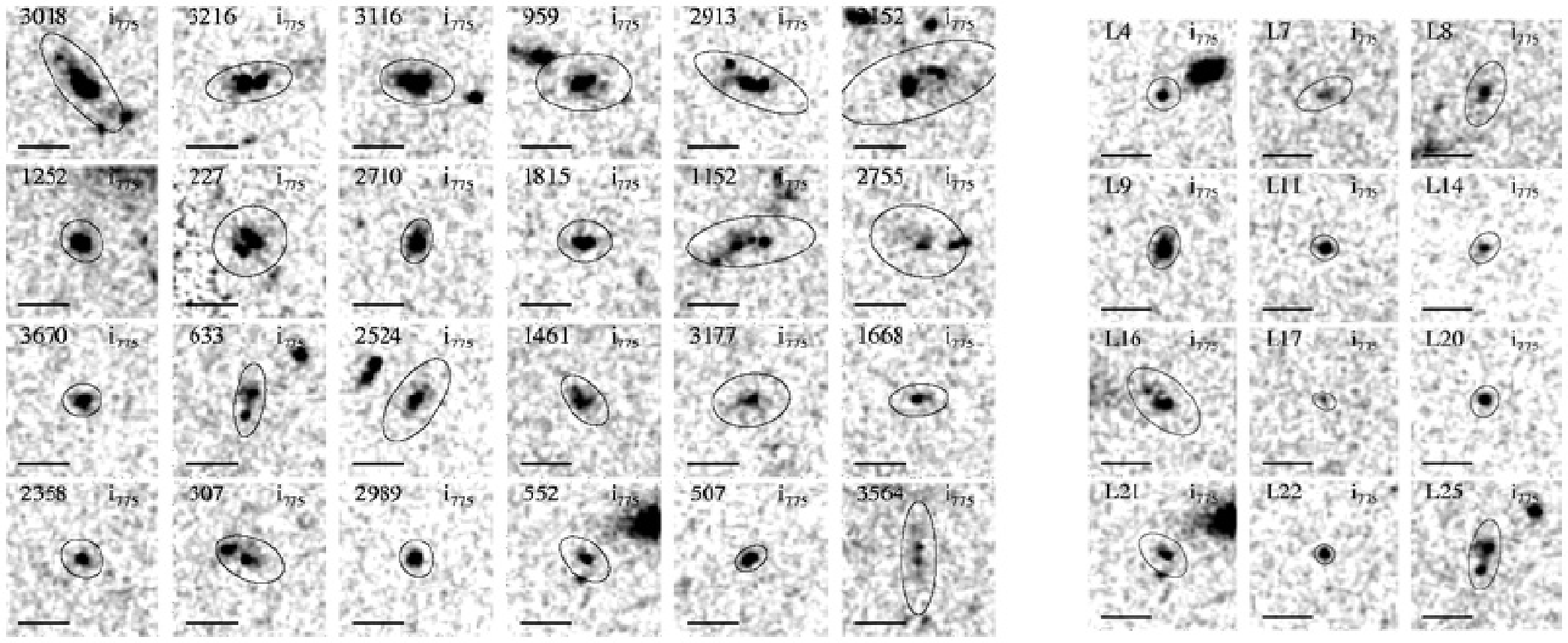}
\caption{Morphologies of galaxies in the protocluster TN J1338-1942 at z = 4.1 with the ACS on the Hubble Space Telescope [From \citet{ove07}]. The filter is $i_{775}$ and each image measures $3'' \times 3''$, corresponding to $\sim$ 20 kpc $\times$ 20 kpc at z $\sim$ 4. 
Left are images for a selection of the overdense $g_{475}$-dropout objects that are candidate Lyman break galaxies (LBGs) in the protocluster. Right are images of 12 spectroscopically confirmed Ly$\alpha$ excesses (LAE) protocluster members. There are a  wide  range  of 
 morphologies  present,  with  some  galaxies  showing  clear  evidence  for  small-scale  interactions. 
}
\label{fig: 1338lbg}       
\end{figure*}


\subsection {Properties of radio-selected protoclusters}
\label{propertycluster}

The largest and most comprehensive HzRG protocluster search project was based on a study of 8 radio galaxies, with redshifts ranging from 2.2 to 5.2, using the narrow-band Ly$\alpha$ search technique with the VLT \citep{pen00a,kur00,ven02,ven04,cro05,ven05,ven07}. Followup observations have been made of candidates in several of the targets using a variety of ground-based telescopes as well as the Hubble and Spitzer Telescope \citep{kur04a,kur04b,mil04,ove05,ove06,int06,kod07}. Typically, $\sim 3$ Mpc-scale regions around the HzRGs were covered in the Ly$\alpha$ searches.  Ly$\alpha$ redshifts were determined for 168 objects. Almost all these objects are star-forming galaxies at similar redshift to the HzRG, with typical star formation rates of a few $M\subsun$ per year, derived from the UV continuum and the Ly$\alpha$ luminosities.

Six of the 8 fields were found to be overdense in Lya emitters by a factor 3 - 5 as compared to the field density of Ly$\alpha$ emitters at similar redshifts. The 6 included {\it all} of the most radio-luminous objects (i.e. all HRzGs with radio luminosities L$_{2.7GHz}$ $>$ $6 \times 10^{33}$ erg s$^{-1}$ Hz$^{-1}$ sr$^{-1}$) \citep{ven07}. Because these targets were chosen arbitrarily, (redshifts that place Ly$\alpha $ in an available narrow-band filter), it is possible that all HzRGs with such large radio luminosities are embedded
in such overdensities.

Measured velocity dispersions  are in the range $\sim$ 300 - 1000 km s$^{-1}$, centered within a few hundred km/s of the mean velocity of the radio galaxies. Taking account of the relatively narrow peaks in the velocity distributions of the Ly$\alpha$ emitters compared with the widths of the imaging filters, the measured galaxy overdensities range from $\sim$ 5 - 15. Estimates for the sizes of the protocluster structures are limited by the typical size of the imaging fields used in the searches to date ($\sim$ 8' equivalent to $\sim$ 3 Mpc). Nevertheless, the protocluster sizes are estimated to be in the range $\sim$ 2 to 5 Mpc \citep{int06,ven07}.

Because cluster-size overdense structures at high-redshifts cannot be old enough to have become bound, the usual method of calculating cluster masses using the virial theorem cannot be applied.
However, estimates for the masses of the protoclusters can be obtained from the volume occupied by the overdensity, the
mean density of the Universe at the redshift of the protocluster,
the measured galaxy overdensity, and the bias parameter \citep[e.g.][]{ste98,ven07}. The masses obtained (a few times $10^{14}$ M$_{\sun}$ - $10^{15}$
M$_{\sun}$, \citep{kur04b,ven05}) are comparable to the masses of
local rich clusters.

It is interesting to inquire what is the relation of the protocluster structures to the general large-scale structure of the Universe. Is the topology of the protocluster filamentary, or do HzRG-selected protoclusters illuminate the densest most tangled regions of the cosmic web. For TN J1338-1942 at redshift z = 4.1, there are indications that the Mpc-sized protocluster of Ly$\alpha$ excess galaxies may be part of a larger-scale structure. A $25' \times 25'$ survey for Lyman break galaxy candidates (B-band dropouts) showed several significant density enhancements amidst large voids \citep{int06} and \citet{ove07}showed that this large scale structure ties in closely with significant sub-clustering across the $3' \times 3'$ ACS field near the radio galaxy. 

Future wide-field narrow-band imaging and spectroscopy  around HzRG-selected clusters should allow the topology of the Universe in the region of protoclusters to be mapped in detail, providing a glimpse of large-scale structure emerging in the early Universe for comparison with computer simulations.

\subsection{Are radio-selected protoclusters typical?}
\label{rclustyp}

We have seen that protocluster-like structures have been found around almost all of the most luminous radio galaxies with $z > 2$ that have been targeted. However only $\sim$ $50\%$ of powerful radio sources at z $\sim$ 0.5 are located in rich clusters and radio sources appear to avoid clusters at low redshift \citep{hil91}. What is known about the environment of radio galaxies having redshifts $1 < z < 2$? This range is difficult to study spectroscopically because of the redshift desert (Section \ref{findhzrg}). However, there have been several reports of detections of clusters and/ or excesses of red galaxies around radio AGN in this redshift range \citep{hal98,kod03,kad02,bes03a}.
We note that there is a strong correlation between the radio luminosities and redshift of objects in such studies due to Malmquist bias. Taking all the data together, there appears to be a substantial increase in the density of the environment around radio galaxies as a function of redshift and/or luminosity.

The statistics of the luminosity function of radio galaxies are consistent
with every brightest cluster galaxy (BCG) having gone through a luminous radio
phase during its evolution \citep{ven02,ven07}. This statement is based on the facts (i) that the space density of luminous steep-spectrum radio sources decreases by $\sim$ 100 between 2.5 $\geq$  z $>$ 0 (Section \ref{space}) and (ii) that the radio synchrotron lifetimes (few $\times $10$^7$yr) are $\sim$ 100 smaller than the cosmological time interval corresponding to the observed redshift range (few $\times $10$^9$yr).
Distant radio galaxies
may therefore be typical progenitors of galaxies that dominate the cores of
local clusters. Likewise, the ancestor of every rich cluster in the local Universe may have gone through a phase in which it hosted a HzRG. This would imply that HzRG-selected protoclusters are typical ancestors of local galaxy clusters.

\subsection{Protocluster evolution}
\label{protoev}

Radio-selected protoclusters are powerful laboratories for
tracing the emergence of large scale structure and for studying the evolution of galaxies in dense cluster environments. An interesting parameter that can be measured for line-emitting galaxies is the velocity dispersion of the protoclusters.
Although more statistics are needed, the velocity dispersion appears to decrease with increasing
redshift \citep{ven07}, consistent with the predictions from simulations of forming massive
clusters \citep[e.g.][]{eke96}.

A population study of the protocluster around the Spiderweb Galaxy PKS 1138-262 at z = 2.2 was carried out by \citet{kur04b, kur04a}, using deep optical and infrared observations with the VLT. Besides Ly$\alpha$ emitters, the study included objects having apparent H$\alpha$ excesses and extremely red objects, with colours characteristic of old galaxies with 4000 $\AA$ breaks at the redshift of the protocluster. An intriguing result of this study is that candidate H$\alpha$ emitters and 4000 $\AA$ break objects appear more concentrated towards the centre of the protocluster
than the Ly$\alpha$ emitters.  This indicates that the galaxies that are dominated by old stellar populations are more settled into the gravitational potential well of the protocluster, consistent with them being older.

An important diagnostic of galaxy and cluster evolution is the cluster colour-magnitude diagram (CMD). The presence of a narrow 'red sequence" in the cluster CMD is well established out to a redshift of 1.4 \citep{sta05}. The red colours imply that the galaxies are dominated by older stars, whose SED peaks redwards of the $4000 \AA$ break. The presence of this red sequence can be used to set a lower limit to the redshift at which the stellar populations formed. It is extremely difficult to measure the CMD at $z > 1.4$, even with the largest ground-based telescopes.

\begin{figure*}
\centering
\includegraphics[width=0.75\textwidth]{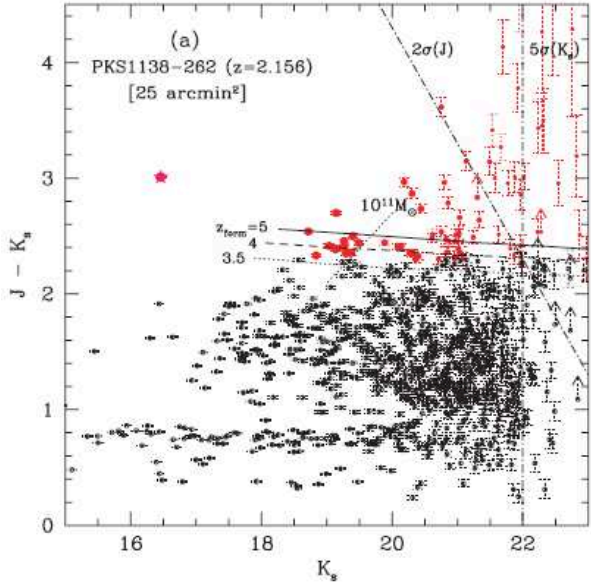}
\caption{Color-magnitude diagram of J - K$_{S}$ plotted against K$_{S}$ for the MRC 1138-262 protocluster at z = 2.2. [From \citet{kod07}].  Filled circles indicate protocluster member candidates selected using colour selection criteria. Large stars mark the targeted radio galaxies. Dotted error bars show 1s photometric errors. Solid, dashed and dotted lines show the expected location of the CMD at the relevant redshift for passive evolution. The iso-stellar mass lines of 10$^{11}$M$_\odot$ and 10$^{10}$M$_\odot$ are shown by thick dotted lines. The dot-dashed lines indicate 5 $\sigma$ (K$_{S}$) and 2$\sigma$ (J) detection limits.
}
\label{fig: 1138cmd}       
\end{figure*}

However, indications of an emergent red sequence has been found in the CMD of the protocluster MRC1138-262 at z = 2.2, both with a wide-field near-IR imager on the Subaru Telescope (\citet{kod07}; Fig. \ref{fig: 1138cmd} and with the much narrow field but more sensitive NICMOS on the HST (Zirm et al. 2007). The galaxy colours indicate that while some relatively quiescent galaxies exist in the protocluster, most of the galaxies are still undergoing star formation  Furthermore, \citet{kod07} studied 3 other radio-selected protoclusters and found that the fraction of red galaxies in the 4 protoclusters increases between z $\sim$ 3 and z $\sim 2$. To summarise, there is strong evidence that most protocluster galaxies undergo substantial star formation between z $\sim$ 3 and z $\sim$ 2 and that the bright end of the red sequence is still being formed during this epoch.

Information about the early evolution of stellar populations in clusters has been obtained at even higher redshifts. In the protocluster surrounding the HzRG TN J1338-1942 at $ z = 4.1$, the distribution of Ly$\alpha$ emitters (LAEs) is highly filamentary and appears to avoid the locations of Lyman break galaxy candidates \citep{ove07}. A similar spatial segregation between LAEs and LBGs was observed in a structure around QSO SDSS J0211-0009 at z=4.87 \citep{kas07}. This indicates that an age- or mass-density relation was emerging little more than 1 Gyr after the Big Bang, when the Universe was only 10\% of its present age.

We note that derived masses of z $>$ 2 LBGs and LAEs are $\gtrsim10$ smaller than the masses of early-type galaxies in local clusters, indicating that a large fraction of the stellar mass still has to
accumulate through merging \citep{ove07}. Detailed observations of protocluster
regions on much larger scales ($\sim$50 co-moving Mpc) are needed to
test if the number density of LBGs is indeed consistent with forming
the cluster red sequence population through merging. Simulations show that clusters with masses of $>10^{14}$ $M_\odot$ can be traced back to regions at z =4 - 5 of 20 - 40 Mpc in size, and that these regions are associated with overdensities of typical dark matter halos hosting LAEs and LBGs of $\delta_g\sim3$ and mass overdensities $\delta_m$ in the range 0.2 - 0.6 \citep{Suw06}. Also, recent numerical simulations of CDM growth predict that quasars at $z\sim6$ may lie in the center of very massive dark matter halos of $\sim4\times10^{12}$ $M_\odot$ \citep{spr05b,li07}. They are surrounded by many fainter galaxies, that will evolve into massive clusters of $\sim 4\times10^{15}$ $M_\odot$ at $z=0$.

The addition of the new IR - oprical camera WFC3 to the HST and the advent of several new wide-field imagers and multi-object spectrographs on ground-based telescopes will allow the detailed evolution of protoclusters to be studied in great detail during the next few years.


\section{CONCLUSIONS AND FUTURE}
\label{Conclusions}

{\it ''Curioser and curiouser!" cried Alice (she was so much surprised that for the moment she quite forgot how to speak good English). "Now I'm opening out like the largest telescope that ever was!"
Alice's Adventures in Wonderland, Lewis Caroll, 1865}
\\
\\
In this review we have shown that HzRGs are fascinating objects in their own right and that they provide important diagnostics for studying the early Universe. Knowledge about HzRGs and their evolution is fundamental to understanding the formation and evolution of galaxies and the large scale structure of the Universe.

There are still many aspects of HzRGs that are not understood. Here are some open questions.

\begin{itemize}

\item  What is the particle acceleration mechanism in the relativistic plasma and why do HzRGs have much steeper radio spectra than nearby radio sources (Section \ref{alphaz}).

\item Why caused the most luminous radio galaxies and quasars to become virtually extinct between z $\sim$ 2 and the present (Sections \ref{space} and \ref{dinosaur})?

\item What is the nature, extent and kinematics of the hot gas, one of the most massive and least studied constituents of HzRGs (Section \ref{hotgas})?

\item What is the origin of the Ly$\alpha$ halo that appears to be falling into the HzRG and how is this warm gas (filling factor $\sim$ 10$^{-5}$) distributed with respect to the hotter and colder gas (filling factors $\sim$ 1) (Sections \ref{infall} and \ref{alphaz})?

\item What are the detailed processes by which the radio jets interact with the gas and trigger starbursts and how important is jet-induced star formation for producing stars in the early Universe (Sections \ref{jetgas} and \ref{alignment})?

\item What are the temperature and densities and composition of the molecular gas and dust? how is this distributed spatially and what are the implications for the star formation histories(Sections \ref{molgas} and \ref{dust})?

\item What is the relative importance of AGN and starburst heating in the mid-IR emission of HzRG?

\item What effect does galaxy merging have on star formation and what physical effects are responsible for downsizing (Section \ref{mass})?

\item What effect does feedback between the AGN and the galaxy have on the evolution of HzRGs and the general evolution of massive galaxies (Section \ref{mass})?

\item What is the detailed mechanism by which the SMBHs produce quasars and luminous collimated jets?

\item How do the AGN/quasars vary on long time-scales and what are the causes of the variability
(Sections \ref{qso} and \ref{mbh})?

\item How is the SMBH built up and what role does merging play in this evolution (Section \ref{mbh})?

\item What is the size distribution of radio-selected protoclusters and what is is the topology of the cosmic web in the neighbourhood of the HzRGS (Section \ref{Protoclusters})?

\item How do the various populations and constituents of distant protoclusters evolve and eventually become virialised clusters (Section \ref{protoev})?

\item What is the radio luminosity function of galaxies in radio-selected protoclusters?  Is the radio emission from the protocluster galaxies an important contributor to forming the radio halos, that are seen at the centres of many nearby rich clusters?  What is the strength and configuration of the protocluster magnetic fields (Sections \ref{Protoclusters} and \ref{relativistic})?

\end{itemize}

There are good prospects for making progress during the next decade. Since its inception, the study of radio galaxies, has been observationally driven.
Several forefront astronomical facilities now being constructed or planned that will give new insights into the nature of HzRGs and their environments.

First, with a combination of sensitivity and spatial resolution, the new low-frequency radio arrays, LOFAR \citep{rot06} and the LWA \citep{kas06}, will open up the frequency window  below $\sim$ 50 MHz for HzRG studies. LOFAR will survey the sky to unprecedented depth at
low-frequencies and will therefore be sensitive to the relatively rare radio sources that have extremely steep spectra. Because of the $\alpha$ vs z relation (Section \ref{alphaz}), LOFAR is likely to detect HzRGs at z $\sim$ 8, if they exist.  Studies of detailed low-frequency spectra and their spatial variations will provide new information about the mechanism responsible for the $\alpha$ vs z relation.
Presently combination of the new radio surveys with planned new deep optical and infrared wide-field surveys, such as PAN-STARRS \citep{hod04} and those with the VST and VISTA \citep{arn07} will be used to identify HzRGs and provide photometric data.

Another task for sensitive radio arrays, such as LOFAR, the EVLA \citep{nap06} and eventually the Square Kilometre Array (SKA), will be to survey the radio emission of galaxies in protoclusters. The new arrays will study radio emission produced by relativistic jets and be able to detect and investigate radio emission from the brightest star forming protocluster galaxies.

Secondly, ALMA \citep{bro04} and the EVLA, with their unprecedented sensitivities and resolutions at millimetre and sub-millimetre wavelengths, will revolutionise the study of molecular gas and dust. Several different CO transitions can be observed, allowing entire ``CO ladders'' to be constructed and the density and temperature structure of the molecular gas to be unraveled. Fainter molecular lines can be used to trace even denser gas than that studied until now. Important information about the dust composition  and the gas to dust ratios is likely to be obtained. 

ALMA's sensitivity at millimetre wavelengths should also facilitate observations of the atomic CI Carbon lines in HzRGs. This would provide an important constraint on the global metalicity of the gas. The fine-structure line of C$^+$ at $\lambda_{\rm rest}$=157.74$\mu$m line is one of the main cooling lines in nearby galaxies, and has now also been detected in several of the most distant quasars known. 

Thirdly, we can expect considerable progress in disentangling the detailed evolutionary history both of HzRGS and of radio-selected protoclusters. This evolutionary detective work will be pursued by combining spectroscopic data from the next generation of spectrographs on 8m-class telescopes with imaging results from the new camera, WFC3 \citep{lec98}, on the Hubble Space Telescope. For example, the detection
of supernovae in z $\sim$ 2 protoclusters will become possible.
On a longer timescale, tracing the detailed history of the formation and evolution of HzRGS and the surrounding protoclusters will be helped enormously with the advent of 30m -class ground-based telescopes in the optical and near-infrared and the James Webb Space Telescope (JWST) \citep{gar06} in the near and mid-infrared.

Fourthly, the next generation X-ray telescope, such as XEUS \citep{bav06} or Constellation-X \citep{tan06} , will make observations of X-rays from HzRGs an important tool for studying galaxy formation. It will have sufficient sensitivity to perform spectroscopic studies of hot gas in HzRGs.  The 0.3 - 10 keV X-ray band contains the inner (K-shell) lines for all of the abundant metals from carbon to zinc as well as many L-shell lines. These atomic transitions provide important new plasma diagnostics of the HzRG hot gas.

Fifthly, and perhaps most exciting, the potential discovery of HzRGS with z $>$ 6 could open up a unique new window for studying the very early Universe during the "Epoch of Reionisation". Recent observational constraints on suggest that cosmic reionization may have taken place between z $\sim$ 11 and  z $\sim$ 6 \citep{fan06}. The existence of HzRGs within the near edge of cosmic reionization could be used as sensitive probes of intermediate- to small-scale structures in the neutral IGM, through redshifted HI absorption observations (Section \ref{h1}) \citep{car02,fur02}, complementary to the very large scale that can be studied in HI emission.

Finally, we point out that in trying to understand HzRGs or in using them to probe the properties of the early Universe, we should exercise humility, remembering that 96\% of the energy density in the universe is in a form of dark matter and dark energy, that are not directly observable.

HzRGS pinpoint the progenitors of cosmic megacities in the local Universe. Studying them is akin  to investigating how human civilisations were urbanised.  During the last few decades we have obtained some glimmerings of how and why this cosmic urbanisation occurred. During the next few decades we shall learn a great deal more about the origin of the megacities, the nature of the first city dwellers, their organisational structure and their successes and failures.

\nocite{*}
\begin{acknowledgements}
We are grateful for suggestions, useful comments and/or  figures from  Wil van Breugel, Ilana Feain, Jaron Kurk, Nicole Nicole Nesvadba, Roderik Overzier, Michiel Reuland, Birgitte Rocca, Huub R{\"o}ttgering, Bram Venemans and Montse Villar-Martin. GM acknowledges support from an Academy Professorship of the Royal Netherlands Academy of Arts and Sciences.
\end{acknowledgements}
\section{APPENDIX: KNOWN HZRGs}
\label{Appendix}
This appendix is a compendium of HzRGs (z $>$ 2,  L$_{500(rest)}$ $>$ 10${^{27.5}}$ erg s${^{-1}}$ Hz${^{-1}}$) known at the time of writing. Table 3 lists the HzRGs, their positions, redshifts, radio luminosities extrapolated to 500MHz rest frequency and relevant references. For convenience the HzRGs are listed in order of right ascencion in Table 4.

The luminosities were calculated using
$$L_{\nu}=4\pi D_L^2 S_{\nu} (1+z)^{-\alpha-1},$$ where D$_{L}$ is the luminosity distance
\footnote{assuming $H_0$=70\,km\,s$^{-1}$\,Mpc$^{-1}$,
$\Omega_{\Lambda}$=0.7 and $\Omega_{\rm M}$=0.3}, $S_{\nu}$ is the
{\it observed} flux density at the rest frequency, and $\alpha$ is the spectral index ($S_{\nu} \propto \nu^{\alpha}$).

Why did we choose a rest-frame frequency of 500 MHz to express the radio luminosities? The radio sources have steep spectra and are generally taken from low-frequency surveys, so it is appropriate to express their luminosities at as low a rest frequency as possible \citep[e.g.][]{blu00}.
For consistency, surveys covering the northern and southern sky with comparable spatial resolution should be used. The most appropriate surveys are the 74\, MHz VLA Low-frequency Sky Survey \citep[VLSS;][]{coh07} and the 1.4\,GHz NVSS \citep{con98}. These contain $\sim$ 75\% of the known HzRGs. For the entire redshift range of the sample, the observed frequency associated to $\nu_{\rm rest}$=500\,MHz is located close to the VLSS frequency, but still located between VLSS and NVSS, thereby avoiding extrapolations to un-observed parts of the radio spectrum.

For HzRGs not included in these catalogues, the following combination of surveys were used in order of preference: WENSS \citep{ren97} and NVSS, Texas \citep{dou96} and NVSS, SUMSS \citep{mau03} and NVSS, or MRC \citep{lar81} and PMN \citep{gri93}. A small number of HzRGs from the CENSORS survey \citep{bro06} do not yet have measured radio spectra. For these we assumed a constant spectral index of $\alpha$ = -1.1, leading to uncertainties in the derived luminosities that could be as much as 0.5 dex. The relevant luminosities are indicated by the daggers.

The table also lists references for the redshifts and relevant followup observations. Although references to the followup work are not complete, they provide a useful pointer for readers who require further information. The key to the references to followup observations are: {\bf CO} = CO,  {\bf G}  = General,  {\bf H} = HST imaging,  {\bf HI} = HI absorption,  {\bf M} = Millimetre/ submillimetre, {\bf N} = Nebular continuum, {\bf OP} = Optical polarisation, {\bf OS} = Optical/infrared spectroscopy,  {\bf P} = Protocluster,  {\bf R} = Radio,  {\bf S} Spitzer IR and {\bf X} = X-rays.

\clearpage
\newpage


\Bibpunct{[}{]}{;}{n}{}{,} %

\begin{sidewaystable}
\begin{center}
{Table 3\par Known radio galaxies with z $>$ 2 and radio luminosities L$_{500(rest)}$ $>$ 10$^{27.5}$W Hz$^{-1}$} 
\label{tablehzrg1}
\\*
\scriptsize{
\begin{tabular}{|p{78pt}|p{38pt}p{38pt}|p{15pt}p{15pt}|p{35pt}|p{245pt}|}
\hline
Name & RA(J2000) & DEC(J2000) & z & Ref. & Log($L_{500}$)& References\\
\hline
MRC 2036-254        & 20:39:24.5 & $-$25:14:31 & 2.000 & \citep{mcc96} & 28.52 & {\bf R}\citep{car97,ath97,ath98a,ath98b} {\bf X}\citep{ove05} \\
MRC 0015-229        & 00:17:58.2 & $-$22:38:04 & 2.01  & \citep{mcc96} & 28.32 & {\bf R}\citep{car97,ath97,ath98a,ath98b}  \\
MRC 0156-252        & 01:58:33.5 & $-$24:59:30 & 2.016 & \citep{mcc90} & 28.46 & {\bf H}\citep{pen01}{\bf R}\citep{car97,ath87,ath98a,ath98b,rei99}  {\bf S}\citep{sey07} {\bf X}\citep{ove05} \\
MG 2348+0507        & 23:48:32.0 & $+$05:07:33 & 2.019 & S2            & 28.58 &  \\
MP J1758-6738       & 17:58:52.9 & $-$67:38:34 & 2.026 & \citep{deb01} & 28.99 &  \\
TXS 0448+091        & 04:51:14.6 & $+$09:14:29 & 2.037 & \citep{rot97} & 28.11 & {\bf R}\citep{car97}  \\
NVSS J002402-325253 & 00:24:02.3 & $-$32:52:54 & 2.043 & \citep{deb06} & 27.93 &  \\
MRC 2048-272        & 20:51:03.4 & $-$27:03:05 & 2.06  & \citep{mcc96} & 28.72 & {\bf H}\citep{pen01} {\bf R}\citep{pen00b}{\bf S}\citep{sey07}{\bf X}\citep{ove05} \\
B3 1204+401         & 12:07:06.2 & $+$39:54:39 & 2.066 & \citep{tho94} & 28.33 & {\bf R}\citep{pen00b} \\
PKS 0011-023        & 00:14:25.5 & $-$02:05:56 & 2.080 & \citep{dun89} & 28.15 & {\bf R}\citep{pen00b} \\
5C 7.223            & 08:24:06.4 & $+$26:28:12 & 2.092 & \citep{wil02} & 27.71 &  \\
MG 1401+0921        & 14:01:18.3 & $+$09:21:24 & 2.093 & \citep{ste99a}& 28.26 &  \\
7C 1740+6640        & 17:40:42.2 & $+$66:38:44 & 2.10  & \citep{lac99} & 27.77 &  \\
7C 1802+6456        & 18:02:22.1 & $+$64:56:45 & 2.110 & \citep{lac99} & 28.18 &  \\
TN J1102-1651       & 11:02:47.1 & $-$16:51:34 & 2.111 & \citep{deb01} & 28.07 &  \\
4C +39.21           & 07:58:08.8 & $+$39:29:28 & 2.119 & \citep{fan01} & 28.51 &  \\
6C 1134+369         & 11:37:07.7 & $+$36:39:54 & 2.125 & \citep{raw90} & 28.31 & {\bf R}\citep{pen00b} \\
TXS 2034+027        & 20:36:34.8 & $+$02:56:55 & 2.129 & \citep{deb01} & 28.45 & {\bf R}\citep{pen00b}\\
TXS 0214+183        & 02:17:25.8 & $+$18:37:03 & 2.131 & \citep{rot97} & 28.46 & {\bf OS}\citep{van97}{\bf R}\citep{car97}  \\
WN J1242+3915       & 12:42:53.1 & $+$39:15:49 & 2.131 & \citep{deb01} & 28.03 &  \\
4C +40.49           & 23:07:53.5 & $+$40:41:49 & 2.140 & \citep{fan01} & 28.46 &  \\
NVSS J103615-321659 & 10:36:15.3 & $-$32:16:57 & 2.144 & \citep{bry07} & 27.82 &    \\
4C +43.31           & 13:52:28.5 & $+$42:59:23 & 2.149 & \citep{fan01} & 28.43 &  \\
NVSS J144932-385657 & 14:49:32.8 & $-$38:56:58 & 2.152 & \citep{bry07} & 28.07 &    \\
TXS 0355-037        & 03:57:48.0 & $-$03:34:08 & 2.153 & \citep{rot97} & 28.34 & {\bf OS}\citep{van97} \\
TNR 2254+1857       & 22:54:53.7 & $+$18:57:04 & 2.154 & \citep{deb01} & 27.80 &  \\
MRC 1138-262        & 11:40:48.3 & $-$26:29:10 & 2.156 & \citep{mcc96} & 29.07 & {\bf G}\citep{pen97}{\bf H}\citep{pen98,pen01,mil06}{\bf M}\citep{ste03,reu04}{\bf OS}\citep{nes06} {\bf P}\citep{pen00a,kur00,pen02,kur04a,kur04b,cro05,kod07,ven07}{\bf R}\citep{car97,ath97,ath98a,ath98b}{\bf S}\citep{sey07}{\bf X}\citep{car98,car02}  \\
6C**0746+5445       & 07:50:24.6 & $+$54:38:07 & 2.156 & \citep{cru06} & 27.64 &  \\
6C* 0024+356        & 00:26:52.0 & $+$35:56:24 & 2.161 & \citep{jar01b}& 28.04 &  \\
MRC 0030-219        & 00:33:23.9 & $-$21:42:01 & 2.168 & \citep{mcc90} & 28.14 & {\bf R}\citep{car97,ath98a}  \\
B2 1056+39          & 10:59:11.5 & $+$39:25:01 & 2.171 & \citep{lil89} & 28.20 & {\bf R}\citep{pen00b} \\
6C* 0135+313        & 01:38:06.6 & $+$31:32:42 & 2.199 & \citep{jar01b}& 27.97 &  \\
5C 7.10             & 08:11:26.5 & $+$26:18:19 & 2.185 & \citep{wil02} & 28.21 &  \\
5C 7.269            & 08:28:39.7 & $+$25:27:30 & 2.218 & \citep{eal96} & 27.78 & {\bf M}\citep{arc01} \\
5C 7.271            & 08:28:59.5 & $+$24:54:00 & 2.224 & \citep{wil02} & 27.97 &  \\
\hline
\end{tabular}
}
\label{listhzrg}
\end{center}
 S2   Spinrad et al MG list (private communication)\\
\end{sidewaystable}
\pagebreak

\begin{sidewaystable}
\begin{center}
{Table 3, continued}
\\*
\scriptsize{
\begin{tabular}{|p{78pt}|p{38pt}p{38pt}|p{15pt}p{15pt}|p{35pt}|p{245pt}|}
\hline
Name & RA(J2000) & DEC(J2000) & z & Ref. & Log($L_{500}$) & References\\
\hline
6C* 0142+427        & 01:45:29.0 & $+$42:57:42 & 2.225 & \citep{jar01b}& 28.10 &  \\
TXS 0200+015        & 02:02:42.9 & $+$01:49:11 & 2.229 & \citep{rot97} & 28.07 & {\bf OA}\citep{van97,jar03,bin06} {\bf R}\citep{car97}  \\
NVSS J101008-383629 & 10:10:08.0 & $-$38:36:29 & 2.236 & \citep{bry07} & 28.30 &    \\
TXS 1113-178        & 11:16:14.7 & $-$18:06:23 & 2.239 & \citep{rot97} & 28.46 & {\bf R}\citep{car97}  \\
6C* 0115+394        & 01:17:55.3 & $+$39:44:33 & 2.241 & \citep{jar01b}& 27.93 &  \\
6C 0629+53          & 06:33:52.0 & $+$53:16:31 & 2.246 & \citep{eal96} & 28.35 &  \\
6C 0901+54          & 09:05:24.2 & $+$54:05:42 & 2.249 & \citep{eal96} & 28.11 &  \\
TN J0452-1737       & 04:52:26.7 & $-$17:37:54 & 2.26  & \citep{deb01} & 28.09 &  \\
4C 40.36            & 18:10:55.7 & $+$40:45:23 & 2.265 & \citep{cha88},\par \citep{ver01a}& 28.79 & {\bf G}\citep{cha88,cha96b}{\bf M}\citep{arc01}{\bf OP}\citep{ver01a} {\bf OS}\citep{eal93b,vil03,hum06}{\bf R}\citep{car97,cai02}{\bf S}\citep{sey07}\\
6C* 0152+463        & 01:55:45.6 & $+$46:37:11 & 2.279 & \citep{jar01b}& 27.97 &  \\
4C 40.22            & 08:59:59.6 & $+$40:24:36 & 2.28  & \citep{pah95} & 28.39 &  \\
MRC 1324-262        & 13:26:54.7 & $-$26:31:43 & 2.28  & \citep{mcc96} & 28.46 & {\bf R}\citep{car97,ath98a,ath98b}  \\
MG 1747+1821        & 17:47:07.0 & $+$18:21:11 & 2.281 & \citep{arc01} & 28.86 & {\bf M}\citep{arc01} {\bf OS}\citep{eal93b} {\bf R}\citep{car97,pen00b}\\
6C* 0106+397        & 01:09:25.4 & $+$40:00:01 & 2.284 & \citep{jar01b}& 28.11 &  \\
MP J0340-6507       & 03:40:44.6 & $-$65:07:12 & 2.289 & \citep{bor07} & 28.82 &  \\
6C 1106+380         & 11:09:28.9 & $+$37:44:31 & 2.290 & \citep{eal96} & 28.44 &  \\
MG 0936+0503        & 09:36:10.1 & $+$05:03:49 & 2.306? & S2            & 28.08 &  \\      
MG 1251+1104        & 12:51:00.1 & $+$11:04:20 & 2.322 & \citep{ste99a}& 28.42 & {\bf M}\citep{arc01}  \\
MRC 0349-211        & 03:51:11.8 & $-$20:58:00 & 2.329 & \citep{mcc91a}& 28.28 & {\bf R}\citep{ath97,ath98a,ath98b}  \\
TXS 0211-122        & 02:14:17.4 & $-$11:58:47 & 2.340 & \citep{van94}\par \citep{ver01a}& 28.48 & {\bf G}\citep{van94}{\bf H}\citep{pen99,pen01}{\bf OP}\citep{ver01a}{\bf OS}\citep{van97,vil03,hum06,hum07} {\bf R}\citep{car97}  \\
4C 48.48            & 19:33:05.6 & $+$48:11:46 & 2.343 & \citep{cha96} & 28.29 & {\bf G}\citep{cha96b}{\bf M}\citep{arc01} {\bf OP}\citep{ver01a} {\bf OS}\citep{vil03,hum06,hum07} {\bf R}\citep{car97} \\
TXS 1707+105        & 17:10:06.9 & $+$10:31:09 & 2.349 & \citep{rot97} & 28.63 & {\bf H}\citep{pen99,pen01} {\bf OS}\citep{van97,hum07} {\bf S}\citep{sey07}\\
BRL 0128-264        & 01:30:27.9 & $-$26:09:57 & 2.348 & \citep{bes99} & 29.13 &   \\
6C* 0118+486        & 01:21:16.3 & $+$48:57:40 & 2.350 & \citep{jar01b}& 27.83 &   \\
NVSS J094949-213432 & 09:49:49.0 & $-$21:34:33 & 2.354 & \citep{bro06} & 26.91$^{\dag}$ &   \\
PKS 1425-148        & 14:28:41.7 & $-$15:02:28 & 2.355 & \citep{deb01} & 28.66 & {\bf R}\citep{pen00b}  \\
NVSS J232651-370909 & 23:26:51.5 & $-$37:09:11 & 2.357 & \citep{deb06} & 28.02 &   \\
4C -00.54           & 14:13:15.1 & $-$00:23:00 & 2.360 & \citep{rot97}\par \citep{ver01a}& 28.41 & {\bf H}\citep{pen99,pen01}{\bf OP}\citep{cim98,ver01a}{\bf OS}\citep{van97,vil03,hum06,hum07} {\bf P}\citep{kaj06} {\bf R}\citep{car97} {\bf S}\citep{sey07}\\
MG 0001+0846        & 00:01:15.5 & $+$08:46:39 & 2.36? & S2            & 28.53 &   \\   
B2 1159+39A         & 12:01:50.0 & $+$39:19:11 & 2.370 & \citep{fan01} & 28.22 &   \\
6C**0854+3500       & 08:57:15.9 & $+$34:48:24 & 2.382 & \citep{cru06} & 27.92 &   \\
LBDS 53W002         & 17:14:14.8 & $+$50:15:30 & 2.390 & \citep{win91} & 27.78 & {\bf CO}\citep{sco97,all00}{\bf M}\citep{arc01} {\bf OS}\citep{eal93b}{\bf S}\citep{sey07} \\
6C 0930+389         & 09:33:06.9 & $+$38:41:48 & 2.395 & \citep{eal96} & 28.41 & {\bf M}\citep{arc01}{\bf R}\citep{pen00b}{\bf S}\citep{sey07} \\
\hline
\end{tabular}
}
\end{center}
{$^{\dag}$ Radio luminosity calculated assuming a spectral index $\alpha$ = -1.1.}\\
S2   Spinrad et al MG list (private communication)\\
\end{sidewaystable}
\pagebreak

\begin{sidewaystable}
\begin{center}
{Table 3, continued}
\\*
\scriptsize{
\begin{tabular}{|p{78pt}|p{38pt}p{38pt}|p{15pt}p{15pt}|p{35pt}|p{245pt}|}
\hline
Name & RA(J2000) & DEC(J2000) & z & Ref. & Log($L_{500}$) & References\\
\hline
4C 34.34            & 11:16:30.4 & $+$34:42:24 & 2.400 & \citep{eal96} & 28.46 & {\bf M}\citep{arc01}  \\
7C 1736+650         & 17:36:37.5 & $+$65:02:28 & 2.400 & \citep{lac99} & 27.65 &   \\
TXS 0748+134        & 07:51:01.1 & $+$13:19:27 & 2.419 & \citep{rot97} & 28.31 & {\bf OS}\citep{van97} {\bf R}\citep{car97}  \\
NVSS J095226-200105 & 09:52:26.5 & $-$20:01:05 & 2.421 & \citep{bro06} & 26.97$^{\dag}$ &   \\
TN J1033-1339       & 10:33:10.7 & $-$13:39:52 & 2.425 & \citep{deb01} & 28.35 &   \\
NVSS J094925-203724 & 09:49:25.9 & $-$20:37:24 & 2.427 & \citep{bro06} & 27.27$^{\dag}$ &   \\
4C 40.02            & 00:30:49.0 & $+$41:10:49 & 2.428 & \citep{eal96} & 28.64 &   \\
B3 0731+438         & 07:35:21.9 & $+$43:44:21 & 2.429 & \citep{mcc91b}& 28.89 & {\bf OP}\citep{ver01a} {\bf OS}\citep{eal93b,vil03,hum06,hum07} {\bf R}\citep{car97} \\
MRC 1106-258        & 11:08:30.3 & $-$26:05:06 & 2.43  & \citep{mcc96} & 28.48 & {\bf R}\citep{car97,ath97,ath98a,ath98b}  \\
5C 7.15             & 08:12:19.2 & $+$26:30:18 & 2.433 & \citep{wil02} & 28.12 &   \\
MRC 0406-244        & 04:08:51.4 & $-$24:18:17 & 2.44  & \citep{mcc91a}& 29.03 & {\bf H}\citep{pen01} {\bf OS}\citep{eal93b} {\bf P}\citep{kaj06}{\bf R}\citep{car97,ath97,ath98a,ath98b}{\bf S}\citep{sey07}{\bf X}\citep{ove05}  \\
6C**0834+4129       & 08:37:49.2 & $+$41:19:54 & 2.442 & \citep{cru06} & 27.80 &   \\
8C 1536+620         & 15:37:09.5 & $+$61:55:36 & 2.450 & LR            & 28.25 &   \\
MG 2308+0336        & 23:08:25.0 & $+$03:37:04 & 2.457 & \citep{ste99a}& 28.51 & {\bf M}\citep{arc01} \\
4C 12.32            & 08:55:21.4 & $+$12:17:27 & 2.468 & \citep{gop95} & 28.56 &   \\
3C 257              & 11:23:09.4 & $+$05:30:18 & 2.474 & \citep{van98} & 29.16 & {\bf M}\citep{arc01}{\bf OS}\citep{eal93b}{\bf S}\citep{sey07} \\
4C 23.56            & 21:07:14.8 & $+$23:31:45 & 2.479 & \citep{cha96}\par \citep{ver01a}& 28.93 & {\bf CO}\citep{eva96}{\bf G}\citep{cha96b} {\bf M}\citep{arc01}{\bf OP}\citep{cim98,ver01a,ver01} {\bf OS}\citep{vil03,hum07}{\bf P}\citep{kaj06}{\bf R}\citep{car97}{\bf S}\citep{sey07}{\bf X}\citep{joh07} \\
TXS 2332+154        & 23:34:58.0 & $+$15:45:50 & 2.481 & \citep{deb01} & 28.36 &   \\
MRC 2104-242        & 21:06:58.2 & $-$24:05:11 & 2.491 & \citep{mcc90} & 28.84 & {\bf H}\citep{pen99,pen01} {\bf OS}\citep{vil03,vil06,hum06,hum07} {\bf P}\citep{kaj06}{\bf R}\citep{ath97,ath98a,ath98b,pen00b}{\bf S}\citep{sey07}\\
WN J0303+3733       & 03:03:26.0 & $+$37:33:42 & 2.505 & \citep{deb01} & 28.44 &   \\
TXS 1647+100        & 16:50:05.0 & $+$09:55:11 & 2.509 & \citep{deb01} & 28.53 & {\bf R}\citep{pen00b} \\
NVSS J213510-333703 & 21:35:10.5 & $-$33:37:04 & 2.518 & \citep{deb06} & 27.90 &   \\
TXS 1558-003        & 16:01:17.4 & $-$00:28:46 & 2.520 & \citep{rot97} & 28.82 & {\bf OS}\citep{van97,vil03,hum06,hum07,vil07b} {\bf P}\citep{kaj06,kod07} {\bf R}\citep{pen00b}{\bf S}\citep{sey07}\\
8C 1039+681         & 10:42:37.4 & $+$67:50:24 & 2.530 & \citep{arc01} & 28.56 & {\bf M}\citep{arc01}{\bf R}\citep{pen00b}\\
NVSS J231144-362215 & 23:11:45.2 & $-$36:22:15 & 2.531 & \citep{deb06} & 27.83 &   \\
6C* 0112+372        & 01:14:50.2 & $+$37:32:33 & 2.535 & \citep{jar01b}& 28.22 &   \\
TXS 1436+157        & 14:39:04.9 & $+$15:31:19 & 2.538 & \citep{rot97} & 28.24 & {\bf OS}\citep{van97} {\bf R}\citep{car97}   \\
WN J1115+5016       & 11:15:06.9 & $+$50:16:24 & 2.54  & \citep{deb01} & 27.82 & {\bf M}\citep{reu04} {\bf S}\citep{sey07} \\
MRC 2139-292        & 21:42:16.7 & $-$28:58:38 & 2.55  & \citep{mcc96} & 28.73 & {\bf R}\citep{car97,ath98a,ath98b} \\
MP J1755-6916       & 17:55:29.8 & $-$69:16:54 & 2.551 & \citep{deb01} & 28.95 & {\bf OA}\citep{wil04}  \\
TXS 2319+223        & 23:21:42.4 & $+$22:37:55 & 2.554 & \citep{deb01} & 28.46 & {\bf R}\citep{pen00b}  \\
6C**1045+4459       & 10:48:32.4 & $+$44:44:28 & 2.571 & \citep{cru06} & 28.05 &   \\
TXS 0828+193       & 08:30:53.4 & +$1$9:13:16 & 2.572 & \citep{rot97}\par \citep{ver01a}& 28.44 & {\bf H}\citep{pen99}{\bf OP}\citep{kno97,ver01a,ver01}{\bf OS}\citep{van97,vil02,vil03,hum06,hum07}  {\bf R}\citep{car97}{\bf S}\citep{sey07} {\bf X}\citep{ove05}  \\
\hline
\end{tabular}
}
\label{listhzrg}
\end{center}
LR   Lacy and Rawlings (private communication)\\
\end{sidewaystable}
\pagebreak

\begin{sidewaystable}
\begin{center}
{Table 3, continued}
\\*
\scriptsize{
\begin{tabular}{|p{78pt}|p{38pt}p{38pt}|p{15pt}p{15pt}|p{35pt}|p{245pt}|}
\hline
Name & RA(J2000) & DEC(J2000) & z & Ref. & Log($L_{\rm 500 MHz}$) & References\\
\hline
PKS 0529-549        & 05:30:25.4 & $-$54:54:21 & 2.575 & \citep{rot97}\par \citep{bro07}& 29.16 & {\bf OS}\citep{van97}{\bf R}\citep{bro07}{\bf S}\citep{sey07}  \\
TXS 2353-003        & 23:55:37.0 & $-$00:02:58 & 2.592 & \citep{deb01} & 28.76 & \\
WNH 1736+6504       & 17:36:37.4 & $+$65:02:35 & 2.6   & \citep{ren99} & 27.74 &   \\
MRC 2052-253        & 20:55:01.5 & $-$25:09:53 & 2.60  & \citep{mcc96} & 28.77 & {\bf R}\citep{pen00b}  \\
WN J0040+3857       & 00:40:56.2 & $+$38:57:30 & 2.606 & \citep{deb01} & 27.80 &   \\
4C 26.38            & 12:32:23.6 & $+$26:04:07 & 2.609 & \citep{cha96} & 28.41 &  {\bf G}\citep{cha96b} \\
B3 0811+391         & 08:14:30.6 & $+$38:58:35 & 2.621 & \citep{owe95} & 27.84 &   \\
PKS 0742+10         & 07:45:33.0 & $+$10:11:13 & 2.624 & \citep{bes03b} & 28.74 &   \\
MRC 2025-218        & 20:27:59.5 & $-$21:40:57 & 2.63  & \citep{mcc90} & 28.74 & {\bf H}\citep{pen99,pen01} {\bf OS}\citep{vil07b}{\bf R}\citep{car97,ath97,ath98a,ath98b}  {\bf S}\citep{sey07} \\6C* 0122+426        & 01:25:52.8 & $+$42:51:51 & 2.635 & \citep{jar01b}& 28.04 &   \\
MRC 0140-257        & 01:42:41.2 & $-$25:30:34 & 2.64  & \citep{mcc96} & 28.44 & {\bf H}\citep{pen01}{\bf OS}\citep{vil07b}{\bf R}\citep{car97,ath97,ath98a,ath98b}{\bf S} \\
TN J1941-1952       & 19:41:00.1 & $-$19:52:14 & 2.667 & \citep{bor07} & 28.85 &   \\
PKS 1357+007        & 14:00:21.4 & $+$00:30:19 & 2.671 & \citep{rot97} & 28.72 & {\bf OS}\citep{van97}{\bf R}\citep{pen00b}  \\
NVSS J233034-330009 & 23:30:34.5 & $-$33:00:12 & 2.675 & \citep{bry07} & 28.23 &    \\
7C 1758+6719        & 17:58:47.2 & $+$67:19:46 & 2.70  & \citep{lac99} & 27.99 &   \\
TXS 2202+128        & 22:05:14.3 & $+$13:05:34 & 2.704 & \citep{rot97} & 28.54 & {\bf H}\citep{pen01}{\bf OA}\citep{wil04}{\bf OS}\citep{van97}{\bf R}\citep{car97}  {\bf S}\citep{sey07} \\
NVSS J094925-200518 & 09:49:25.8 & $-$20:05:18 & 2.706 & \citep{bro06} & 27.25$^{\dag}$ &   \\
6C**1102+4329       & 11:05:43.1 & $+$43:13:25 & 2.734 & \citep{cru06} & 28.17 &   \\
TXS 1545-234        & 15:48:17.5 & $-$23:37:02 & 2.754 & \citep{rot97} & 28.74 & {\bf OA}\citep{wil04} {\bf OS}\citep{van97} {\bf R}\citep{car97}  \\
NVSS J152435-352623 & 15:24:35.4 & $-$35:26:22 & 2.760 & \citep{bry07} & 28.48 &    \\
TN J0920-0712       & 09:20:22.3 & $-$07:12:18 & 2.762 & \citep{deb01} & 28.48 &   \\
MG 1019+0534        & 10:19:33.3 & $+$05:34:35 & 2.765 & \citep{ste99a}& 28.57 & {\bf M}\citep{arc01} {\bf R}\citep{pen00b} {\bf S}\citep{sey07} \\
NVSS J111921-363139 & 11:19:21.8 & $-$36:31:39 & 2.769 & \citep{bry07} & 29.39 &    \\
TXS 0417-181        & 04:19:43.6 & $-$18:01:56 & 2.770 & \citep{rot97} & 28.64 & {\bf R}\citep{car97}{\bf OS}\citep{van97}  \\
NVSS J213637-340318 & 21:36:38.4 & $-$34:03:21 & 2.770 & \citep{bry07} & 28.39 &    \\
WNR J1338+3532      & 13:38:15.1 & $+$35:32:04 & 2.772 & \citep{deb01} & 28.42 & {\bf R}\citep{pen00b}   \\
7C 1807+6719        & 18:07:13.4 & $+$67:19:42 & 2.78  & \citep{lac99} & 27.99 &   \\
4C +44.02           & 00:36:53.5 & $+$44:43:21 & 2.790 & \citep{fan01} & 28.98 &   \\
NVSS J140223-363539 & 14:02:23.6 & $-$36:35:42 & 2.804 & \citep{bry07} & 28.34 &    \\
WNH 1821+6117       & 18:22:18.8 & $+$61:18:37 & 2.81  & \citep{ren99} & 27.94 &   \\
4C +03.16           & 09:00:14.3 & $+$03:24:08 & 2.814 & \citep{rot97} & 28.48 &   \\
6C**0824+5344       & 08:27:58.9 & $+$53:34:15 & 2.824 & \citep{cru06} & 28.27 &   \\
NVSS J094552-201441 & 09:45:51.0 & $-$20:14:47 & 2.829 & \citep{bro06} & 27.12$^{\dag}$ &   \\
NVSS J095357-203652 & 09:53:57.4 & $-$20:36:51 & 2.829 & \citep{bro06} & 27.18$^{\dag}$ &   \\
\hline
\end{tabular}
}
\end{center}
{$^{\dag}$ Radio luminosity calculated assuming a spectral index $\alpha$ = -1.1.}\\
\end{sidewaystable}
\pagebreak

\begin{sidewaystable}
\begin{center}
{Table 3, continued}
\\*
\scriptsize{
\begin{tabular}{|p{78pt}|p{38pt}p{38pt}|p{15pt}p{15pt}|p{35pt}|p{245pt}|}
\hline
Name & RA(J2000) & DEC(J2000) & z & Ref. & Log($L_{500}$) & References\\
\hline
WNH 1717+6827       & 17:16:56.4 & $+$68:24:02 & 2.84  & \citep{ren99} & 27.98 &   \\
MG 0148+1028        & 01:48:28.9 & $+$10:28:21 & 2.845 & \citep{ste99a}& 28.56 &   \\
MRC 0052-241        & 00:54:29.8 & $-$23:51:30 & 2.86  & \citep{mcc96} & 28.77 & {\bf P}\citep{ven07}  \\
B2 1132+37          & 11:35:05.9 & $+$37:08:41 & 2.88  & \citep{eal96} & 28.71 &  {\bf R}\citep{pen00b}\\
4C 24.28            & 13:48:14.8 & $+$24:15:50 & 2.889 & \citep{cha96} & 29.05 & {\bf G}\citep{cha96b}{\bf H}\citep{pen99} {\bf M}\citep{arc01}{\bf R}\citep{car97,cai02} {\bf S}\citep{sey07} \\
TXS 0134+251        & 01:37:07.0 & $+$25:21:19 & 2.897 & \citep{deb01} & 28.57 &   \\
NVSS J213238-335318 & 21:32:39.0 & $-$33:53:19 & 2.900 & \citep{bry07} & 28.60 &    \\
4C 28.58            & 23:51:59.1 & $+$29:10:29 & 2.905 & \citep{cha96} & 28.91 & {\bf G}\citep{cha96b}{\bf R} \citep{cai02}  {\bf S}\citep{sey07}\\
MRC 0943-242        & 09:45:32.8 & $-$24:28:50 & 2.922 & \citep{mcc96}\par \citep{ver01a}& 28.62 &{\bf H}\citep{pen99,pen01}{\bf OA}\citep{bin00,jar03,bin06}  {\bf OP}\citep{ver01a}  {\bf OS}\citep{vil03,hum06,hum07}{\bf P}\citep{ven07,kod07} {\bf R}\citep{car97,ath98a}{\bf S}\citep{sey07} \\
B3 0744+464         & 07:47:43.7 & $+$46:18:58 & 2.926 & \citep{mcc91b}& 28.87 & {\bf R}\citep{car97} {\bf OS}\citep{eal93b}  \\
NVSS J151020-352803 & 15:10:20.8 & $-$35:28:03 & 2.937 & \citep{bry07} & 28.35 &    \\
TXS 0952-217        & 09:54:29.5 & $-$21:56:53 & 2.950 & \citep{bro06} & 28.08$^{\dag}$ &   \\
WNH 1758+5821       & 17:58:48.0 & $+$58:21:41 & 2.96  & \citep{ren99} & 27.98 &   \\
6C* 0020+440        & 00:23:08.5 & $+$44:48:23 & 2.988 & \citep{jar01b}& 28.49 &   \\
WN J0747+3654       & 07:47:29.4 & $+$36:54:38 & 2.992 & \citep{deb01} & 28.14 & {\bf M}\citep{reu04}{\bf S}\citep{sey07} \\
6C**0743+5019       & 07:58:06.0 & $+$50:11:03 & 2.996 & \citep{cru06} & 28.41 &   \\
WN J0231+3600       & 02:31:11.5 & $+$36:00:27 & 3.080 & \citep{deb01} & 28.20 & {\bf M}\citep{reu04} \\
WN J1053+5424       & 10:53:36.3 & $+$54:24:42 & 3.083 & \citep{bor07} & 28.55 &   \\
B3 J2330+3927       & 23:30:24.9 & $+$39:27:11 & 3.086 & \citep{deb03} & 28.33 & {\bf CO}\citep{deb03} {\bf HI}\citep{deb03}{\bf M}\citep{ste03,reu04} {\bf R}\citep{per05}{\bf S}\citep{sey07} \\
TN J1112-2948       & 11:12:23.9 & $-$29:48:06 & 3.09  & \citep{deb01} & 28.76 & {\bf M}\citep{reu04} \\
NVSS J095751-213321 & 09:57:51.3 & $-$21:33:21 & 3.126 & \citep{bro06} & 28.11$^{\dag}$ &   \\
MRC 0316-257        & 03:18:12.1 & $-$25:35:10 & 3.142 & \citep{mcc90} & 28.95 & {\bf H}\citep{pen01}{\bf M}\citep{reu04}{\bf OS}\citep{eal93b} {\bf P}\citep{lef96,ven05,kod07,ven07} {\bf R}\citep{car97,ath97,ath98a,ath98b}{\bf S}\citep{sey07}\\
WN J0617+5012       & 06:17:39.4 & $+$50:12:55 & 3.153 & \citep{deb01} & 28.02 & {\bf M}\citep{reu04} {\bf S}\citep{sey07} \\
MRC 0251-273        & 02:53:16.7 & $-$27:09:10 & 3.16  & \citep{mcc96} & 28.54 & {\bf M}\citep{reu04} {\bf S}\citep{sey07}\\
WN J1123+3141       & 11:23:55.9 & $+$31:41:26 & 3.216 & \citep{deb01} & 28.51 & {\bf M}\citep{reu04}{\bf S}\citep{sey07} \\
NVSS J230123-364656 & 23:01:23.5 & $-$36:46:56 & 3.220 & \citep{deb06} & 28.34 & {\bf M}\citep{reu04}  \\
6CE 1232+3942  	    & 12:35:04.8 & $+$39:25:39 & 3.221 & \citep{raw90} & 28.99 & {\bf M}\citep{chi94,arc01}{\bf R}\citep{car97} {\bf OS}\citep{eal93b}{\bf S}\citep{sey07} \\
WNH 1702+6042       & 17:03:36.2 & $+$60:38:52 & 3.223 & \citep{ren99} & 28.23 & {\bf M}\citep{reu04}  \\
NVSS J232100-360223 & 23:21:00.6 & $-$36:02:25 & 3.320 & \citep{deb06} & 28.22 &   \\
6C**0832+5443       & 08:36:09.9 & $+$54:33:26 & 3.341 & \citep{cru06} & 28.24 &   \\
B2 0902+34          & 09:05:30.1 & $+$34:07:57 & 3.395 & \citep{lil88} & 28.78 & {\bf H}\citep{pen99}{\bf HI}\citep{uso91,bri93,deb96,cod03,cha04} {\bf M}\citep{chi94,arc01} {\bf OS}\citep{eal93,eal93b,reu07}{\bf S}\citep{sey07} \\
NVSS J094724-210505 & 09:47:24.5 & $-$21:05:06 & 3.377 & \citep{bro06} & 27.43$^{\dag}$ &   \\
NVSS J095438-210425 & 09:54:38.4 & $-$21:04:25 & 3.431 & \citep{bro06} & 28.13$^{\dag}$ &   \\
\hline
\end{tabular}
}
\end{center}
{$^{\dag}$ Radio luminosity calculated assuming a spectral index $\alpha$ = -1.1.}\\
\end{sidewaystable}
\pagebreak

\begin{sidewaystable}
\begin{center}
{Table 3, continued}
\\*
\scriptsize{
\begin{tabular}{|p{78pt}|p{38pt}p{38pt}|p{15pt}p{15pt}|p{35pt}|p{245pt}|}
\hline
Name & RA(J2000) & DEC(J2000) & z & Ref. & Log($L_{500}$) & References\\
\hline
NVSS J231402-372925 & 23:14:02.4 & $-$37:29:27 & 3.450 & \citep{deb06} & 28.83 &   \\
TN J0205+2242       & 02:05:10.7 & $+$22:42:50 & 3.506 & \citep{deb01} & 28.46 & {\bf M}\citep{reu04} {\bf S}\citep{sey07}\\
TN J0121+1320       & 01:21:42.7 & $+$13:20:58 & 3.517 & \citep{deb01} & 28.49 & {\bf CO}\citep{deb03b} {\bf M}\citep{reu04}{\bf OA}\citep{wil04}{\bf S}\citep{sey07}\\
6C 1909+72          & 19:08:23.7 & $+$72:20:12 & 3.537 & \citep{deb01} & 29.12 & {\bf CO}\citep{pap00}{\bf M}\citep{ste03,reu04} {\bf R}\citep{pen00b} {\bf S}\citep{sey07}\\
4C 1243+036         & 12:45:38.4 & $+$03:23:20 & 3.560 & \citep{van96}\par \citep{ver01a}& 29.23 & {\bf H}\citep{pen99} {\bf M}\citep{arc01}{\bf OP}\citep{ver01a}{\bf OS}\citep{van96,hum07} {\bf S}\citep{sey07}\\
WN J1911+6342       & 19:11:49.5 & $+$63:42:10 & 3.590 & \citep{deb01} & 28.14 & {\bf M}\citep{reu04}{\bf S}\citep{sey07}  \\
MG 2141+192         & 21:44:07.5 & $+$19:29:15 & 3.592 & \citep{ste99a}& 29.08 & {\bf H}\citep{pen99}{\bf M}\citep{arc01,reu04}{\bf R}\citep{car97} {\bf S}\citep{sey07}   \\
6C* 0032+412        & 00:34:53.1 & $+$41:31:32 & 3.658 & \citep{jar01b}& 28.75 & {\bf M}\citep{arc01} {\bf S}\citep{sey07} \\
TN J1049-1258       & 10:49:06.2 & $-$12:58:19 & 3.697 & \citep{bor07} & 28.94 &   \\
4C 60.07            & 05:12:54.8 & $+$60:30:51 & 3.791 & \citep{cha96} & 29.20 & {\bf CO}\citep{pap00,gre04}{\bf G}\citep{cha96b}{\bf M}\citep{arc01,ste03,reu04}{\bf OS}\citep{reu07} {\bf S}\citep{sey07}\\
4C 41.17            & 06:50:52.2 & $+$41:30:31 & 3.792 & \citep{cha90}\par \citep{dey97}& 29.18 & {\bf CO}\citep{deb05a,pap05,eva96,ivi96}{\bf G}\citep{cha90,cha96b}{\bf H}\citep{bic00} {\bf M}\citep{dun94,chi94,arc01,ste03}{\bf N}\citep{dic95}{\bf OS}\citep{eal93b,dey97,reu07} {\bf P}\citep{ivi00,gre07} {\bf R}\citep{car97,gur97}{\bf S}\citep{sey07}{\bf X}\citep{sch03}\\
TN J2007-1316       & 20:07:53.2 & $-$13:16:44 & 3.837 & \citep{bor07} & 29.13 & {\bf M}\citep{reu04}{\bf S}\citep{sey07}  \\
NVSS J231727-352606 & 23:17:27.4 & $-$35:26:07 & 3.874 & \citep{deb06} & 28.71 &   \\
NVSS J021308-322338 & 02:13:08.0 & $-$32:23:40 & 3.976 & \citep{deb06} & 28.40 &   \\
TN J1338-1942       & 13:38:26.1 & $-$19:42:30 & 4.105 & \citep{deb99}\par \citep{wil04}& 28.70 &{\bf M}\citep{reu04} {\bf OA}\citep{wil04}{\bf OS}\citep{deb99} {\bf P}\citep{ven02,mil04,deb04b,int06,ven07,ove07}{\bf R}\citep{pen00b} {\bf S}\citep{sey07}\\
TN J1123-2154       & 11:23:10.2 & $-$21:54:05 & 4.109 & \citep{deb01} & 28.45 & {\bf M}\citep{reu04}  \\
8C 1435+63          & 14:36:39.0 & $+$63:19:04 & 4.251 & \citep{lac94} & 29.40 & {\bf G}\citep{lac94}{\bf M}\citep{arc01,ivi95,ste03}{\bf R}\citep{car97}{\bf S}\citep{sey07}  \\
6C 0140+326         & 01:43:43.8 & $+$32:53:49 & 4.413 & \citep{raw96}\par \citep{deb01}& 28.73 & {\bf M}\citep{arc01}{\bf S}\citep{sey07} \\
RC J0311+0507       & 03:11:48.0 & $+$05:08:03 & 4.514 & \citep{kop06} & 29.49 &   \\
TN J0924-2201       & 09:24:19.9 & $-$22:01:42 & 5.197 & \citep{van99}\par \citep{deb01}& 29.51 & {\bf CO}\citep{kla05}{\bf M}\citep{reu04} {\bf P}\citep{ven04,ove06,ven07}{\bf S}\citep{sey07}\\
\hline
\end{tabular}
}
\label{listhzrg}
\end{center}
{$^{\dag}$ Radio luminosity calculated assuming a spectral index $\alpha$ = -1.1.}\\
\end{sidewaystable}

\begin{table}
\tiny{
\begin{center}
{Table 4. HzRGs in order of right ascencion}
\\*
\end{center}
\begin{tabular}{|ll|ll|ll|}
\hline
Name & z & Name & z & Name & z \\
\hline
MG 0001+0846        & 2.360 & 5C 7.271            & 2.224 & MG 1401+0921        & 2.093  \\
PKS 0011-023        & 2.080 & TXS 0828+193        & 2.572 & NVSS J140223-363539 & 2.804  \\
MRC 0015-229        & 2.010 & 6C**0832+5443       & 3.341 & 4C -00.54           & 2.360  \\
6C* 0020+440        & 2.988 & 6C**0834+4129       & 2.442 & PKS 1425-148        & 2.355  \\
NVSS J002402-325253 & 2.043 & LBDS 65W172         & 2.551 & 8C 1435+63          & 4.251  \\
6C* 0024+356        & 2.161 & 4C 12.32            & 2.468 & TXS 1436+157        & 2.538  \\
4C 40.02            & 2.428 & 6C**0854+3500       & 2.382 & NVSS J144932-385657 & 2.152  \\
MRC 0030-219        & 2.168 & 4C 40.22            & 2.280 & NVSS J151020-352803 & 2.937  \\
6C* 0032+412        & 3.658 & 4C +03.16           & 2.814 & NVSS J152435-352623 & 2.760  \\
4C +44.02           & 2.790 & 6C 0901+54          & 2.249 & 8C 1536+620         & 2.450  \\
WN J0040+3857       & 2.606 & B2 0902+34          & 3.395 & TXS 1545-234        & 2.754  \\
MRC 0052-241        & 2.860 & TN J0920-0712       & 2.762 & TXS 1558-003        & 2.520  \\
6C* 0106+397        & 2.284 & TN J0924-2201       & 5.197 & TXS 1647+100        & 2.509  \\
6C* 0112+372        & 2.535 & 6C 0930+389         & 2.395 & WNH 1702+6042       & 3.223  \\
6C* 0115+394        & 2.241 & MG 0936+0503        & 2.306 & TXS 1707+105        & 2.349  \\
6C* 0118+486        & 2.350 & MRC 0943-242        & 2.922 & LBDS 53W002         & 2.390  \\
TN J0121+1320       & 3.517 & NVSS J094552-201441 & 2.829 & WNH 1717+6827       & 2.840  \\
6C* 0122+426        & 2.635 & NVSS J094724-210505 & 3.377 & WNH 1736+6504       & 2.600  \\
BRL 0128-264        & 2.348 & NVSS J094925-200518 & 2.706 & 7C 1736+650         & 2.400  \\
TXS 0134+251        & 2.897 & NVSS J094925-203724 & 2.427 & 7C 1740+6640        & 2.100  \\
6C* 0135+313        & 2.199 & NVSS J094949-213432 & 2.354 & MG 1747+1821        & 2.281  \\
MRC 0140-257        & 2.640 & NVSS J095226-200105 & 2.421 & MP J1755-6916       & 2.551  \\
6C 0140+326         & 4.413 & NVSS J095357-203652 & 2.829 & 7C 1758+6719        & 2.700  \\
6C* 0142+427        & 2.225 & TXS 0952-217        & 2.950 & WNH 1758+5821       & 2.960  \\
MG 0148+1028        & 2.845 & NVSS J095438-210425 & 3.431 & MP J1758-6738       & 2.026  \\
6C* 0152+463        & 2.279 & NVSS J095751-213321 & 3.126 & 7C 1802+6456        & 2.110  \\
MRC 0156-252        & 2.016 & NVSS J101008-383629 & 2.236 & 7C 1807+6719        & 2.780  \\
TXS 0200+015        & 2.229 & MG 1019+0534        & 2.765 & 4C 40.36            & 2.265  \\
TN J0205+2242       & 3.506 & TN J1033-1339       & 2.425 & WNH 1821+6117       & 2.810  \\
NVSS J021308-322338 & 3.976 & NVSS J103615-321659 & 2.144 & 6C 1909+72          & 3.537  \\
TXS 0211-122        & 2.340 & 8C 1039+681         & 2.530 & WN J1911+6342       & 3.590  \\
TXS 0214+183        & 2.131 & 6C**1045+4459       & 2.571 & 4C 48.48            & 2.343  \\
WN J0231+3600       & 3.080 & TN J1049-1258       & 3.697 & TN J1941-1952       & 2.667  \\
MRC 0251-273        & 3.160 & WN J1053+5424       & 3.083 & TN J2007-1316       & 3.837  \\
WN J0303+3733       & 2.505 & B2 1056+39          & 2.171 & MRC 2025-218        & 2.630  \\
RC J0311+0507       & 4.514 & TN J1102-1651       & 2.111 & TXS 2034+027        & 2.129  \\
MRC 0316-257        & 3.142 & 6C**1102+4329       & 2.734 & MRC 2036-254        & 2.000  \\
MP J0340-6507       & 2.289 & MRC 1106-258        & 2.430 & MRC 2048-272        & 2.060  \\
MRC 0349-211        & 2.329 & 6C 1106+380         & 2.290 & MRC 2052-253        & 2.600  \\
TXS 0355-037        & 2.153 & TN J1112-2948       & 3.090 & MRC 2104-242        & 2.491  \\
MRC 0406-244        & 2.440 & WN J1115+5016       & 2.540 & 4C 23.56            & 2.479  \\
TXS 0417-181        & 2.770 & TXS 1113-178        & 2.239 & NVSS J213238-335318 & 2.900  \\
TXS 0448+091        & 2.037 & 4C 34.34            & 2.400 & NVSS J213510-333703 & 2.518  \\
TN J0452-1737       & 2.260 & NVSS J111921-363139 & 2.769 & NVSS J213637-340318 & 2.770  \\
4C 60.07            & 3.791 & 3C 257              & 2.474 & MRC 2139-292        & 2.550  \\
PKS 0529-549        & 2.575 & TN J1123-2154       & 4.109 & MG 2141+192         & 3.592  \\
WN J0617+5012       & 3.153 & WN J1123+3141       & 3.216 & TXS 2202+128        & 2.704  \\
6C 0629+53          & 2.246 & B2 1132+37          & 2.880 & TNR 2254+1857       & 2.154  \\
4C 41.17            & 3.792 & 6C 1134+369         & 2.125 & NVSS J230123-364656 & 3.220  \\
B3 0731+438         & 2.429 & MRC 1138-262        & 2.156 & 4C +40.49           & 2.140  \\
PKS 0742+10         & 2.624 & B2 1159+39A         & 2.370 & MG 2308+0336        & 2.457  \\
WN J0747+3654       & 2.992 & B3 1204+401         & 2.066 & NVSS J231144-362215 & 2.531 \\
B3 0744+464         & 2.926 & 4C 26.38            & 2.609 & NVSS J231402-372925 & 3.450 \\
6C**0746+5445       & 2.156 & 6C 1232+3942        & 3.221 & NVSS J231727-352606 & 3.874 \\
TXS 0748+134        & 2.419 & WN J1242+3915       & 2.131 & NVSS J232100-360223 & 3.320 \\
6C**0743+5019       & 2.996 & 4C 1243+036         & 3.560 & TXS 2319+223        & 2.554 \\
4C +39.21           & 2.119 & MG 1251+1104        & 2.322 & NVSS J232651-370909 & 2.357 \\
5C 7.10             & 2.185 & MRC 1324-262        & 2.280 & B3 J2330+3927       & 3.086 \\
5C 7.15             & 2.433 & WNR J1338+3532      & 2.772 & NVSS J233034-330009 & 2.675 \\
B3 0811+391         & 2.621 & TN J1338-1942       & 4.105 & TXS 2332+154        & 2.481 \\
5C 7.223            & 2.092 & 4C 24.28            & 2.889 & MG 2348+0507        & 2.019 \\
6C**0824+5344       & 2.824 & 4C +43.31           & 2.149 & 4C 28.58            & 2.905 \\
5C 7.269            & 2.218 & PKS 1357+007        & 2.671 & TXS 2353-003        & 2.592 \\
\hline
\end{tabular}
}
\end{table}

\clearpage
\newpage
\bibliographystyle{spbasic}
\bibliography{mileydebreuck}   

\end{document}